\documentclass[longauth,twocolumn,traditabstract]{aa}

\def\setsymbol#1#2{\expandafter\def\csname #1\endcsname{#2}}
\def\getsymbol#1{\csname #1\endcsname}

\def\Planck{\textit{Planck}}

\def\allearlypapers{\nocite{planck2011-1.1, planck2011-1.3, planck2011-1.4, planck2011-1.5, planck2011-1.6, planck2011-1.7, planck2011-1.10, planck2011-1.10sup, planck2011-5.1a, planck2011-5.1b, planck2011-5.2a, planck2011-5.2b, planck2011-5.2c, planck2011-6.1, planck2011-6.2, planck2011-6.3a, planck2011-6.4a, planck2011-6.4b, planck2011-6.6, planck2011-7.0, planck2011-7.2, planck2011-7.3, planck2011-7.7a, planck2011-7.7b, planck2011-7.12, planck2011-7.13}}

\def\all2013resultspapers{\nocite{planck2013-p01, planck2013-p02, planck2013-p02a, planck2013-p02d, planck2013-p02b, planck2013-p03, planck2013-p03c, planck2013-p03f, planck2013-p03d, planck2013-p03e, planck2013-p01a, planck2013-p06, planck2013-p03a, planck2013-pip88, planck2013-p08, planck2013-p11, planck2013-p12, planck2013-p13, planck2013-p14, planck2013-p15, planck2013-p05b, planck2013-p17, planck2013-p09, planck2013-p09a, planck2013-p20, planck2013-p19, planck2013-pipaberration, planck2013-p05, planck2013-p05a, planck2013-pip56, planck2013-p06b}}

\newbox\tablebox    \newdimen\tablewidth
\def\leaderfil{\leaders\hbox to 5pt{\hss.\hss}\hfil}
\def\tablenote#1 #2\par{\begingroup \parindent=0.8em
    \abovedisplayshortskip=0pt\belowdisplayshortskip=0pt
    \noindent
    $$\hss\vbox{\hsize\tablewidth \hangindent=\parindent \hangafter=1 \noindent
    \hbox to \parindent{$^#1$\hss}\strut#2\strut\par}\hss$$
    \endgroup}

\def\L2{\ifmmode L_2\else $L_2$\fi}

\def\DeltaT{\ifmmode \Delta T\else $\Delta T$\fi}
\def\deltat{\ifmmode \Delta t\else $\Delta t$\fi}
\def\fknee{\ifmmode f_{\rm knee}\else $f_{\rm knee}$\fi}
\def\Fmax{\ifmmode F_{\rm max}\else $F_{\rm max}$\fi}
\def\solar{\ifmmode{\rm M}_{\mathord\odot}\else${\rm M}_{\mathord\odot}$\fi}
\def\Msolar{\ifmmode{\rm M}_{\mathord\odot}\else${\rm M}_{\mathord\odot}$\fi}
\def\Lsolar{\ifmmode{\rm L}_{\mathord\odot}\else${\rm L}_{\mathord\odot}$\fi}

\def\inv{\ifmmode^{-1}\else$^{-1}$\fi}
\def\mo{\ifmmode^{-1}\else$^{-1}$\fi}
\def\sup#1{\ifmmode ^{\rm #1}\else $^{\rm #1}$\fi}
\def\expo#1{\ifmmode \times 10^{#1}\else $\times 10^{#1}$\fi}
\def\,{\thinspace}
\def\lsim{\mathrel{\raise .4ex\hbox{\rlap{$<$}\lower 1.2ex\hbox{$\sim$}}}}
\def\gsim{\mathrel{\raise .4ex\hbox{\rlap{$>$}\lower 1.2ex\hbox{$\sim$}}}}

\def\simprop{\mathrel{\raise .4ex\hbox{\rlap{$\propto$}\lower 1.2ex\hbox{$\sim$}}}}
\def\deg{\ifmmode^\circ\else$^\circ$\fi}
\def\pdeg{\ifmmode $\setbox0=\hbox{$^{\circ}$}\rlap{\hskip.11\wd0 .}$^{\circ}
          \else \setbox0=\hbox{$^{\circ}$}\rlap{\hskip.11\wd0 .}$^{\circ}$\fi}
\def\arcs{\ifmmode {^{\scriptstyle\prime\prime}}
          \else $^{\scriptstyle\prime\prime}$\fi}
\def\arcm{\ifmmode {^{\scriptstyle\prime}}
          \else $^{\scriptstyle\prime}$\fi}
\newdimen\sa  \newdimen\sb
\def\parcs{\sa=.07em \sb=.03em
     \ifmmode \hbox{\rlap{.}}^{\scriptstyle\prime\kern -\sb\prime}\hbox{\kern -\sa}
     \else \rlap{.}$^{\scriptstyle\prime\kern -\sb\prime}$\kern -\sa\fi}
\def\parcm{\sa=.08em \sb=.03em
     \ifmmode \hbox{\rlap{.}\kern\sa}^{\scriptstyle\prime}\hbox{\kern-\sb}
     \else \rlap{.}\kern\sa$^{\scriptstyle\prime}$\kern-\sb\fi}
\def\ra[#1 #2 #3.#4]{#1\sup{h}#2\sup{m}#3\sup{s}\llap.#4}
\def\dec[#1 #2 #3.#4]{#1\deg#2\arcm#3\arcs\llap.#4}
\def\deco[#1 #2 #3]{#1\deg#2\arcm#3\arcs}
\def\rra[#1 #2]{#1\sup{h}#2\sup{m}}

\def\dots{\relax\ifmmode \ldots\else $\ldots$\fi}
\def\WHzsr{\ifmmode $W\,Hz\mo\,sr\mo$\else W\,Hz\mo\,sr\mo\fi}
\def\mHz{\ifmmode $\,mHz$\else \,mHz\fi}
\def\GHz{\ifmmode $\,GHz$\else \,GHz\fi}
\def\mKs{\ifmmode $\,mK\,s$^{1/2}\else \,mK\,s$^{1/2}$\fi}
\def\muKs{\ifmmode \,\mu$K\,s$^{1/2}\else \,$\mu$K\,s$^{1/2}$\fi}
\def\muKRJs{\ifmmode \,\mu$K$_{\rm RJ}$\,s$^{1/2}\else \,$\mu$K$_{\rm RJ}$\,s$^{1/2}$\fi}
\def\muKHz{\ifmmode \,\mu$K\,Hz$^{-1/2}\else \,$\mu$K\,Hz$^{-1/2}$\fi}
\def\MJysr{\ifmmode \,$MJy\,sr\mo$\else \,MJy\,sr\mo\fi}
\def\MJysrmK{\ifmmode \,$MJy\,sr\mo$\,mK$_{\rm CMB}\mo\else \,MJy\,sr\mo\,mK$_{\rm CMB}\mo$\fi}
\def\microns{\ifmmode \,\mu$m$\else \,$\mu$m\fi}

\def\muK{\ifmmode \,\mu$K$\else \,$\mu$\hbox{K}\fi}
\def\microK{\ifmmode \,\mu$K$\else \,$\mu$\hbox{K}\fi}
\def\muW{\ifmmode \,\mu$W$\else \,$\mu$\hbox{W}\fi}
\def\kms{\ifmmode $\,km\,s$^{-1}\else \,km\,s$^{-1}$\fi}
\def\kmsMpc{\ifmmode $\,\kms\,Mpc\mo$\else \,\kms\,Mpc\mo\fi}
\setsymbol{LFI:center:frequency:70GHz:units}{70.3\,GHz}
\setsymbol{LFI:center:frequency:44GHz:units}{44.1\,GHz}
\setsymbol{LFI:center:frequency:30GHz:units}{28.5\,GHz}

\setsymbol{LFI:center:frequency:70GHz}{70.3}
\setsymbol{LFI:center:frequency:44GHz}{44.1}
\setsymbol{LFI:center:frequency:30GHz}{28.5}

\setsymbol{LFI:center:frequency:LFI18:Rad:M:units}{71.7\GHz}
\setsymbol{LFI:center:frequency:LFI19:Rad:M:units}{67.5\GHz}
\setsymbol{LFI:center:frequency:LFI20:Rad:M:units}{69.2\GHz}
\setsymbol{LFI:center:frequency:LFI21:Rad:M:units}{70.4\GHz}
\setsymbol{LFI:center:frequency:LFI22:Rad:M:units}{71.5\GHz}
\setsymbol{LFI:center:frequency:LFI23:Rad:M:units}{70.8\GHz}
\setsymbol{LFI:center:frequency:LFI24:Rad:M:units}{44.4\GHz}
\setsymbol{LFI:center:frequency:LFI25:Rad:M:units}{44.0\GHz}
\setsymbol{LFI:center:frequency:LFI26:Rad:M:units}{43.9\GHz}
\setsymbol{LFI:center:frequency:LFI27:Rad:M:units}{28.3\GHz}
\setsymbol{LFI:center:frequency:LFI28:Rad:M:units}{28.8\GHz}
\setsymbol{LFI:center:frequency:LFI18:Rad:S:units}{70.1\GHz}
\setsymbol{LFI:center:frequency:LFI19:Rad:S:units}{69.6\GHz}
\setsymbol{LFI:center:frequency:LFI20:Rad:S:units}{69.5\GHz}
\setsymbol{LFI:center:frequency:LFI21:Rad:S:units}{69.5\GHz}
\setsymbol{LFI:center:frequency:LFI22:Rad:S:units}{72.8\GHz}
\setsymbol{LFI:center:frequency:LFI23:Rad:S:units}{71.3\GHz}
\setsymbol{LFI:center:frequency:LFI24:Rad:S:units}{44.1\GHz}
\setsymbol{LFI:center:frequency:LFI25:Rad:S:units}{44.1\GHz}
\setsymbol{LFI:center:frequency:LFI26:Rad:S:units}{44.1\GHz}
\setsymbol{LFI:center:frequency:LFI27:Rad:S:units}{28.5\GHz}
\setsymbol{LFI:center:frequency:LFI28:Rad:S:units}{28.2\GHz}

\setsymbol{LFI:center:frequency:LFI18:Rad:M}{71.7}
\setsymbol{LFI:center:frequency:LFI19:Rad:M}{67.5}
\setsymbol{LFI:center:frequency:LFI20:Rad:M}{69.2}
\setsymbol{LFI:center:frequency:LFI21:Rad:M}{70.4}
\setsymbol{LFI:center:frequency:LFI22:Rad:M}{71.5}
\setsymbol{LFI:center:frequency:LFI23:Rad:M}{70.8}
\setsymbol{LFI:center:frequency:LFI24:Rad:M}{44.4}
\setsymbol{LFI:center:frequency:LFI25:Rad:M}{44.0}
\setsymbol{LFI:center:frequency:LFI26:Rad:M}{43.9}
\setsymbol{LFI:center:frequency:LFI27:Rad:M}{28.3}
\setsymbol{LFI:center:frequency:LFI28:Rad:M}{28.8}
\setsymbol{LFI:center:frequency:LFI18:Rad:S}{70.1}
\setsymbol{LFI:center:frequency:LFI19:Rad:S}{69.6}
\setsymbol{LFI:center:frequency:LFI20:Rad:S}{69.5}
\setsymbol{LFI:center:frequency:LFI21:Rad:S}{69.5}
\setsymbol{LFI:center:frequency:LFI22:Rad:S}{72.8}
\setsymbol{LFI:center:frequency:LFI23:Rad:S}{71.3}
\setsymbol{LFI:center:frequency:LFI24:Rad:S}{44.1}
\setsymbol{LFI:center:frequency:LFI25:Rad:S}{44.1}
\setsymbol{LFI:center:frequency:LFI26:Rad:S}{44.1}
\setsymbol{LFI:center:frequency:LFI27:Rad:S}{28.5}
\setsymbol{LFI:center:frequency:LFI28:Rad:S}{28.2}

\setsymbol{LFI:white:noise:sensitivity:70GHz:units}{134.7\muKs}
\setsymbol{LFI:white:noise:sensitivity:44GHz:units}{164.7\muKs}
\setsymbol{LFI:white:noise:sensitivity:30GHz:units}{143.4\muKs}

\setsymbol{LFI:white:noise:sensitivity:70GHz}{134.7}
\setsymbol{LFI:white:noise:sensitivity:44GHz}{164.7}
\setsymbol{LFI:white:noise:sensitivity:30GHz}{143.4}

\setsymbol{LFI:white:noise:sensitivity:LFI18:Rad:M:units}{512.0\muKs}
\setsymbol{LFI:white:noise:sensitivity:LFI19:Rad:M:units}{581.4\muKs}
\setsymbol{LFI:white:noise:sensitivity:LFI20:Rad:M:units}{590.8\muKs}
\setsymbol{LFI:white:noise:sensitivity:LFI21:Rad:M:units}{455.2\muKs}
\setsymbol{LFI:white:noise:sensitivity:LFI22:Rad:M:units}{492.0\muKs}
\setsymbol{LFI:white:noise:sensitivity:LFI23:Rad:M:units}{507.7\muKs}
\setsymbol{LFI:white:noise:sensitivity:LFI24:Rad:M:units}{462.2\muKs}
\setsymbol{LFI:white:noise:sensitivity:LFI25:Rad:M:units}{413.6\muKs}
\setsymbol{LFI:white:noise:sensitivity:LFI26:Rad:M:units}{478.6\muKs}
\setsymbol{LFI:white:noise:sensitivity:LFI27:Rad:M:units}{277.7\muKs}
\setsymbol{LFI:white:noise:sensitivity:LFI28:Rad:M:units}{312.3\muKs}
\setsymbol{LFI:white:noise:sensitivity:LFI18:Rad:S:units}{465.7\muKs}
\setsymbol{LFI:white:noise:sensitivity:LFI19:Rad:S:units}{555.6\muKs}
\setsymbol{LFI:white:noise:sensitivity:LFI20:Rad:S:units}{623.2\muKs}
\setsymbol{LFI:white:noise:sensitivity:LFI21:Rad:S:units}{564.1\muKs}
\setsymbol{LFI:white:noise:sensitivity:LFI22:Rad:S:units}{534.4\muKs}
\setsymbol{LFI:white:noise:sensitivity:LFI23:Rad:S:units}{542.4\muKs}
\setsymbol{LFI:white:noise:sensitivity:LFI24:Rad:S:units}{399.2\muKs}
\setsymbol{LFI:white:noise:sensitivity:LFI25:Rad:S:units}{392.6\muKs}
\setsymbol{LFI:white:noise:sensitivity:LFI26:Rad:S:units}{418.6\muKs}
\setsymbol{LFI:white:noise:sensitivity:LFI27:Rad:S:units}{302.9\muKs}
\setsymbol{LFI:white:noise:sensitivity:LFI28:Rad:S:units}{285.3\muKs}

\setsymbol{LFI:white:noise:sensitivity:LFI18:Rad:M}{512.0}
\setsymbol{LFI:white:noise:sensitivity:LFI19:Rad:M}{581.4}
\setsymbol{LFI:white:noise:sensitivity:LFI20:Rad:M}{590.8}
\setsymbol{LFI:white:noise:sensitivity:LFI21:Rad:M}{455.2}
\setsymbol{LFI:white:noise:sensitivity:LFI22:Rad:M}{492.0}
\setsymbol{LFI:white:noise:sensitivity:LFI23:Rad:M}{507.7}
\setsymbol{LFI:white:noise:sensitivity:LFI24:Rad:M}{462.2}
\setsymbol{LFI:white:noise:sensitivity:LFI25:Rad:M}{413.6}
\setsymbol{LFI:white:noise:sensitivity:LFI26:Rad:M}{478.6}
\setsymbol{LFI:white:noise:sensitivity:LFI27:Rad:M}{277.7}
\setsymbol{LFI:white:noise:sensitivity:LFI28:Rad:M}{312.3}
\setsymbol{LFI:white:noise:sensitivity:LFI18:Rad:S}{465.7}
\setsymbol{LFI:white:noise:sensitivity:LFI19:Rad:S}{555.6}
\setsymbol{LFI:white:noise:sensitivity:LFI20:Rad:S}{623.2}
\setsymbol{LFI:white:noise:sensitivity:LFI21:Rad:S}{564.1}
\setsymbol{LFI:white:noise:sensitivity:LFI22:Rad:S}{534.4}
\setsymbol{LFI:white:noise:sensitivity:LFI23:Rad:S}{542.4}
\setsymbol{LFI:white:noise:sensitivity:LFI24:Rad:S}{399.2}
\setsymbol{LFI:white:noise:sensitivity:LFI25:Rad:S}{392.6}
\setsymbol{LFI:white:noise:sensitivity:LFI26:Rad:S}{418.6}
\setsymbol{LFI:white:noise:sensitivity:LFI27:Rad:S}{302.9}
\setsymbol{LFI:white:noise:sensitivity:LFI28:Rad:S}{285.3}

\setsymbol{LFI:knee:frequency:70GHz:units}{29.5\mHz}
\setsymbol{LFI:knee:frequency:44GHz:units}{56.2\mHz}
\setsymbol{LFI:knee:frequency:30GHz:units}{113.7\mHz}

\setsymbol{LFI:knee:frequency:70GHz}{29.5}
\setsymbol{LFI:knee:frequency:44GHz}{56.2}
\setsymbol{LFI:knee:frequency:30GHz}{113.7}

\setsymbol{LFI:knee:frequency:LFI18:Rad:M:units}{16.3\mHz}
\setsymbol{LFI:knee:frequency:LFI19:Rad:M:units}{15.1\mHz}
\setsymbol{LFI:knee:frequency:LFI20:Rad:M:units}{18.7\mHz}
\setsymbol{LFI:knee:frequency:LFI21:Rad:M:units}{37.2\mHz}
\setsymbol{LFI:knee:frequency:LFI22:Rad:M:units}{12.7\mHz}
\setsymbol{LFI:knee:frequency:LFI23:Rad:M:units}{34.6\mHz}
\setsymbol{LFI:knee:frequency:LFI24:Rad:M:units}{46.2\mHz}
\setsymbol{LFI:knee:frequency:LFI25:Rad:M:units}{24.9\mHz}
\setsymbol{LFI:knee:frequency:LFI26:Rad:M:units}{67.6\mHz}
\setsymbol{LFI:knee:frequency:LFI27:Rad:M:units}{187.4\mHz}
\setsymbol{LFI:knee:frequency:LFI28:Rad:M:units}{122.2\mHz}
\setsymbol{LFI:knee:frequency:LFI18:Rad:S:units}{17.7\mHz}
\setsymbol{LFI:knee:frequency:LFI19:Rad:S:units}{22.0\mHz}
\setsymbol{LFI:knee:frequency:LFI20:Rad:S:units}{8.7\mHz}
\setsymbol{LFI:knee:frequency:LFI21:Rad:S:units}{25.9\mHz}
\setsymbol{LFI:knee:frequency:LFI22:Rad:S:units}{15.8\mHz}
\setsymbol{LFI:knee:frequency:LFI23:Rad:S:units}{129.8\mHz}
\setsymbol{LFI:knee:frequency:LFI24:Rad:S:units}{100.9\mHz}
\setsymbol{LFI:knee:frequency:LFI25:Rad:S:units}{38.9\mHz}
\setsymbol{LFI:knee:frequency:LFI26:Rad:S:units}{58.9\mHz}
\setsymbol{LFI:knee:frequency:LFI27:Rad:S:units}{104.4\mHz}
\setsymbol{LFI:knee:frequency:LFI28:Rad:S:units}{40.7\mHz}

\setsymbol{LFI:knee:frequency:LFI18:Rad:M}{16.3}
\setsymbol{LFI:knee:frequency:LFI19:Rad:M}{15.1}
\setsymbol{LFI:knee:frequency:LFI20:Rad:M}{18.7}
\setsymbol{LFI:knee:frequency:LFI21:Rad:M}{37.2}
\setsymbol{LFI:knee:frequency:LFI22:Rad:M}{12.7}
\setsymbol{LFI:knee:frequency:LFI23:Rad:M}{34.6}
\setsymbol{LFI:knee:frequency:LFI24:Rad:M}{46.2}
\setsymbol{LFI:knee:frequency:LFI25:Rad:M}{24.9}
\setsymbol{LFI:knee:frequency:LFI26:Rad:M}{67.6}
\setsymbol{LFI:knee:frequency:LFI27:Rad:M}{187.4}
\setsymbol{LFI:knee:frequency:LFI28:Rad:M}{122.2}
\setsymbol{LFI:knee:frequency:LFI18:Rad:S}{17.7}
\setsymbol{LFI:knee:frequency:LFI19:Rad:S}{22.0}
\setsymbol{LFI:knee:frequency:LFI20:Rad:S}{8.7}
\setsymbol{LFI:knee:frequency:LFI21:Rad:S}{25.9}
\setsymbol{LFI:knee:frequency:LFI22:Rad:S}{15.8}
\setsymbol{LFI:knee:frequency:LFI23:Rad:S}{129.8}
\setsymbol{LFI:knee:frequency:LFI24:Rad:S}{100.9}
\setsymbol{LFI:knee:frequency:LFI25:Rad:S}{38.9}
\setsymbol{LFI:knee:frequency:LFI26:Rad:S}{58.9}
\setsymbol{LFI:knee:frequency:LFI27:Rad:S}{104.4}
\setsymbol{LFI:knee:frequency:LFI28:Rad:S}{40.7}

\setsymbol{LFI:slope:70GHz:units}{$-1.03$\mHz}
\setsymbol{LFI:slope:44GHz:units}{$-0.89$\mHz}
\setsymbol{LFI:slope:30GHz:units}{$-0.87$\mHz}

\setsymbol{LFI:slope:70GHz}{$-1.03$}
\setsymbol{LFI:slope:44GHz}{$-0.89$}
\setsymbol{LFI:slope:30GHz}{$-0.87$}

\setsymbol{LFI:slope:LFI18:Rad:M:units}{$-1.04$\mHz}
\setsymbol{LFI:slope:LFI19:Rad:M:units}{$-1.09$\mHz}
\setsymbol{LFI:slope:LFI20:Rad:M:units}{$-0.69$\mHz}
\setsymbol{LFI:slope:LFI21:Rad:M:units}{$-1.56$\mHz}
\setsymbol{LFI:slope:LFI22:Rad:M:units}{$-1.01$\mHz}
\setsymbol{LFI:slope:LFI23:Rad:M:units}{$-0.96$\mHz}
\setsymbol{LFI:slope:LFI24:Rad:M:units}{$-0.83$\mHz}
\setsymbol{LFI:slope:LFI25:Rad:M:units}{$-0.91$\mHz}
\setsymbol{LFI:slope:LFI26:Rad:M:units}{$-0.95$\mHz}
\setsymbol{LFI:slope:LFI27:Rad:M:units}{$-0.87$\mHz}
\setsymbol{LFI:slope:LFI28:Rad:M:units}{$-0.88$\mHz}
\setsymbol{LFI:slope:LFI18:Rad:S:units}{$-1.15$\mHz}
\setsymbol{LFI:slope:LFI19:Rad:S:units}{$-1.00$\mHz}
\setsymbol{LFI:slope:LFI20:Rad:S:units}{$-0.95$\mHz}
\setsymbol{LFI:slope:LFI21:Rad:S:units}{$-0.92$\mHz}
\setsymbol{LFI:slope:LFI22:Rad:S:units}{$-1.01$\mHz}
\setsymbol{LFI:slope:LFI23:Rad:S:units}{$-0.95$\mHz}
\setsymbol{LFI:slope:LFI24:Rad:S:units}{$-0.73$\mHz}
\setsymbol{LFI:slope:LFI25:Rad:S:units}{$-1.16$\mHz}
\setsymbol{LFI:slope:LFI26:Rad:S:units}{$-0.79$\mHz}
\setsymbol{LFI:slope:LFI27:Rad:S:units}{$-0.82$\mHz}
\setsymbol{LFI:slope:LFI28:Rad:S:units}{$-0.91$\mHz}

\setsymbol{LFI:slope:LFI18:Rad:M}{$-1.04$}
\setsymbol{LFI:slope:LFI19:Rad:M}{$-1.09$}
\setsymbol{LFI:slope:LFI20:Rad:M}{$-0.69$}
\setsymbol{LFI:slope:LFI21:Rad:M}{$-1.56$}
\setsymbol{LFI:slope:LFI22:Rad:M}{$-1.01$}
\setsymbol{LFI:slope:LFI23:Rad:M}{$-0.96$}
\setsymbol{LFI:slope:LFI24:Rad:M}{$-0.83$}
\setsymbol{LFI:slope:LFI25:Rad:M}{$-0.91$}
\setsymbol{LFI:slope:LFI26:Rad:M}{$-0.95$}
\setsymbol{LFI:slope:LFI27:Rad:M}{$-0.87$}
\setsymbol{LFI:slope:LFI28:Rad:M}{$-0.88$}
\setsymbol{LFI:slope:LFI18:Rad:S}{$-1.15$}
\setsymbol{LFI:slope:LFI19:Rad:S}{$-1.00$}
\setsymbol{LFI:slope:LFI20:Rad:S}{$-0.95$}
\setsymbol{LFI:slope:LFI21:Rad:S}{$-0.92$}
\setsymbol{LFI:slope:LFI22:Rad:S}{$-1.01$}
\setsymbol{LFI:slope:LFI23:Rad:S}{$-0.95$}
\setsymbol{LFI:slope:LFI24:Rad:S}{$-0.73$}
\setsymbol{LFI:slope:LFI25:Rad:S}{$-1.16$}
\setsymbol{LFI:slope:LFI26:Rad:S}{$-0.79$}
\setsymbol{LFI:slope:LFI27:Rad:S}{$-0.82$}
\setsymbol{LFI:slope:LFI28:Rad:S}{$-0.91$}

\setsymbol{LFI:FWHM:70GHz:units}{13\parcm01}
\setsymbol{LFI:FWHM:44GHz:units}{27\parcm92}
\setsymbol{LFI:FWHM:30GHz:units}{32\parcm65}

\setsymbol{LFI:FWHM:70GHz}{13.01}
\setsymbol{LFI:FWHM:44GHz}{27.92}
\setsymbol{LFI:FWHM:30GHz}{32.65}

\setsymbol{LFI:FWHM:LFI18:units}{13\parcm39}
\setsymbol{LFI:FWHM:LFI19:units}{13\parcm01}
\setsymbol{LFI:FWHM:LFI20:units}{12\parcm75}
\setsymbol{LFI:FWHM:LFI21:units}{12\parcm74}
\setsymbol{LFI:FWHM:LFI22:units}{12\parcm87}
\setsymbol{LFI:FWHM:LFI23:units}{13\parcm27}
\setsymbol{LFI:FWHM:LFI24:units}{22\parcm98}
\setsymbol{LFI:FWHM:LFI25:units}{30\parcm46}
\setsymbol{LFI:FWHM:LFI26:units}{30\parcm31}
\setsymbol{LFI:FWHM:LFI27:units}{32\parcm65}
\setsymbol{LFI:FWHM:LFI28:units}{32\parcm66}

\setsymbol{LFI:FWHM:LFI18}{13.39}
\setsymbol{LFI:FWHM:LFI19}{13.01}
\setsymbol{LFI:FWHM:LFI20}{12.75}
\setsymbol{LFI:FWHM:LFI21}{12.74}
\setsymbol{LFI:FWHM:LFI22}{12.87}
\setsymbol{LFI:FWHM:LFI23}{13.27}
\setsymbol{LFI:FWHM:LFI24}{22.98}
\setsymbol{LFI:FWHM:LFI25}{30.46}
\setsymbol{LFI:FWHM:LFI26}{30.31}
\setsymbol{LFI:FWHM:LFI27}{32.65}
\setsymbol{LFI:FWHM:LFI28}{32.66}

\setsymbol{LFI:FWHM:uncertainty:LFI18:units}{0.170\arcm}
\setsymbol{LFI:FWHM:uncertainty:LFI19:units}{0.174\arcm}
\setsymbol{LFI:FWHM:uncertainty:LFI20:units}{0.170\arcm}
\setsymbol{LFI:FWHM:uncertainty:LFI21:units}{0.156\arcm}
\setsymbol{LFI:FWHM:uncertainty:LFI22:units}{0.164\arcm}
\setsymbol{LFI:FWHM:uncertainty:LFI23:units}{0.171\arcm}
\setsymbol{LFI:FWHM:uncertainty:LFI24:units}{0.652\arcm}
\setsymbol{LFI:FWHM:uncertainty:LFI25:units}{1.075\arcm}
\setsymbol{LFI:FWHM:uncertainty:LFI26:units}{1.131\arcm}
\setsymbol{LFI:FWHM:uncertainty:LFI27:units}{1.266\arcm}
\setsymbol{LFI:FWHM:uncertainty:LFI28:units}{1.287\arcm}

\setsymbol{LFI:FWHM:uncertainty:LFI18}{0.170}
\setsymbol{LFI:FWHM:uncertainty:LFI19}{0.174}
\setsymbol{LFI:FWHM:uncertainty:LFI20}{0.170}
\setsymbol{LFI:FWHM:uncertainty:LFI21}{0.156}
\setsymbol{LFI:FWHM:uncertainty:LFI22}{0.164}
\setsymbol{LFI:FWHM:uncertainty:LFI23}{0.171}
\setsymbol{LFI:FWHM:uncertainty:LFI24}{0.652}
\setsymbol{LFI:FWHM:uncertainty:LFI25}{1.075}
\setsymbol{LFI:FWHM:uncertainty:LFI26}{1.131}
\setsymbol{LFI:FWHM:uncertainty:LFI27}{1.266}
\setsymbol{LFI:FWHM:uncertainty:LFI28}{1.287}

\setsymbol{HFI:center:frequency:100GHz:units}{100\,GHz}
\setsymbol{HFI:center:frequency:143GHz:units}{143\,GHz}
\setsymbol{HFI:center:frequency:217GHz:units}{217\,GHz}
\setsymbol{HFI:center:frequency:353GHz:units}{353\,GHz}
\setsymbol{HFI:center:frequency:545GHz:units}{545\,GHz}
\setsymbol{HFI:center:frequency:857GHz:units}{857\,GHz}

\setsymbol{HFI:center:frequency:100GHz}{100}
\setsymbol{HFI:center:frequency:143GHz}{143}
\setsymbol{HFI:center:frequency:217GHz}{217}
\setsymbol{HFI:center:frequency:353GHz}{353}
\setsymbol{HFI:center:frequency:545GHz}{545}
\setsymbol{HFI:center:frequency:857GHz}{857}

\setsymbol{HFI:Ndetectors:100GHz}{8}
\setsymbol{HFI:Ndetectors:143GHz}{11}
\setsymbol{HFI:Ndetectors:217GHz}{12}
\setsymbol{HFI:Ndetectors:353GHz}{12}
\setsymbol{HFI:Ndetectors:545GHz}{3}
\setsymbol{HFI:Ndetectors:857GHz}{4}

\setsymbol{HFI:FWHM:Maps:100GHz:units}{9\parcm88}
\setsymbol{HFI:FWHM:Maps:143GHz:units}{7\parcm18}
\setsymbol{HFI:FWHM:Maps:217GHz:units}{4\parcm87}
\setsymbol{HFI:FWHM:Maps:353GHz:units}{4\parcm65}
\setsymbol{HFI:FWHM:Maps:545GHz:units}{4\parcm72}
\setsymbol{HFI:FWHM:Maps:857GHz:units}{4\parcm39}
\setsymbol{HFI:FWHM:Maps:100GHz}{9.88}
\setsymbol{HFI:FWHM:Maps:143GHz}{7.18}
\setsymbol{HFI:FWHM:Maps:217GHz}{4.87}
\setsymbol{HFI:FWHM:Maps:353GHz}{4.65}
\setsymbol{HFI:FWHM:Maps:545GHz}{4.72}
\setsymbol{HFI:FWHM:Maps:857GHz}{4.39}

\setsymbol{HFI:beam:ellipticity:Maps:100GHz}{1.15}
\setsymbol{HFI:beam:ellipticity:Maps:143GHz}{1.01}
\setsymbol{HFI:beam:ellipticity:Maps:217GHz}{1.06}
\setsymbol{HFI:beam:ellipticity:Maps:353GHz}{1.05}
\setsymbol{HFI:beam:ellipticity:Maps:545GHz}{1.14}
\setsymbol{HFI:beam:ellipticity:Maps:857GHz}{1.19}

\setsymbol{HFI:FWHM:Mars:100GHz:units}{9\parcm37}
\setsymbol{HFI:FWHM:Mars:143GHz:units}{7\parcm04}
\setsymbol{HFI:FWHM:Mars:217GHz:units}{4\parcm68}
\setsymbol{HFI:FWHM:Mars:353GHz:units}{4\parcm43}
\setsymbol{HFI:FWHM:Mars:545GHz:units}{3\parcm80}
\setsymbol{HFI:FWHM:Mars:857GHz:units}{3\parcm67}

\setsymbol{HFI:FWHM:Mars:100GHz}{9.37}
\setsymbol{HFI:FWHM:Mars:143GHz}{7.04}
\setsymbol{HFI:FWHM:Mars:217GHz}{4.68}
\setsymbol{HFI:FWHM:Mars:353GHz}{4.43}
\setsymbol{HFI:FWHM:Mars:545GHz}{3.80}
\setsymbol{HFI:FWHM:Mars:857GHz}{3.67}

\setsymbol{HFI:beam:ellipticity:Mars:100GHz}{1.18}
\setsymbol{HFI:beam:ellipticity:Mars:143GHz}{1.03}
\setsymbol{HFI:beam:ellipticity:Mars:217GHz}{1.14}
\setsymbol{HFI:beam:ellipticity:Mars:353GHz}{1.09}
\setsymbol{HFI:beam:ellipticity:Mars:545GHz}{1.25}
\setsymbol{HFI:beam:ellipticity:Mars:857GHz}{1.03}

\setsymbol{HFI:CMB:relative:calibration:100GHz}{$\lsim 1\%$}
\setsymbol{HFI:CMB:relative:calibration:143GHz}{$\lsim 1\%$}
\setsymbol{HFI:CMB:relative:calibration:217GHz}{$\lsim 1\%$}
\setsymbol{HFI:CMB:relative:calibration:353GHz}{$\lsim 1\%$}
\setsymbol{HFI:CMB:relative:calibration:545GHz}{}
\setsymbol{HFI:CMB:relative:calibration:857GHz}{}

\setsymbol{HFI:CMB:absolute:calibration:100GHz}{$\lsim 2\%$}
\setsymbol{HFI:CMB:absolute:calibration:143GHz}{$\lsim 2\%$}
\setsymbol{HFI:CMB:absolute:calibration:217GHz}{$\lsim 2\%$}
\setsymbol{HFI:CMB:absolute:calibration:353GHz}{$\lsim 2\%$}
\setsymbol{HFI:CMB:absolute:calibration:545GHz}{}
\setsymbol{HFI:CMB:absolute:calibration:857GHz}{}

\setsymbol{HFI:FIRAS:gain:calibration:accuracy:statistical:100GHz}{}
\setsymbol{HFI:FIRAS:gain:calibration:accuracy:statistical:143GHz}{}
\setsymbol{HFI:FIRAS:gain:calibration:accuracy:statistical:217GHz}{}
\setsymbol{HFI:FIRAS:gain:calibration:accuracy:statistical:353GHz}{2.5\%}
\setsymbol{HFI:FIRAS:gain:calibration:accuracy:statistical:545GHz}{1\%}
\setsymbol{HFI:FIRAS:gain:calibration:accuracy:statistical:857GHz}{0.5\%}

\setsymbol{HFI:FIRAS:gain:calibration:accuracy:systematic:100GHz}{}
\setsymbol{HFI:FIRAS:gain:calibration:accuracy:systematic:143GHz}{}
\setsymbol{HFI:FIRAS:gain:calibration:accuracy:systematic:217GHz}{}
\setsymbol{HFI:FIRAS:gain:calibration:accuracy:systematic:353GHz}{}
\setsymbol{HFI:FIRAS:gain:calibration:accuracy:systematic:545GHz}{7\%}
\setsymbol{HFI:FIRAS:gain:calibration:accuracy:systematic:857GHz}{7\%}

\setsymbol{HFI:FIRAS:zero:point:accuracy:100GHz:units}{0.8\MJysr}
\setsymbol{HFI:FIRAS:zero:point:accuracy:143GHz:units}{}
\setsymbol{HFI:FIRAS:zero:point:accuracy:217GHz:units}{}
\setsymbol{HFI:FIRAS:zero:point:accuracy:353GHz:units}{1.4\MJysr}
\setsymbol{HFI:FIRAS:zero:point:accuracy:545GHz:units}{2.2\MJysr}
\setsymbol{HFI:FIRAS:zero:point:accuracy:857GHz:units}{1.7\MJysr}

\setsymbol{HFI:FIRAS:zero:point:accuracy:100GHz}{0.8}
\setsymbol{HFI:FIRAS:zero:point:accuracy:143GHz}{}
\setsymbol{HFI:FIRAS:zero:point:accuracy:217GHz}{}
\setsymbol{HFI:FIRAS:zero:point:accuracy:353GHz}{1.4}
\setsymbol{HFI:FIRAS:zero:point:accuracy:545GHz}{2.2}
\setsymbol{HFI:FIRAS:zero:point:accuracy:857GHz}{1.7}

\setsymbol{HFI:unit:conversion:100GHz:units}{0.2415\MJysrmK}
\setsymbol{HFI:unit:conversion:143GHz:units}{0.3694\MJysrmK}
\setsymbol{HFI:unit:conversion:217GHz:units}{0.4811\MJysrmK}
\setsymbol{HFI:unit:conversion:353GHz:units}{0.2883\MJysrmK}
\setsymbol{HFI:unit:conversion:545GHz:units}{0.05826\MJysrmK}
\setsymbol{HFI:unit:conversion:857GHz:units}{0.002238\MJysrmK}

\setsymbol{HFI:unit:conversion:100GHz}{0.2415}
\setsymbol{HFI:unit:conversion:143GHz}{0.3694}
\setsymbol{HFI:unit:conversion:217GHz}{0.4811}
\setsymbol{HFI:unit:conversion:353GHz}{0.2883}
\setsymbol{HFI:unit:conversion:545GHz}{0.05826}
\setsymbol{HFI:unit:conversion:857GHz}{0.002238}

\setsymbol{HFI:colour:correction:alpha=-2:V1.01:100GHz}{0.9893}
\setsymbol{HFI:colour:correction:alpha=-2:V1.01:143GHz}{0.9759}
\setsymbol{HFI:colour:correction:alpha=-2:V1.01:217GHz}{1.0007}
\setsymbol{HFI:colour:correction:alpha=-2:V1.01:353GHz}{1.0028}
\setsymbol{HFI:colour:correction:alpha=-2:V1.01:545GHz}{1.0019}
\setsymbol{HFI:colour:correction:alpha=-2:V1.01:857GHz}{0.9889}

\setsymbol{HFI:colour:correction:alpha=0:V1.01:100GHz}{1.0008}
\setsymbol{HFI:colour:correction:alpha=0:V1.01:143GHz}{1.0148}
\setsymbol{HFI:colour:correction:alpha=0:V1.01:217GHz}{0.9909}
\setsymbol{HFI:colour:correction:alpha=0:V1.01:353GHz}{0.9888}
\setsymbol{HFI:colour:correction:alpha=0:V1.01:545GHz}{0.9878}
\setsymbol{HFI:colour:correction:alpha=0:V1.01:857GHz}{1.0014}

\providecommand{\sorthelp}[1]{}

\usepackage[draft,breaklinks,colorlinks,citecolor=blue]{hyperref}
\usepackage{amsmath}
\usepackage{natbib}
\usepackage{graphicx}
\usepackage{txfonts}
\usepackage{natbib}
\usepackage{fixltx2e}

\bibpunct{(}{)}{;}{a}{}{,}

\def\psiell{\psi_{\mathrm{ell}}}
\def\BeamNu{B_\nu}
\def\BeamA{\bar{B}_{\alpha}}
\def\Pointing{\vec{n}}
\def\bandpass{\tau}

\def\deg{^{\circ}}

\def\all2103resultspapers{\nocite{planck2013-p01, planck2013-p02, planck2013-p02a, planck2013-p02d, planck2013-p02b, planck2013-p03, planck2013-p03c, planck2013-p03f, planck2013-p03d, planck2013-p03e, planck2013-p01a, planck2013-p06, planck2013-p03a, planck2013-pip88, planck2013-p08, planck2013-p11, planck2013-p12, planck2013-p13, planck2013-p14, planck2013-p15, planck2013-p05b, planck2013-p17, planck2013-p09, planck2013-p09a, planck2013-p20, planck2013-p19, planck2013-pipaberration, planck2013-p05, planck2013-p05a, planck2013-pip56, planck2013-p06b}}

\begin{document}

\title{\Planck{} 2013 results. IV. Low Frequency Instrument beams and window functions}

\authorrunning{Planck Collaboration}
\titlerunning{LFI beams and window functions}

\author{\small
Planck Collaboration:
N.~Aghanim\inst{59}
\and
C.~Armitage-Caplan\inst{90}
\and
M.~Arnaud\inst{73}
\and
M.~Ashdown\inst{70, 6}
\and
F.~Atrio-Barandela\inst{17}
\and
J.~Aumont\inst{59}
\and
C.~Baccigalupi\inst{84}
\and
A.~J.~Banday\inst{93, 8}
\and
R.~B.~Barreiro\inst{66}
\and
E.~Battaner\inst{94}
% ...
}
\institute{\small
APC, AstroParticule et Cosmologie, Universit\'{e} Paris Diderot, CNRS/IN2P3, CEA/lrfu, Observatoire de Paris, Sorbonne Paris Cit\'{e}, 10, rue Alice Domon et L\'{e}onie Duquet, 75205 Paris Cedex 13, France\\
\and
Aalto University Mets\"{a}hovi Radio Observatory, Mets\"{a}hovintie 114, FIN-02540 Kylm\"{a}l\"{a}, Finland\\
\and
African Institute for Mathematical Sciences, 6-8 Melrose Road, Muizenberg, Cape Town, South Africa\\
\and
Agenzia Spaziale Italiana Science Data Center, Via del Politecnico snc, 00133, Roma, Italy\\
\and
Agenzia Spaziale Italiana, Viale Liegi 26, Roma, Italy\\
\and
Astrophysics Group, Cavendish Laboratory, University of Cambridge, J J Thomson Avenue, Cambridge CB3 0HE, U.K.\\
\and
CITA, University of Toronto, 60 St. George St., Toronto, ON M5S 3H8, Canada\\
\and
CNRS, IRAP, 9 Av. colonel Roche, BP 44346, F-31028 Toulouse cedex 4, France\\
\and
California Institute of Technology, Pasadena, California, U.S.A.\\
\and
Centre for Theoretical Cosmology, DAMTP, University of Cambridge, Wilberforce Road, Cambridge CB3 0WA, U.K.\\
\and
Centro de Estudios de F\'{i}sica del Cosmos de Arag\'{o}n (CEFCA), Plaza San Juan, 1, planta 2, E-44001, Teruel, Spain\\
\and
Computational Cosmology Center, Lawrence Berkeley National Laboratory, Berkeley, California, U.S.A.\\
\and
Consejo Superior de Investigaciones Cient\'{\i}ficas (CSIC), Madrid, Spain\\
\and
DSM/Irfu/SPP, CEA-Saclay, F-91191 Gif-sur-Yvette Cedex, France\\
\and
DTU Space, National Space Institute, Technical University of Denmark, Elektrovej 327, DK-2800 Kgs. Lyngby, Denmark\\
\and
D\'{e}partement de Physique Th\'{e}orique, Universit\'{e} de Gen\`{e}ve, 24, Quai E. Ansermet,1211 Gen\`{e}ve 4, Switzerland\\
\and
Departamento de F\'{\i}sica Fundamental, Facultad de Ciencias, Universidad de Salamanca, 37008 Salamanca, Spain\\
\and
Departamento de F\'{\i}sica, Universidad de Oviedo, Avda. Calvo Sotelo s/n, Oviedo, Spain\\
\and
Department of Astronomy and Astrophysics, University of Toronto, 50 Saint George Street, Toronto, Ontario, Canada\\
\and
Department of Astrophysics/IMAPP, Radboud University Nijmegen, P.O. Box 9010, 6500 GL Nijmegen, The Netherlands\\
\and
Department of Electrical Engineering and Computer Sciences, University of California, Berkeley, California, U.S.A.\\
\and
Department of Physics \& Astronomy, University of British Columbia, 6224 Agricultural Road, Vancouver, British Columbia, Canada\\
\and
Department of Physics and Astronomy, Dana and David Dornsife College of Letter, Arts and Sciences, University of Southern California, Los Angeles, CA 90089, U.S.A.\\
\and
Department of Physics and Astronomy, University College London, London WC1E 6BT, U.K.\\
\and
Department of Physics, Florida State University, Keen Physics Building, 77 Chieftan Way, Tallahassee, Florida, U.S.A.\\
\and
Department of Physics, Gustaf H\"{a}llstr\"{o}min katu 2a, University of Helsinki, Helsinki, Finland\\
\and
Department of Physics, Princeton University, Princeton, New Jersey, U.S.A.\\
\and
Department of Physics, University of California, One Shields Avenue, Davis, California, U.S.A.\\
\and
Department of Physics, University of California, Santa Barbara, California, U.S.A.\\
\and
Department of Physics, University of Illinois at Urbana-Champaign, 1110 West Green Street, Urbana, Illinois, U.S.A.\\
\and
Dipartimento di Fisica e Astronomia G. Galilei, Universit\`{a} degli Studi di Padova, via Marzolo 8, 35131 Padova, Italy\\
\and
Dipartimento di Fisica e Scienze della Terra, Universit\`{a} di Ferrara, Via Saragat 1, 44122 Ferrara, Italy\\
\and
Dipartimento di Fisica, Universit\`{a} La Sapienza, P. le A. Moro 2, Roma, Italy\\
\and
Dipartimento di Fisica, Universit\`{a} degli Studi di Milano, Via Celoria, 16, Milano, Italy\\
\and
Dipartimento di Fisica, Universit\`{a} degli Studi di Trieste, via A. Valerio 2, Trieste, Italy\\
\and
Dipartimento di Fisica, Universit\`{a} di Roma Tor Vergata, Via della Ricerca Scientifica, 1, Roma, Italy\\
\and
Discovery Center, Niels Bohr Institute, Blegdamsvej 17, Copenhagen, Denmark\\
\and
Dpto. Astrof\'{i}sica, Universidad de La Laguna (ULL), E-38206 La Laguna, Tenerife, Spain\\
\and
European Space Agency, ESAC, Planck Science Office, Camino bajo del Castillo, s/n, Urbanizaci\'{o}n Villafranca del Castillo, Villanueva de la Ca\~{n}ada, Madrid, Spain\\
\and
European Space Agency, ESTEC, Keplerlaan 1, 2201 AZ Noordwijk, The Netherlands\\
\and
Haverford College Astronomy Department, 370 Lancaster Avenue, Haverford, Pennsylvania, U.S.A.\\
\and
Helsinki Institute of Physics, Gustaf H\"{a}llstr\"{o}min katu 2, University of Helsinki, Helsinki, Finland\\
\and
INAF - Osservatorio Astrofisico di Catania, Via S. Sofia 78, Catania, Italy\\
\and
INAF - Osservatorio Astronomico di Padova, Vicolo dell'Osservatorio 5, Padova, Italy\\
\and
INAF - Osservatorio Astronomico di Roma, via di Frascati 33, Monte Porzio Catone, Italy\\
\and
INAF - Osservatorio Astronomico di Trieste, Via G.B. Tiepolo 11, Trieste, Italy\\
\and
INAF Istituto di Radioastronomia, Via P. Gobetti 101, 40129 Bologna, Italy\\
\and
INAF/IASF Bologna, Via Gobetti 101, Bologna, Italy\\
\and
INAF/IASF Milano, Via E. Bassini 15, Milano, Italy\\
\and
INFN, Sezione di Bologna, Via Irnerio 46, I-40126, Bologna, Italy\\
\and
INFN, Sezione di Roma 1, Universit\`{a} di Roma Sapienza, Piazzale Aldo Moro 2, 00185, Roma, Italy\\
\and
IPAG: Institut de Plan\'{e}tologie et d'Astrophysique de Grenoble, Universit\'{e} Joseph Fourier, Grenoble 1 / CNRS-INSU, UMR 5274, Grenoble, F-38041, France\\
\and
ISDC Data Centre for Astrophysics, University of Geneva, ch. d'Ecogia 16, Versoix, Switzerland\\
\and
IUCAA, Post Bag 4, Ganeshkhind, Pune University Campus, Pune 411 007, India\\
\and
Imperial College London, Astrophysics group, Blackett Laboratory, Prince Consort Road, London, SW7 2AZ, U.K.\\
\and
Infrared Processing and Analysis Center, California Institute of Technology, Pasadena, CA 91125, U.S.A.\\
\and
Institut N\'{e}el, CNRS, Universit\'{e} Joseph Fourier Grenoble I, 25 rue des Martyrs, Grenoble, France\\
\and
Institut Universitaire de France, 103, bd Saint-Michel, 75005, Paris, France\\
\and
Institut d'Astrophysique Spatiale, CNRS (UMR8617) Universit\'{e} Paris-Sud 11, B\^{a}timent 121, Orsay, France\\
\and
Institut d'Astrophysique de Paris, CNRS (UMR7095), 98 bis Boulevard Arago, F-75014, Paris, France\\
\and
Institute for Space Sciences, Bucharest-Magurale, Romania\\
\and
Institute of Astronomy and Astrophysics, Academia Sinica, Taipei, Taiwan\\
\and
Institute of Astronomy, University of Cambridge, Madingley Road, Cambridge CB3 0HA, U.K.\\
\and
Institute of Theoretical Astrophysics, University of Oslo, Blindern, Oslo, Norway\\
\and
Instituto de Astrof\'{\i}sica de Canarias, C/V\'{\i}a L\'{a}ctea s/n, La Laguna, Tenerife, Spain\\
\and
Instituto de F\'{\i}sica de Cantabria (CSIC-Universidad de Cantabria), Avda. de los Castros s/n, Santander, Spain\\
\and
Istituto di Fisica del Plasma, CNR-ENEA-EURATOM Association, Via R. Cozzi 53, Milano, Italy\\
\and
Jet Propulsion Laboratory, California Institute of Technology, 4800 Oak Grove Drive, Pasadena, California, U.S.A.\\
\and
Jodrell Bank Centre for Astrophysics, Alan Turing Building, School of Physics and Astronomy, The University of Manchester, Oxford Road, Manchester, M13 9PL, U.K.\\
\and
Kavli Institute for Cosmology Cambridge, Madingley Road, Cambridge, CB3 0HA, U.K.\\
\and
LAL, Universit\'{e} Paris-Sud, CNRS/IN2P3, Orsay, France\\
\and
LERMA, CNRS, Observatoire de Paris, 61 Avenue de l'Observatoire, Paris, France\\
\and
Laboratoire AIM, IRFU/Service d'Astrophysique - CEA/DSM - CNRS - Universit\'{e} Paris Diderot, B\^{a}t. 709, CEA-Saclay, F-91191 Gif-sur-Yvette Cedex, France\\
\and
Laboratoire Traitement et Communication de l'Information, CNRS (UMR 5141) and T\'{e}l\'{e}com ParisTech, 46 rue Barrault F-75634 Paris Cedex 13, France\\
\and
Laboratoire de Physique Subatomique et de Cosmologie, Universit\'{e} Joseph Fourier Grenoble I, CNRS/IN2P3, Institut National Polytechnique de Grenoble, 53 rue des Martyrs, 38026 Grenoble cedex, France\\
\and
Laboratoire de Physique Th\'{e}orique, Universit\'{e} Paris-Sud 11 \& CNRS, B\^{a}timent 210, 91405 Orsay, France\\
\and
Lawrence Berkeley National Laboratory, Berkeley, California, U.S.A.\\
\and
Max-Planck-Institut f\"{u}r Astrophysik, Karl-Schwarzschild-Str. 1, 85741 Garching, Germany\\
\and
McGill Physics, Ernest Rutherford Physics Building, McGill University, 3600 rue University, Montr\'{e}al, QC, H3A 2T8, Canada\\
\and
MilliLab, VTT Technical Research Centre of Finland, Tietotie 3, Espoo, Finland\\
\and
Niels Bohr Institute, Blegdamsvej 17, Copenhagen, Denmark\\
\and
Observational Cosmology, Mail Stop 367-17, California Institute of Technology, Pasadena, CA, 91125, U.S.A.\\
\and
SB-ITP-LPPC, EPFL, CH-1015, Lausanne, Switzerland\\
\and
SISSA, Astrophysics Sector, via Bonomea 265, 34136, Trieste, Italy\\
\and
School of Physics and Astronomy, Cardiff University, Queens Buildings, The Parade, Cardiff, CF24 3AA, U.K.\\
\and
School of Physics and Astronomy, University of Nottingham, Nottingham NG7 2RD, U.K.\\
\and
Space Sciences Laboratory, University of California, Berkeley, California, U.S.A.\\
\and
Special Astrophysical Observatory, Russian Academy of Sciences, Nizhnij Arkhyz, Zelenchukskiy region, Karachai-Cherkessian Republic, 369167, Russia\\
\and
Stanford University, Dept of Physics, Varian Physics Bldg, 382 Via Pueblo Mall, Stanford, California, U.S.A.\\
\and
Sub-Department of Astrophysics, University of Oxford, Keble Road, Oxford OX1 3RH, U.K.\\
\and
Theory Division, PH-TH, CERN, CH-1211, Geneva 23, Switzerland\\
\and
UPMC Univ Paris 06, UMR7095, 98 bis Boulevard Arago, F-75014, Paris, France\\
\and
Universit\'{e} de Toulouse, UPS-OMP, IRAP, F-31028 Toulouse cedex 4, France\\
\and
University of Granada, Departamento de F\'{\i}sica Te\'{o}rica y del Cosmos, Facultad de Ciencias, Granada, Spain\\
% ...
}


\abstract{

This paper presents the characterization of the in-flight beams, the beam window functions and the associated {uncertainties} for the \Planck\ Low Frequency Instrument (LFI).
Knowledge of the beam profiles is {necessary for determining the transfer function to go} from the observed to the actual sky anisotropy power spectrum.  
The main beam distortions affect the beam window function, complicating the reconstruction of the anisotropy power spectrum at high multipoles, whereas the sidelobes affect the low and intermediate multipoles.
The in-flight assessment of the LFI main beams {relies} on the measurements performed during Jupiter observations. 
By stacking the data from {multiple} Jupiter transits, the main beam profiles are measured down to --20 dB at 30 and 44\,GHz, and down to --25 dB at 70\,GHz.
The main beam solid angles are determined to better than 0.2\% at each LFI frequency band.
{The \Planck\ pre-launch optical model is conveniently tuned to characterize the main beams independently of any noise effects}.
This approach provides an optical model whose beams fully reproduce the measurements in the main beam region, but also allows {a description of the}
beams at power levels lower than can be achieved by the Jupiter measurements themselves.
The agreement between the simulated beams and the {measured} beams is better than 1\% at each LFI frequency band.
The simulated beams are used for the computation of the window functions for the effective beams.
The error budget {for} the window functions {is} estimated from both main beam and sidelobe contributions, and accounts for the radiometer bandshapes.
The total uncertainties in the effective beam window functions are: {2\% and 1.2\% at 30 and 44\,GHz, respectively (at $\ell \approx 600$), and 0.7\% at 70\,GHz (at $\ell \approx 1000$) }.

}

\maketitle

\section{Introduction}
\label{introduction}
This paper, one of a set \all2103resultspapers associated with the 2013 release of data from the \Planck\footnote{\Planck\ (\url{http://www.esa.int/Planck}) is a project of the European Space Agency (ESA) with instruments provided by two scientific consortia funded by ESA member states (in particular the lead countries France and Italy), with contributions from NASA (USA) and telescope reflectors provided by a collaboration between ESA and a scientific consortium led and funded by Denmark.} mission~(Planck Collaboration I-XXXI, 2014), describes the beams and window functions of the Low Frequency Instrument (LFI).

Detailed knowledge of the instrumental angular response is a key re\-quire\-ment for the anal\-y\-sis of high precision measurements of the cosmic microwave background (CMB). 
{Modern} experiments employ multi-frequency focal plane arrays whose off-axis beams necessarily deviate, to some extent, from an ideal, axisymmetric {(circular)}, Gaussian shape. 
The radiation patterns of {the} individual detector and their projected angular locations need to be reconstructed with great precision to avoid significant systematic effects in the data \citep{hill2009,nolta2009,huffenberger2010}. 

The \Planck\ optical system is designed to ensure high image quality over a wide field of view, for detectors spanning over 1.5 decades in wavelength \citep{tauber2010b}.
The LFI optical layout is composed of an array of 11 corrugated feed horns, each coupled to an orthomode transducer which splits the incoming electromagnetic wave into two orthogonal, linearly polarized components.  
Thus, the LFI {observed} the sky with 11 pairs of beams associated with the 22 pseudo-correlation radiometers. 
Each beam of the pair is named \texttt{LFIXXM} or \texttt{LFIXXS} for the two polarization states (Main Arm and Side Arm of the orthomode transducer, respectively). 
Here \texttt{XX} is the radiometer chain assembly number, ranging from \texttt{18} to \texttt{28}. 
The beams from \texttt{LFI18} to \texttt{LFI23} are in the V--band (nominally from 63 to 77\,GHz); we refer to them as 70\,GHz.
The beams from \texttt{LFI24} to \texttt{LFI26} are in the Q--band (from 39.6 to 48.4\,GHz); we refer to them as 44\,GHz.
The beams \texttt{LFI27} and \texttt{LFI28} are in the Ka--band (from 27 to 33\,GHz); we refer to them as 30\,GHz.
The optimization of the LFI optical system leading to the focal plane configuration used in flight is described in \cite{sandri2010}, while the preliminary characterization of the LFI beams based on the first in-flight data are reported in \cite{planck2011-1.4} and \cite{planck2011-1.6}. 

\begin{figure}
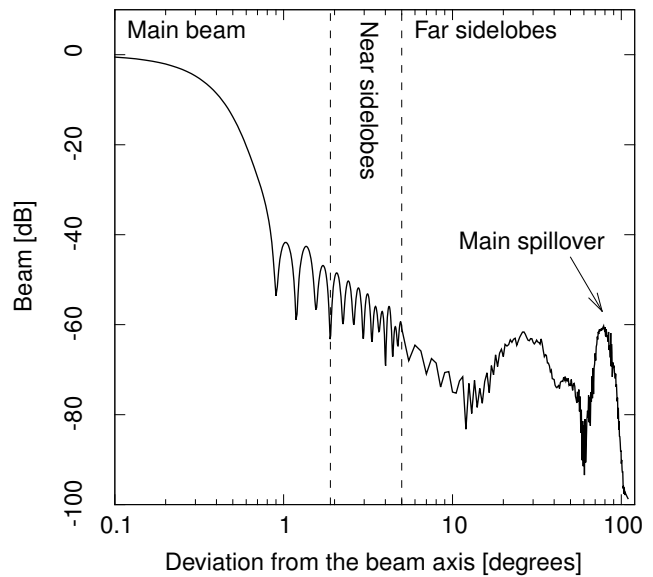

	\centering
	\input beam_plot.tex
	\caption{\label{fig:beamShape} Typical shape of a 30\,GHz beam 
(\texttt{LFI27M}). The plot shows the distinction between the main beam, near 
sidelobes and far sidelobes. The distinction between ``near'' and 
``far'' sidelobes is of course arbitrary:  {their boundary is marked at} 
5$^\circ$. The peak of the spillover of the primary mirror is clearly 
visible, at an angle of roughly 90$^\circ$.}
\label{fig:beamplot}
\end{figure}

The LFI map-making procedure does not take into account the beam profile, which is effectively assumed to be a pencil beam.
To correct for the beam shape, the angular power spectrum computed from the observed map is divided by the \textit{beam window function} to reveal the intrinsic angular power spectrum of the sky. 
For this reason, beam knowledge directly affects the cosmological analysis.
Typically, the beam should be mapped to less than --30 dB of the peak to achieve 1\% accuracy on the angular power spectrum \citep{page2003a}.
By stacking the data from the first four Jupiter transits, the LFI beams have been measured down to --20 dB at 30 and 44\,GHz, and down to --25 dB at 70\,GHz with an uncertainty of about 0.3\% on the angular resolution and about 0.5\% on the main beam ellipticity.
In order to achieve the beam knowledge at lower power levels and improve the accuracy on the angular power spectrum, a substantial effort has been made to tune the \Planck\ optical model, presented in \cite{tauber2010b}, to fit the in-flight measurements of the LFI beams. This ensures a good representation of the LFI optics, for both the {main beam and sidelobes}.
The separation of the instrumental angular response into main beam and sidelobes can be somewhat arbitrary.
In the framework of this paper, we consider three regions defined with respect to the beam {boresight} and shown in Fig.~\ref{fig:beamplot}:
\begin{enumerate}
\item the $main$ $beam$, which is defined as extending to 1.9, 1.3, and 0.9$^\circ$ at 30, 44, and 70\,GHz, respectively;
\item the $near$ $sidelobes$, which are defined as extending between the main beam angular limit and 5$^\circ$;
\item the $far$ $sidelobes$, which are defined as the beam response greater than 5$^\circ$ from the {boresight}.
\end{enumerate}

{More than 99\% of all the power falls into the main beam region}. 
The collected power coming from the region outside the main beam is called \textit{straylight} and it is a major source of systematic effects in \Planck\ observations, and in CMB experiments in general.
Straylight impacts the measured signal in two ways: (i) through direct contamination, and (ii) in the photometric calibration of the {detected} radiometer signal. 
The {modeled} straylight contamination has been detected in the LFI maps, and is reported in the companion paper \citet{planck2013-p02a}.
{We emphasize} that, since no direct measurement of LFI sidelobes was performed in-flight, an accurate knowledge of the main beams provides a crucial means, though indirect, to quantify the straylight contamination, as the sidelobes can be estimated by fitting the electromagnetic model to the main beam data. 

For the clarity of the present paper and for consistency with the \Planck\ companion papers, we make three important definitions: 
\begin{enumerate}
\item The \textit{optical beam} is the optical response of the feed horn coupled to the telescope. It is independent of both the radiometer response (bandshape and non-linearity) and of the satellite motion (spinning and scanning strategy). It represents the pure optical transfer function. The main beam properties of the optical beams can be evaluated using optical simulations performed with methods largely validated by ground measurements.
\item The \textit{scanning beam} is the beam that can be directly measured in-flight using planet observations. It stems from the optical beam coupled with the radiometer response, and smeared by the satellite motion. So, with respect to the optical beams, the scanning beams have slightly larger angular resolution, ellipticity, and solid angle.
\item The \textit{effective beam} is a beam defined in the map-domain, and is obtained by averaging the scanning beams pointing at a given pixel of the sky map taking into account the scanning strategy and the orientation of the beams themselves when they point along the direction to that pixel.   Therefore, whereas for each radiometer there is one corresponding optical and scanning beam, the same radiometer has the same number of effective beams as there are pixels in the observed sky map. The importance of the effective beams is {twofold}: they are used in the window function computation, and their solid angles are needed for the estimation of the flux density of point sources. 
\end{enumerate}


The data analysis pipeline, {starting} from Jupiter observations and flowing down to the window function characterization, is discussed in this paper {as follows}: Sect.~\ref{scanning_beams} describes the scanning beams as measured in the first four Jupiter transits, and the simulations which provide their best-fit model;
Sect.~\ref{effective_beams} describes the effective beams, calculated using the simulated beams and taking into account the \Planck\ scanning strategy;
In Sect.~\ref{window_function} we present the LFI window functions. 
An estimate of the propagation of beam uncertainties to the beam window functions is reported in Sect.~\ref{error_propagation}.
In this section we also report the impact of the near and far sidelobes {on} the window function.
For the present data release we do not correct the beam window function for the sidelobes.
Instead their effect is added {to} the total error budget. 
For the next data release, we plan to include a detailed analysis carried out with the in-band integrated beams (main beam and sidelobes) that will be included in the data reduction pipeline, both in the calibration and in the window function estimation.  Sect.~\ref{conclusions} summarizes the conclusions.

\section{Scanning Beams}
\label{scanning_beams}
Jupiter is the best compact source in the sky for mapping the LFI beams with a high signal to noise ratio. 
The brightness temperature of the planet is close to 150\,K and gives an antenna temperature from 40 to 350\,mK depending on frequency, when the dilution factor of the beams is accounted for. {The angular response of the detector in antenna temperature ($T_{\rm A}$)  to an unpolarized source is proportional to the power function of the beam as follows}:   
\begin{eqnarray}
T_{\rm A}(\theta,\phi)^{{\rm M}} &\propto&  | E(\theta,\phi)^{{\rm M}}_{{\rm cp}} | ^2 + | E(\theta,\phi)^{{\rm M}}_{{\rm xp}} | ^2 + \nonumber \\
&+& \left \{\chi^{{\rm OMT}} \cdot \left[   | E(\theta,\phi)^{{\rm S}}_{{\rm cp}} | ^2 + | E(\theta,\phi)^{{\rm S}}_{{\rm xp}} | ^2 \right]\right\} 
\label{eq:TM}
\end{eqnarray}

\begin{eqnarray}
\label{eq:TS}
T_{\rm A}(\theta,\phi)^{{\rm S}} &\propto&  | E(\theta,\phi)^{{\rm S}}_{{\rm cp}} | ^2 + | E(\theta,\phi)^{{\rm S}}_{{\rm xp}} | ^2 +  \nonumber \\
&+& \left\{ \chi^{{\rm OMT}} \cdot \left[   | E(\theta,\phi)^{{\rm M}}_{{\rm cp}} | ^2 + | E(\theta,\phi)^{{\rm M}}_{{\rm xp}} | ^2 \right]\right\}, 
\end{eqnarray}

\noindent
where $E(\theta,\phi)^{{\rm M,S}}_{{\rm cp}}$ and $E(\theta,\phi)^{{\rm M,S}}_{{\rm xp}}$ are respectively the co-polar and cross-polar electric field components of the beam in the M-radiometer and S-radiometer, computed in the main beam frame ($\theta = \theta_{{\rm MB}}$ and $\phi = \phi_{{\rm MB}}$); and $\chi^{{\rm omt}}$ is the orthomode transducer (OMT) cross-polarization. {The main beam frame is the one aligned with the main beam polarization direction. The OMT cross-polarization} was measured during the hardware development \citep{darcangelo2009b} and {was} always less than $-25$ dB over the operational bandwidth, so that the terms between the curly brackets are considered negligible. {In case of a polarized source the response is slightly different, and is calculated in Appendix \ref{appendix}. Even if the emission from Jupiter is polarized, the effect is well below the noise level. A level of 1\%  of polarization, for instance, results in an effect $-45~$dB below the beam peak.}
To assess the beam properties, we {use} four Jupiter transits named ``J1'', ``J2'', ``J3'', and ``J4''. Table \ref{tab:ods} reports the date and the corresponding observational days (OD) of each transit. 

\begin{table}
\centering
\caption{Approximate dates of the Jupiter observations. The ranges include the scan by the entire LFI field of view.}
\begin{tabular}{c c c}
\hline
\hline
\noalign{\vskip 2pt}
Jupiter transit & Date & OD \\
\hline
\noalign{\vskip 2pt}
Scan 1 (J1)& 21/10/09 -- 05/11/09 & 161 -- 176 \\ 
Scan 2 (J2)& 27/06/10 -- 12/07/10 & 410 -- 425 \\
Scan 3 (J3)& 03/12/10 -- 18/12/10 & 569 -- 584 \\
Scan 4 (J4)& 30/07/11 -- 08/08/11 & 808 -- 817 \\
\hline  
\end{tabular}
\label{tab:ods}
\end{table}

\subsection{Planet Data Handling}

\begin{figure}[hptb]
\centering
\includegraphics[width=9cm]{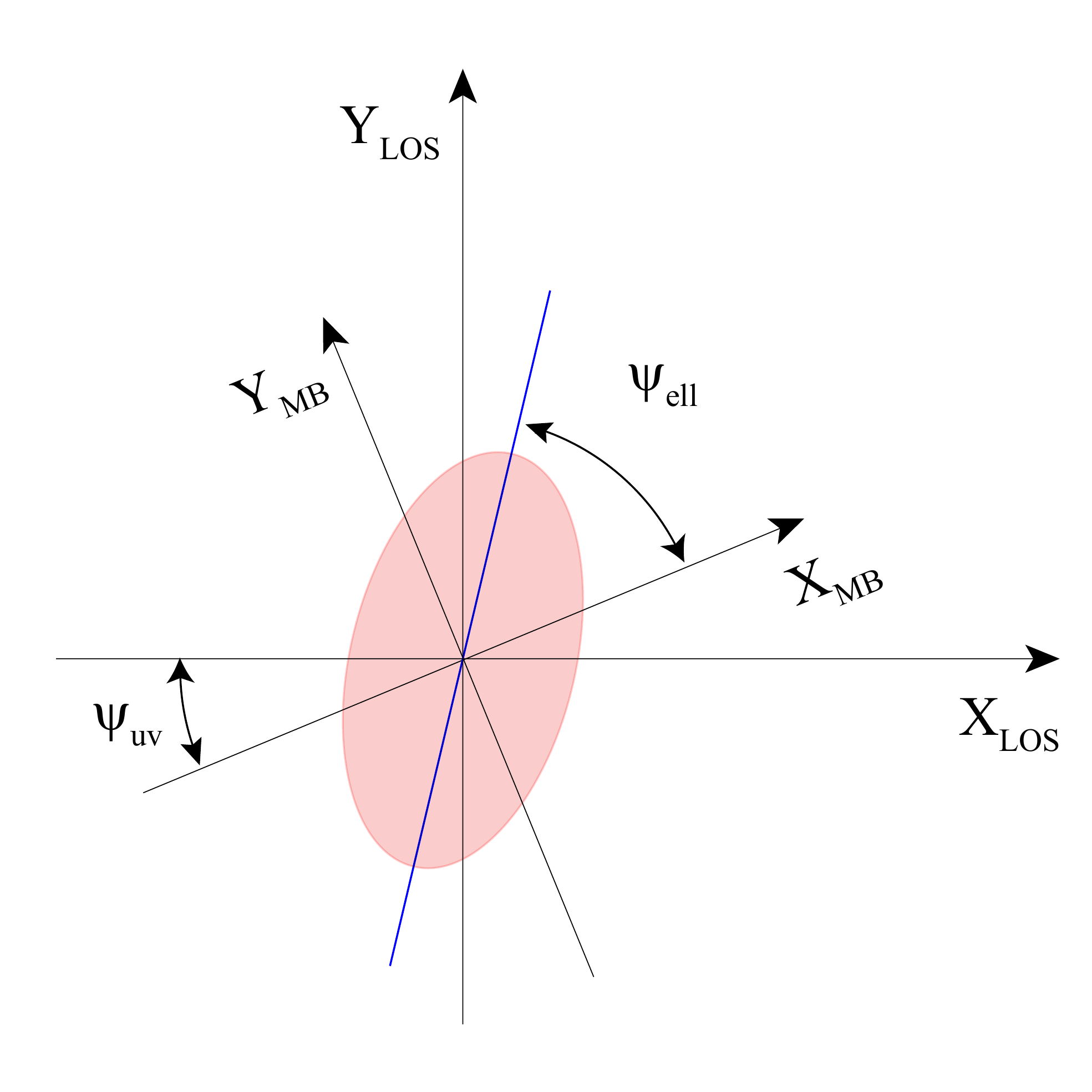} 
\caption{Graphic representation of $\psi_{{\rm ell}}$ defined as the angle between the major axis of the fitted elliptical Gaussian beam and the x-axis of the main beam frame, (XY)$_{\rm{MB}}$, which is aligned with the main beam polarization direction. In the figure also the LOS frame is reported. The angle between the main beam polarization direction and the x-axis of the LOS frame is named $\psi_{{\rm uv}}$ and  is described in \cite{planck2013-p02}.}
\label{fig:psiell}
\end{figure}

\begin{figure*}[hptb]
\centering
\begin{tabular}{c c}
\includegraphics[width=8.5cm]{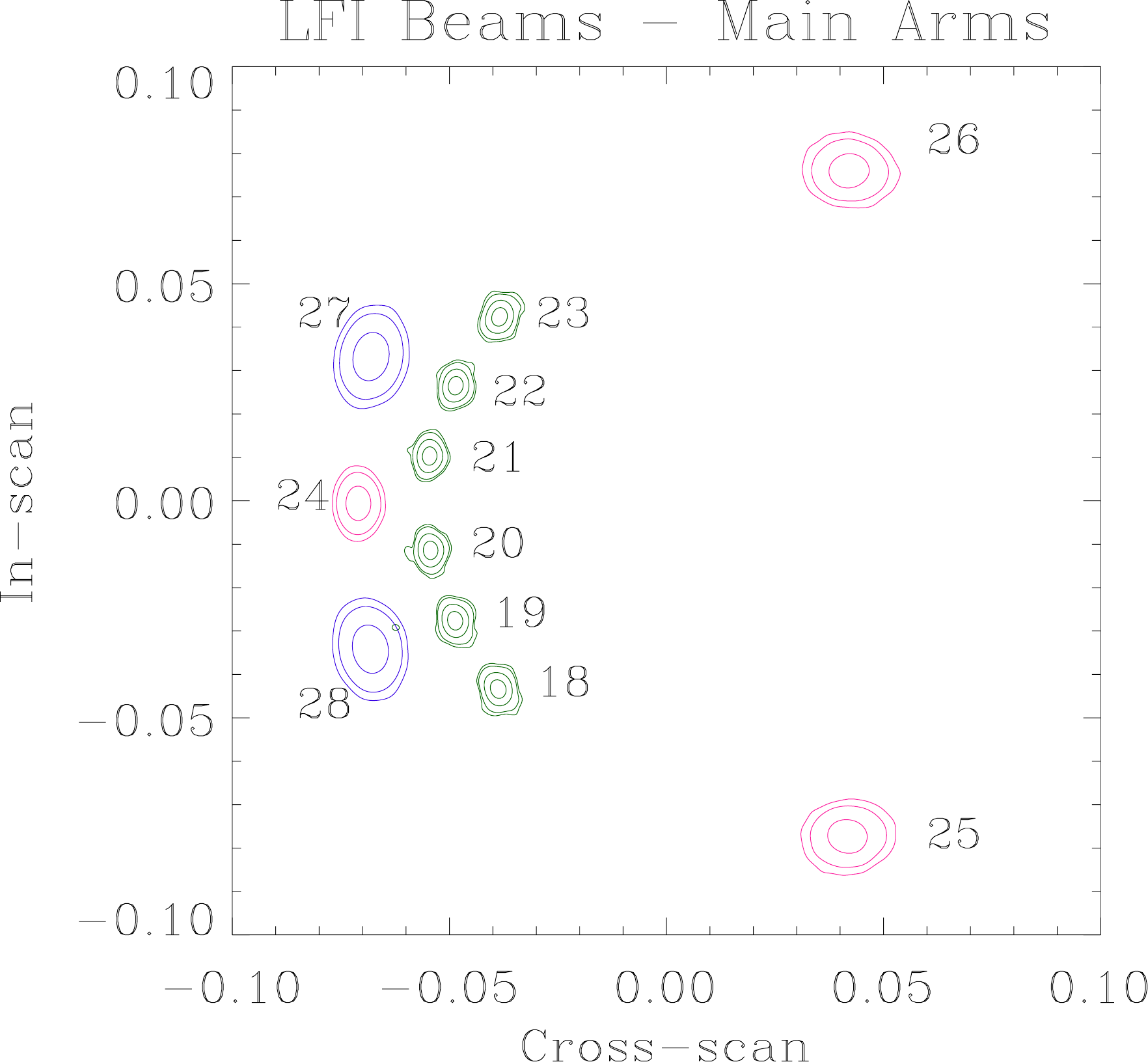} &
\includegraphics[width=8.5cm]{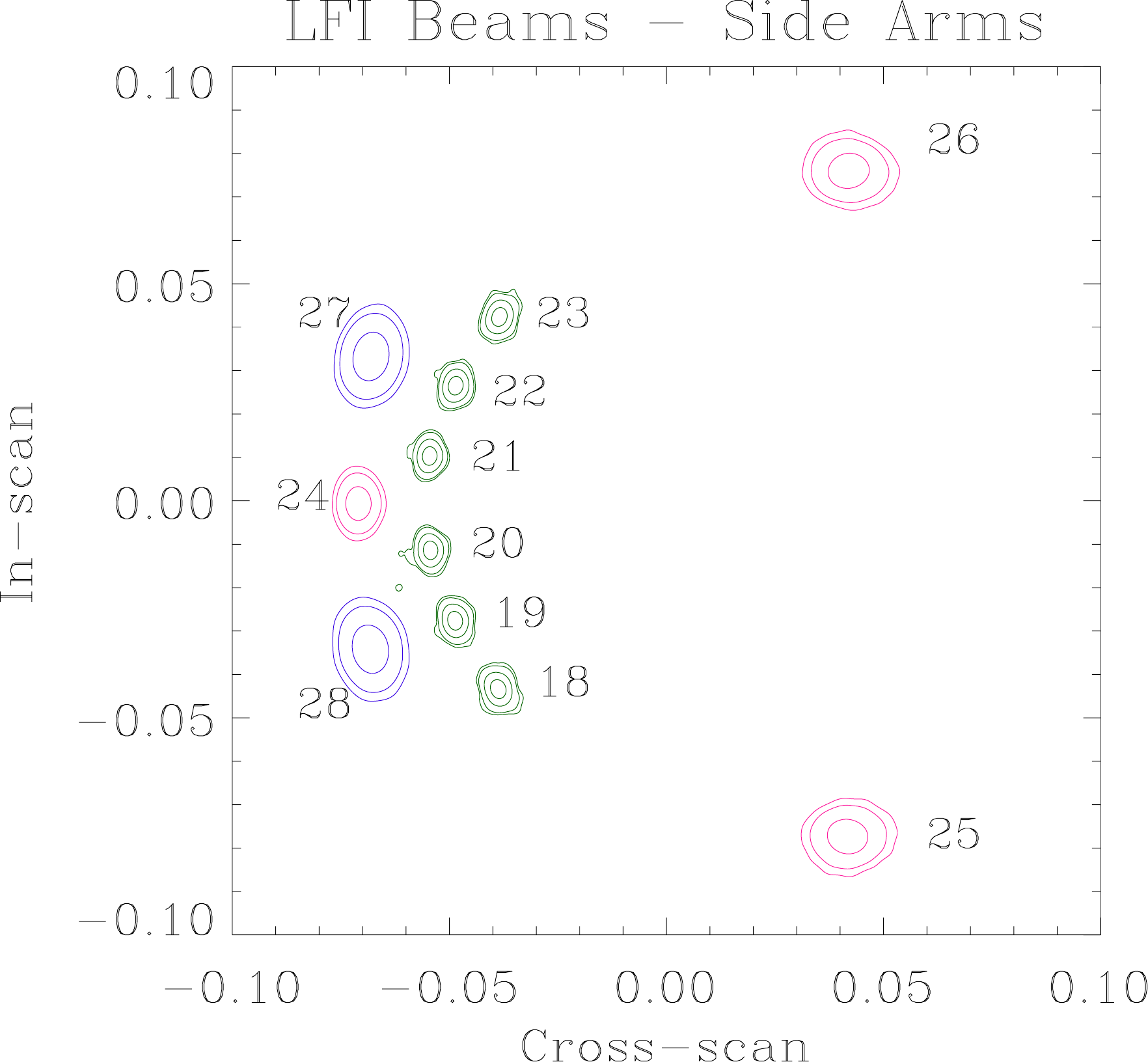} \\
\end{tabular}
\caption{Scanning beam profiles for both polarization arms, reconstructed from the first four Jupiter transits. The beams are plotted in contours of --3, --10, --20, and --25 dB from the peak at 70\,GHz (green), and --3, --10, --20 at 30\,GHz (blue) and 44\,GHz (pink).}
\label{fig:uv}
\end{figure*}

The LFI in-flight main beam reconstruction is based on a {minimization} code described in \cite{burigana2001} and incorporated into the Level 2 \Planck\ LFI DPC pipeline. 
The code {uses the calibrated timelines of Jupiter transits observed by the LFI beams, to fit the beam shape to an elliptical Gaussian function. }
With this Gaussian approximation, the angular resolution {is defined} in terms of the full width half maximum (FWHM), the beam ellipticity ($e$), and the beam orientation ($\psi_{{\rm ell}}$, see Fig.~\ref{fig:psiell}). 
Moreover, this fit {is} used to define the beam center so that the beam pointing directions agree with the convention adopted in \cite{planck2013-p02}.
The fit is performed in the plane of the \Planck\ field of view, centered along the nominal line of sight (LOS) defined in \cite{tauber2010b}.
In Fig.~\ref{fig:uv} the LFI footprint on the sky is reported for both polarization arms.
The data selection is done using the pointing information contained in the satellite Attitude History File \citep{planck2013-p02}, which in turn is used to infer the nominal LOS direction synchronously with the sampled data. 
The data selected for fits to Jupiter lie on square grids centered with respect to the main beam pointing direction, of about 1.7$^\circ$ in total size at 70\,GHz, 2.6$^\circ$ at 44\,GHz, and 3.8$^\circ$ at 30\,GHz.

For each radiometer arm, the selected data are characterized by an array of data samples specified by the signal amplitude (in thermodynamic temperature) during the transit, the positions ($x$, corresponding to the scan circles, and $y$, corresponding to the positions along each scan circle) of Jupiter during the transit, and the distances between \Planck\ and the planet itself during the transit. 
An initial guess for the main beam input parameters and their possible ranges has been evaluated directly on the measured timelines, together with an estimate of the noise corresponding to the sensitivity of the ensemble of signal data.
Since the {average noise value is not negligible}  (at 30 GHz, it is about 0.2\% of the peak power), an offset has been applied in order to have a noise characterized by a null average value. 
Furthermore, only the data with a signal above the 3\,$\sigma$ noise level from the noise have been considered in the minimisation routine of the fitting code. 
This implies slightly higher error bars, but guarantees a negligible effect due to the background.
No destriping was performed on the timelines because it was found that the 1/$f$ noise does not affect the reconstructed beam shape above --20 dB. 
In particular, the 1/$f$ noise will not affect our later estimate of the window functions because these are obtained from simulated beams derived from a {specific} optical model, as described later in this section.
The fit procedure gives an analytical description of the LFI beams, through the parameters that characterize the elliptical Gaussian profile and the corresponding statistical uncertainties; the latter are computed using the Minuit processor {\tt MINOS}\footnote{\url{http://seal.web.cern.ch}} which calculates the parameter errors by taking into account both parameter correlations and non-linearities.

\begin{figure}[htpb]
\centering
\includegraphics[width=9cm]{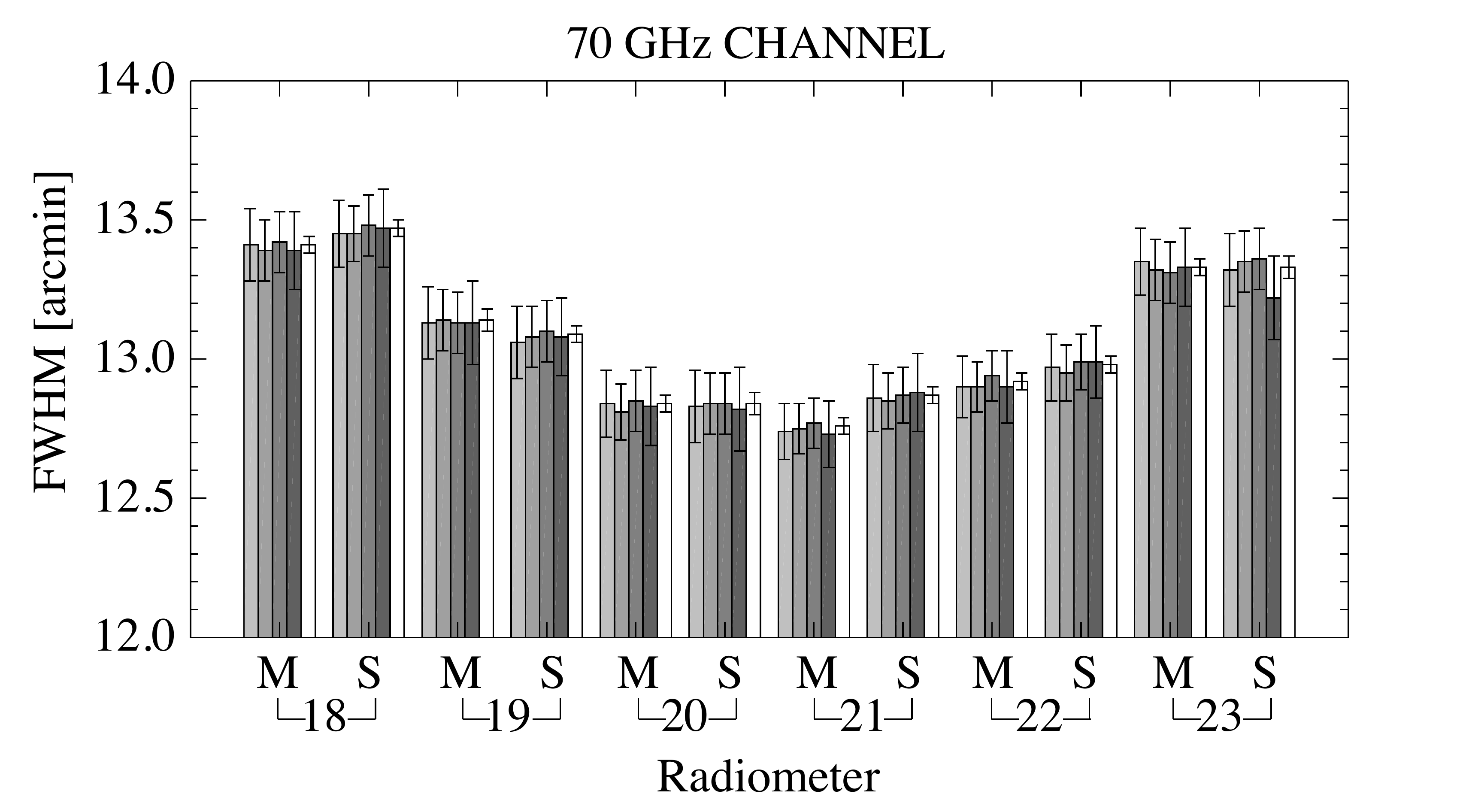} 
\includegraphics[width=9cm]{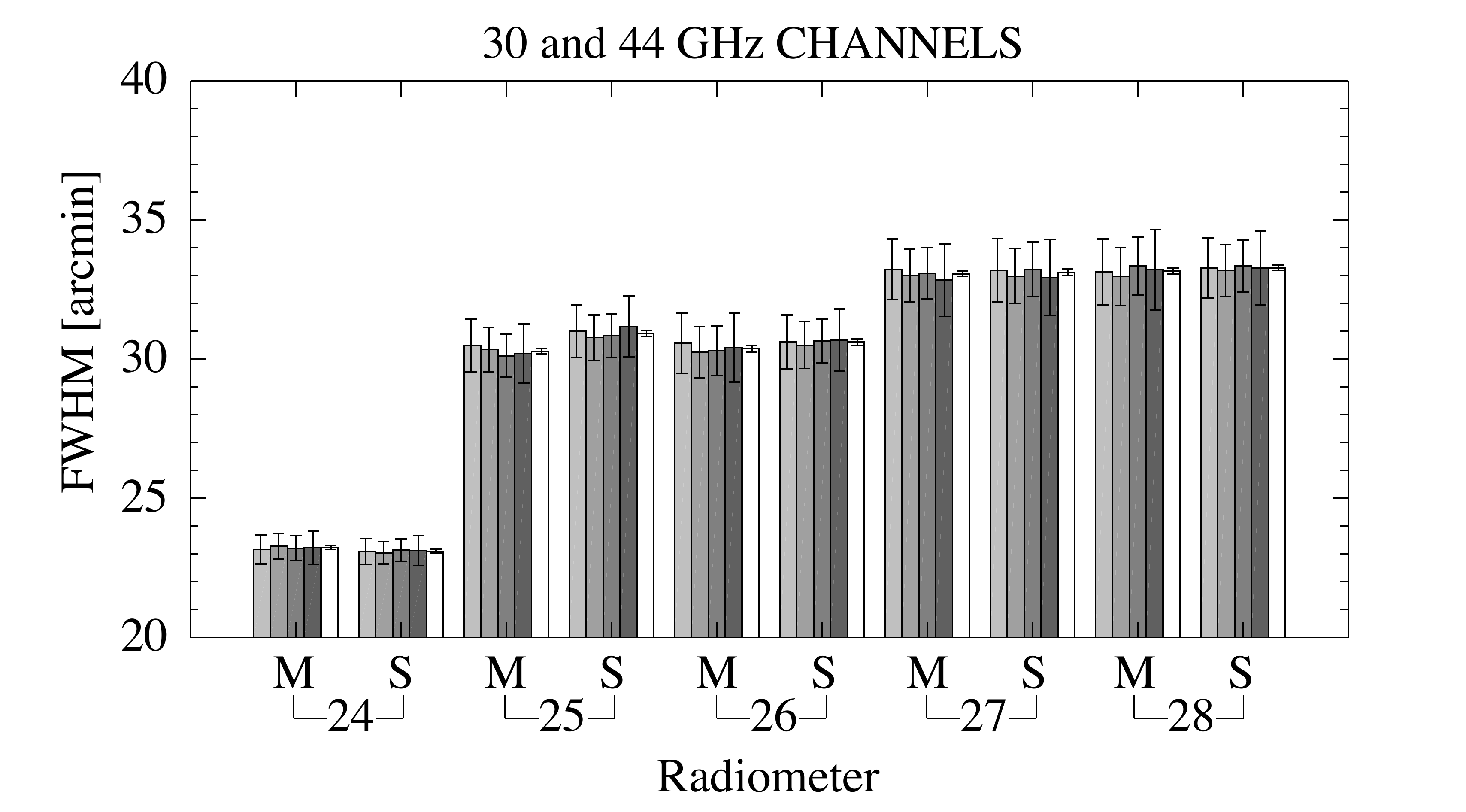}
\caption{FWHM at 70 GHz (upper panel) and 30/44\,GHz (lower panel) for the four Jupiter scans (grey bars) and for the stacked beams (white bars), {in which the four scans are considered together}.}
\label{fig:histofwhm}
\end{figure}

\begin{figure}[htpb]
\centering
\includegraphics[width=9cm]{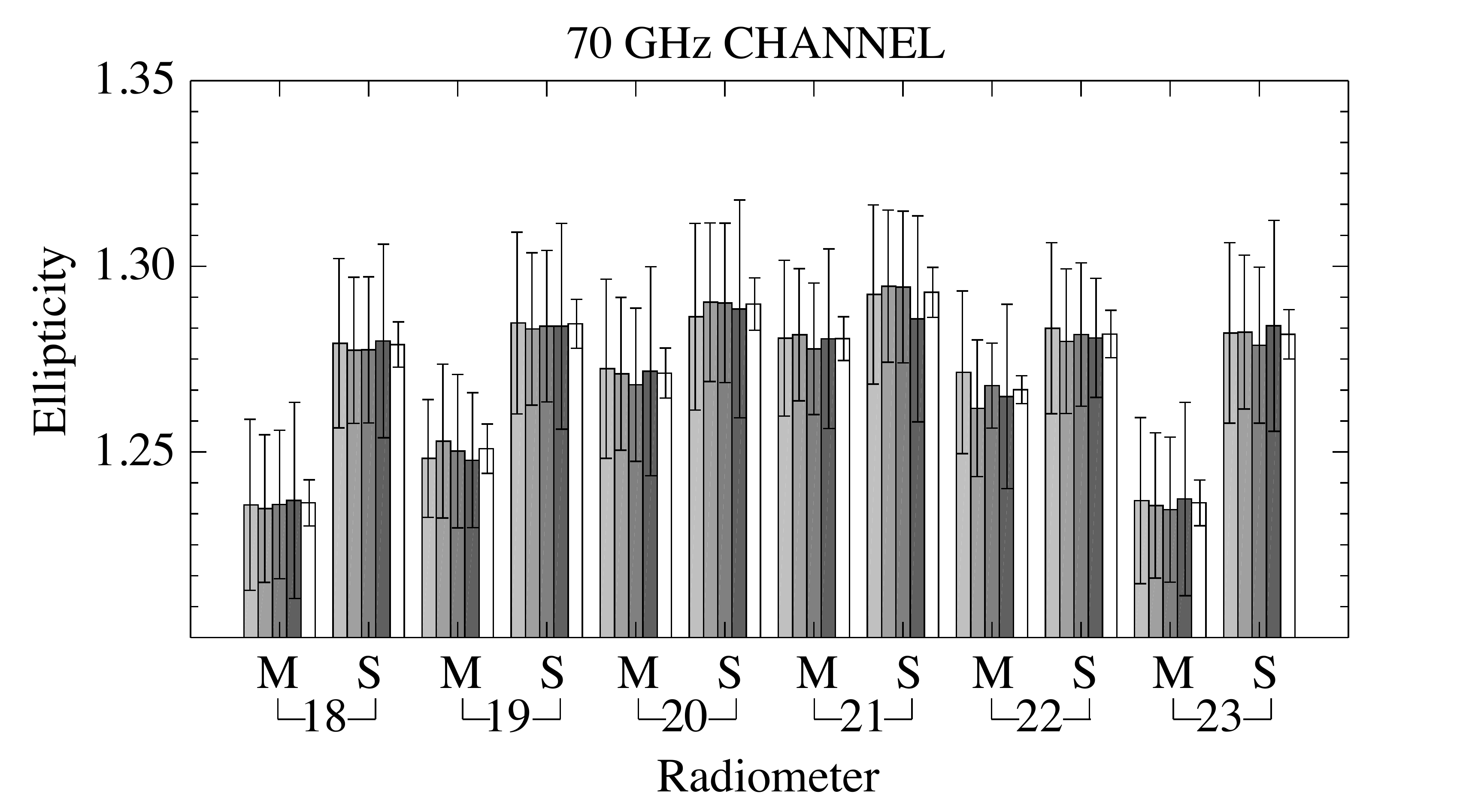} 
\includegraphics[width=9cm]{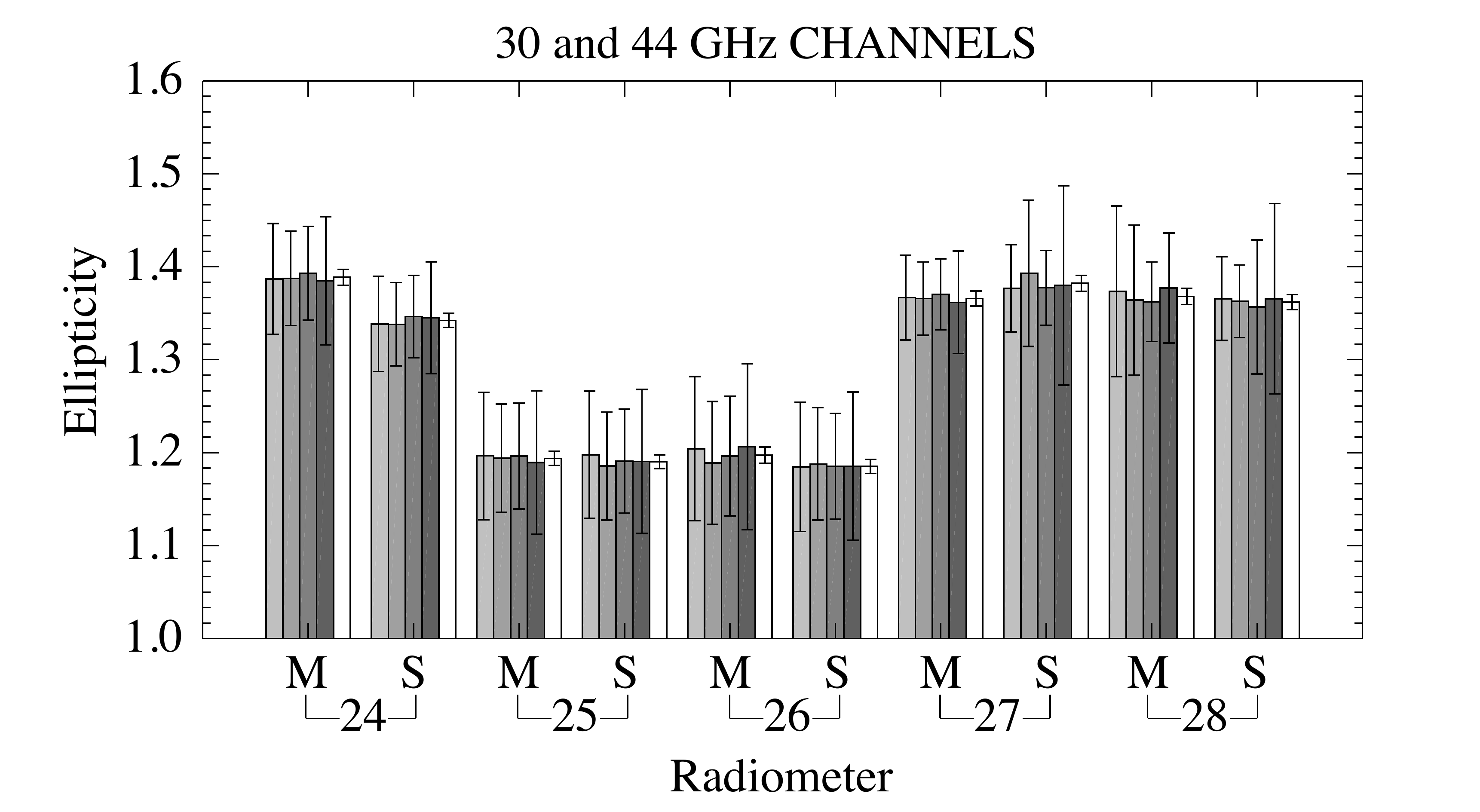}
\caption{Ellipticity at 70 GHz (upper panel) and 30/44\,GHz (lower panel) for the four Jupiter scans (grey bars) and for the stacked beams (white bars), {in which the four scans are considered together}.}
\label{fig:histoe}
\end{figure}

Table \ref{tab:imo} reports the main beam descriptive parameters with the estimated uncertainties evaluated from the stacked beams obtained {considering the four Jupiter transit data together}.
In the bar charts, shown in Figs.~\ref{fig:histofwhm} and \ref{fig:histoe}, the four transits are considered separately and then stacked\footnote{FWHM = $\sqrt{8 \times {\rm ln}(2) \times \sigma_{{\rm max}}^{{\rm b}} \times \sigma_{{\rm min}}^{{\rm b}}}$; $e = \sigma_{{\rm max}}^{{\rm b}}/\sigma_{{\rm min}}^{{\rm b}}$; $\psi_{{\rm ell}}$ is defined as the angle between the major axis of the ellipse and the x-axis of the {main beam frame.}}.
It is evident that the four measurements give basically the same results.
Thus, no {time-dependent} optical effects are evident in these data, which were taken from October 2009 to August 2011. 
The improvement in terms of the uncertainties obtained using the four scans together is remarkable. 

\begin{table}[hptb]
\footnotesize
\centering
\caption{Main beam de\-scrip\-tive pa\-ram\-e\-ters of the scanning beams, with  uncertainties (1\,$\sigma$)  .}
\begin{tabular}{c c c r@{.}l}
\hline
\hline
\noalign{\vskip 2pt}
Beam & FWHM & Ellipticity & \multicolumn{2}{c}{$\psi_{{\rm ell}}$} \\
 & (arcmin) &  & \multicolumn{2}{c}{(degrees)} \\
\hline
\noalign{\vskip 2pt}
\multicolumn{5}{l}{70\,GHz} \\
\noalign{\vskip 4pt}
\texttt{18M} 	& 13.41	$\pm$ 0.03  &	1.24  $\pm$ 0.01	& 85&51	  $\pm$ 0.68 \\  
\texttt{18S} 	& 13.47	$\pm$ 0.03	& 1.28	$\pm$ 0.01	& 86&35	  $\pm$ 0.55 \\  
\texttt{19M} 	& 13.14	$\pm$ 0.04	& 1.25	$\pm$ 0.01	& 78&94	  $\pm$ 0.67 \\  
\texttt{19S} 	& 13.09	$\pm$ 0.03	& 1.28	$\pm$ 0.01	& 79&12	  $\pm$ 0.58 \\  
\texttt{20M}	& 12.84	$\pm$ 0.03	& 1.27	$\pm$ 0.01	& 71&62	  $\pm$ 0.62 \\  
\texttt{20S}	& 12.84	$\pm$ 0.04	& 1.29	$\pm$ 0.01	& 72&61	  $\pm$ 0.61 \\  
\texttt{21M}	& 12.76	$\pm$ 0.03	& 1.28	$\pm$ 0.01	& 108&00	$\pm$ 0.52 \\  
\texttt{21S}	& 12.87	$\pm$ 0.03	& 1.29	$\pm$ 0.01	& 106&98	$\pm$ 0.57 \\  
\texttt{22M}	& 12.92	$\pm$ 0.03	& 1.27	$\pm$ 0.01	& 102&05	$\pm$ 0.57 \\  
\texttt{22S}	& 12.98	$\pm$ 0.03	& 1.28	$\pm$ 0.01	& 101&74	$\pm$ 0.57 \\  
\texttt{23M}	& 13.33	$\pm$ 0.03	& 1.24	$\pm$ 0.01	& 93&48	  $\pm$ 0.67 \\  
\texttt{23S}	& 13.33	$\pm$ 0.04	& 1.28	$\pm$ 0.01	& 93&60  	$\pm$ 0.59 \\ 
\hline
\noalign{\vskip 2pt}
\multicolumn{5}{l}{44\,GHz} \\
\noalign{\vskip 4pt} 
\texttt{24M}	& 23.23	$\pm$ 0.07	& 1.39	$\pm$ 0.01	& 89&85	  $\pm$ 0.53 \\  
\texttt{24S}	& 23.10	$\pm$ 0.07	& 1.34	$\pm$ 0.01	& 89&98	  $\pm$ 0.53 \\  
\texttt{25M}	& 30.28	$\pm$ 0.10	& 1.19	$\pm$ 0.01	& 115&41	$\pm$ 1.02 \\  
\texttt{25S}	& 30.92	$\pm$ 0.10	& 1.19	$\pm$ 0.01	& 117&34	$\pm$ 1.02 \\  
\texttt{26M}	& 30.37	$\pm$ 0.12	& 1.20	$\pm$ 0.01	& 62&13  	$\pm$ 1.14 \\  
\texttt{26S}	& 30.61	$\pm$ 0.11	& 1.19	$\pm$ 0.01	& 61&42	  $\pm$ 1.09 \\
\hline
\noalign{\vskip 2pt}
\multicolumn{5}{l}{30\,GHz} \\
\noalign{\vskip 4pt}  
\texttt{27M}	& 33.06	$\pm$ 0.10	& 1.37	$\pm$ 0.01	& 101&24	$\pm$ 0.53 \\  
\texttt{27S}	& 33.12	$\pm$ 0.11	& 1.38	$\pm$ 0.01	& 101&37	$\pm$ 0.54 \\  
\texttt{28M}	& 33.17	$\pm$ 0.11	& 1.37	$\pm$ 0.01	& 78&53  	$\pm$ 0.57 \\  
\texttt{28S}	& 33.28	$\pm$ 0.10	& 1.36	$\pm$ 0.01	& 78&87	  $\pm$ 0.54 \\  
\hline
\end{tabular}
\label{tab:imo}
\end{table}                                                                  

\subsection{From optical beams to scanning beams}
As noted earlier, the optical beams are the optical response of the feed horns coupled to the \Planck\ telescope independent of both the radiometer response (bandshape and non-linearity) and the satellite motion (spinning and scanning strategy). {The calculation of the optical beams is described in subsection 2.2.1 below.  This calculation is then extended to the scanning beams taking into account the satellite motion.  These model results are valuable for two reasons.  First they allow us to extend our estimates of the beam pattern to lower levels, where the signal from Jupiter is lost in the noise.  This in turn allows a calculation of the main beam efficiency.  Second, these models permit estimates of cross-polarization response.  In Subsection 2.2.2, we describe a test of these models and in Subsection 2.2.3, the small corrections needed to account for the finite bandwidth of the LFI receivers are described.}

\subsubsection{Main beams}
In the main beam region, the optical beams have been evaluated from simulations carried out by the application of physical optics and the physical theory of diffraction using {\tt GRASP}\footnote{The {\tt GRASP} software was de\-vel\-oped by TICRA (Co\-pen\-hagen, DK) for analysing gen\-eral re\-flec\-tor antennas (\url{http://www.ticra.it}).}.
A dedicated optical study has been carried out with the goal of fitting the simulated beams to the in-flight measurements.
The optical model was tuned to minimize the binned residual maps down to --15\,dB from the power peak, as described in \citep{planck2013-p28}.
This approach is preferable to {the use of} polynomial fits because it is less affected by the noise and the background: the optical model {turns out to be more stable than polynomial fits, so that the full focal plane can be simultaneously fitted with a single optical model.
This procedure yields} an ensemble of noise-free beams that are representative of the \Planck\ LFI flight optical beams, including both beam aberrations at very low levels and the cross-polarization response, which was not measured in flight.
Of course, before the comparison with the data, the optical beams are properly smeared to take into account the satellite motion.
Beam smearing comes from the fact that, while integrating toward a particular direction in the sky, the satellite moves and the optical beam is convolved with a top hat along the scanning direction. Since during the scanning the beam {is just shifting}, the convolution is equivalent to an average.
Whereas this effect is negligible in the calibration step \citep{planck2013-p02b}, this is not the case for the main beam measurements with planets, for which this effect smears the optical beam along the scan direction, increasing the beam asymmetries in a non-negligible way.

\begin{figure}[htpb]
\centering
\begin{tabular}{c c}
\includegraphics[width=4cm]{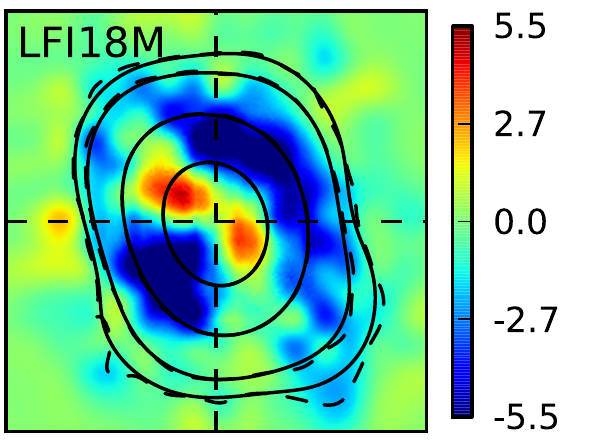} &
\includegraphics[width=4cm]{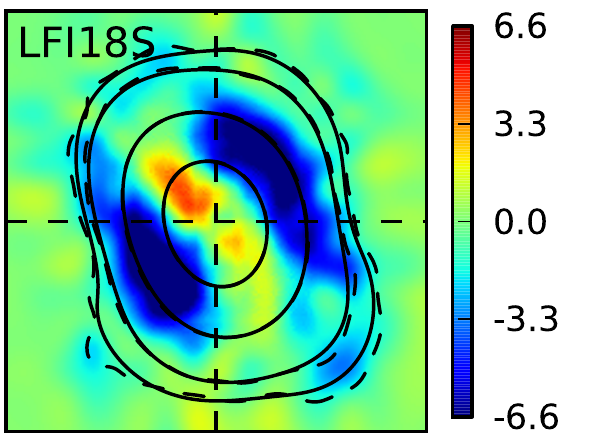} \\
\includegraphics[width=4cm]{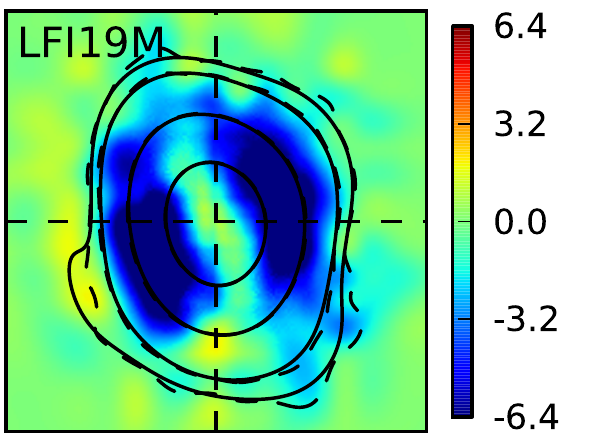} &
\includegraphics[width=4cm]{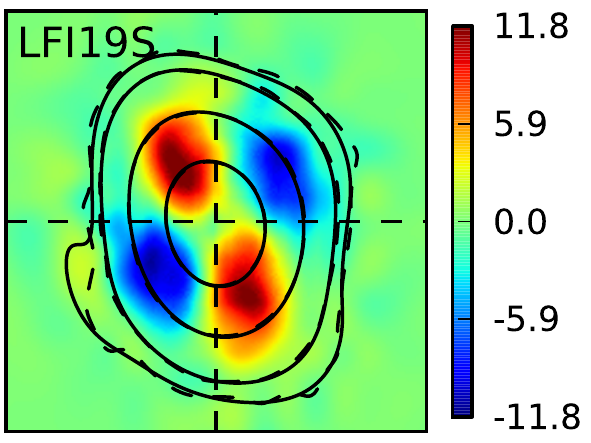} \\
\includegraphics[width=4cm]{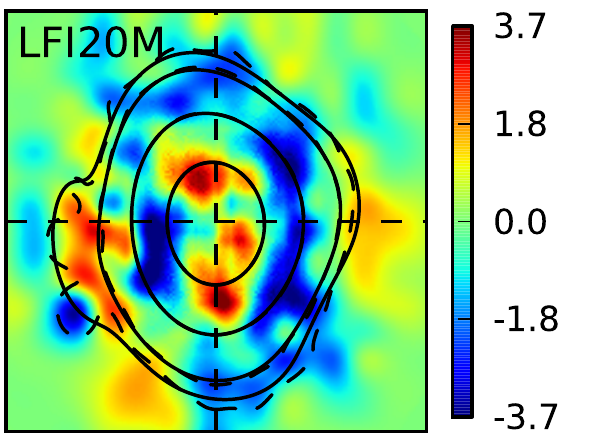} &
\includegraphics[width=4cm]{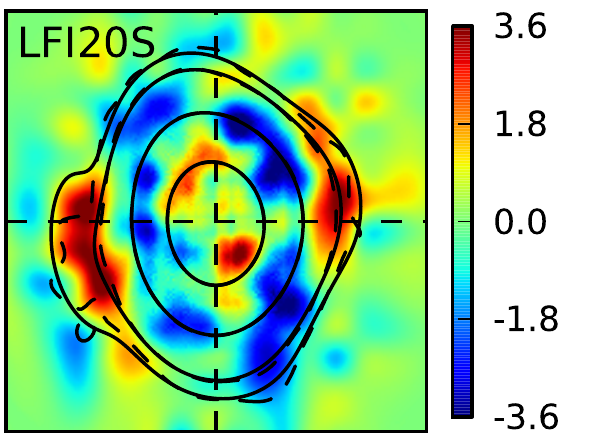} \\
\includegraphics[width=4cm]{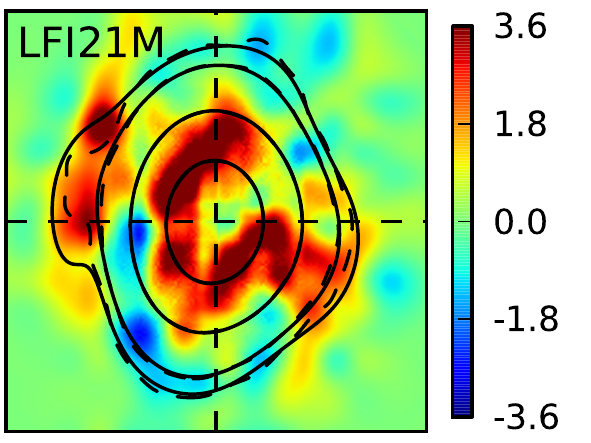} &
\includegraphics[width=4cm]{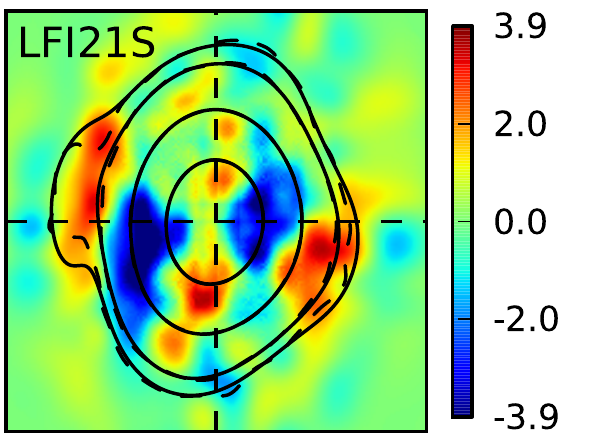} \\
\includegraphics[width=4cm]{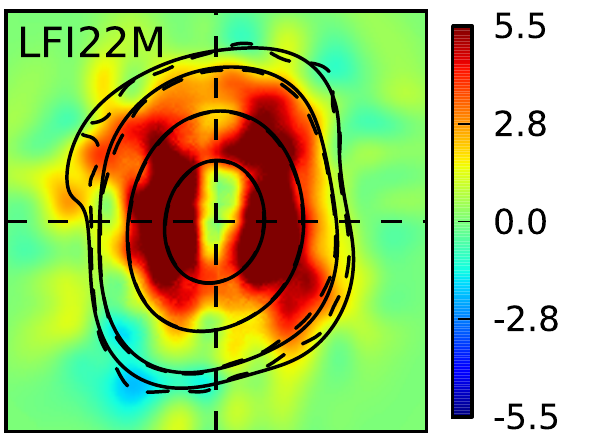} &
\includegraphics[width=4cm]{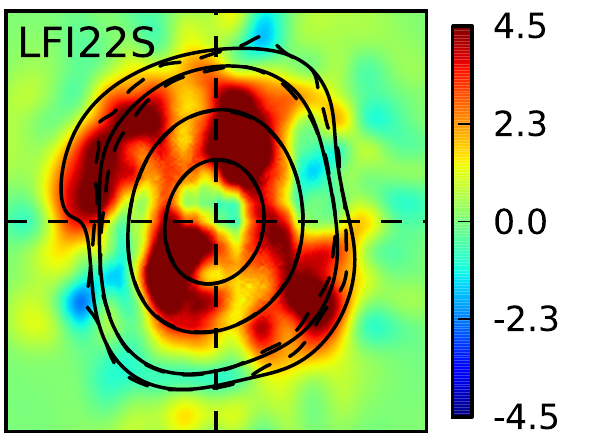} \\
\includegraphics[width=4cm]{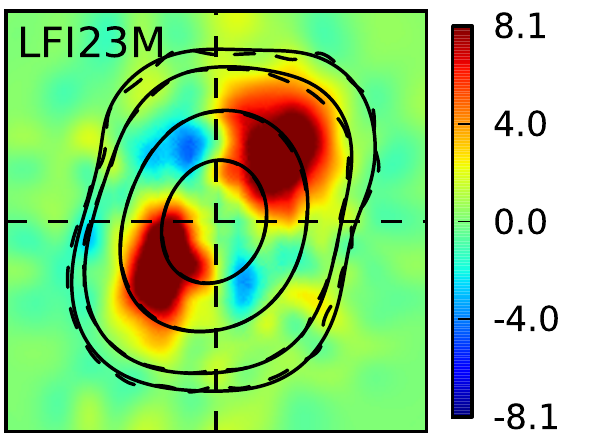} &
\includegraphics[width=4cm]{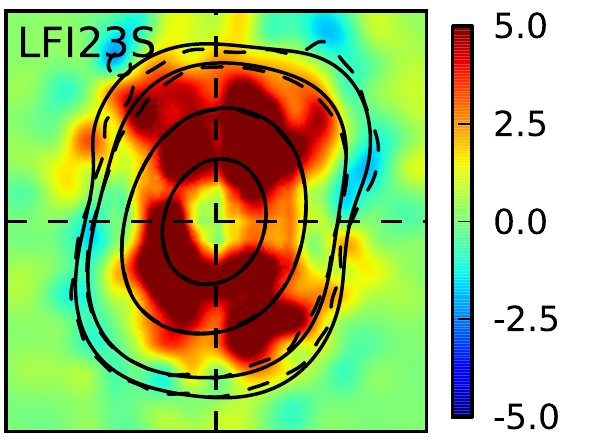} \\
\end{tabular}
\caption{
Difference between measured (dashed line) scanning beams and simulated (solid line) beams (70\,GHz channel). 
The color scale spans 2.25 times the rms of the beam difference and the units of the color bar are in thousandths of the peak height, i.e., 0.1\% of the beam maximum.
The contours correspond to --3, --10, --20, and --25 dB from the peak.
The size of each patch is 50'$\times$50', centered along the beam line of sight.}
\label{fig:comp70}
\end{figure}

\begin{figure}[htpb]
\centering
\begin{tabular}{c c}
\includegraphics[width=4cm]{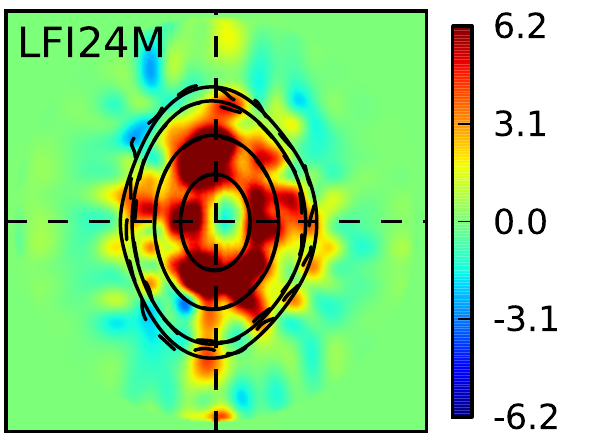} &
\includegraphics[width=4cm]{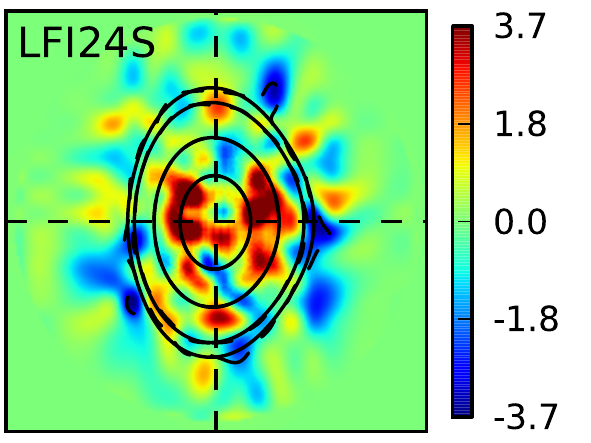} \\
\includegraphics[width=4cm]{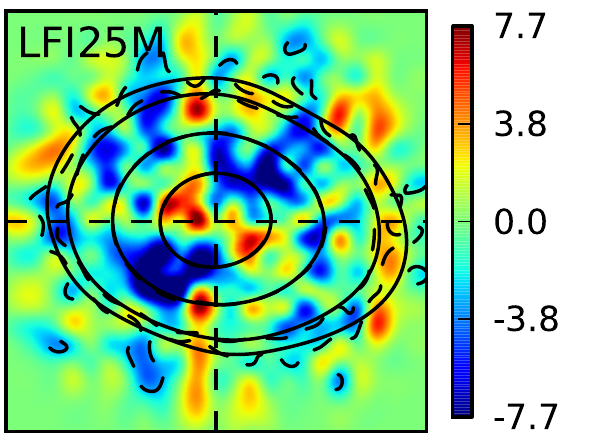} &
\includegraphics[width=4cm]{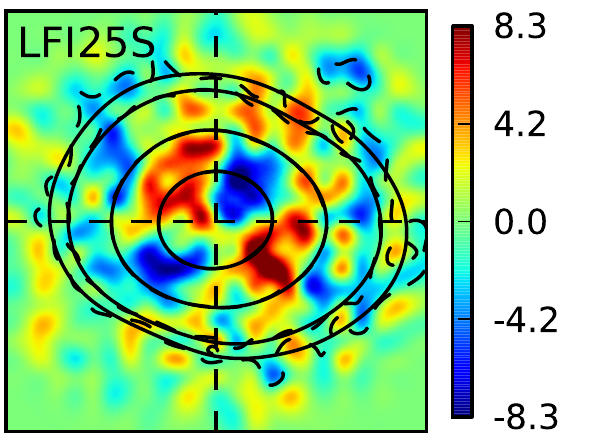} \\
\includegraphics[width=4cm]{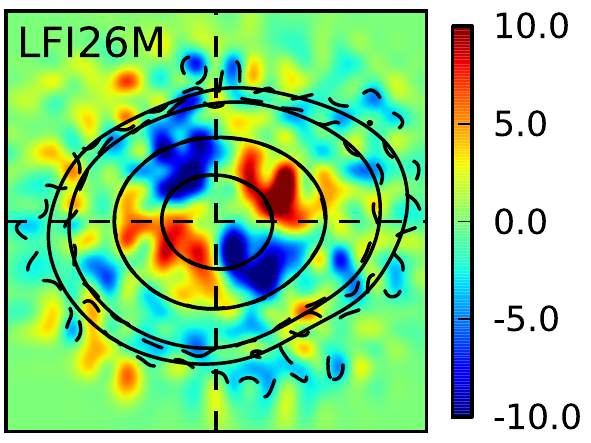} &
\includegraphics[width=4cm]{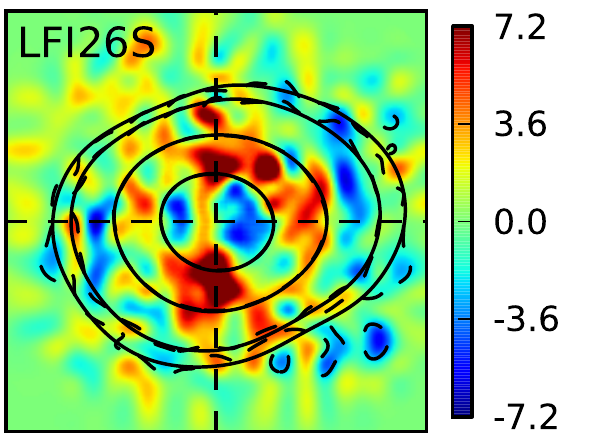} \\
\end{tabular}
\caption{
Difference between measured (dashed line) scanning beams and simulated (solid line) beams (44\,GHz channel). 
The color scale spans 2.25 times the rms of the beam difference and the units of the color bar are in thousandths of the peak height, i.e., 0.1\% of the beam maximum.
The contours correspond to --3, --10, --20, and --25 dB from the peak.
The size of each patch is 120'$\times$120', centered along the beam line of sight.}
\label{fig:comp44}
\end{figure}

\begin{figure}[htpb]
\centering
\begin{tabular}{c c}
\includegraphics[width=4cm]{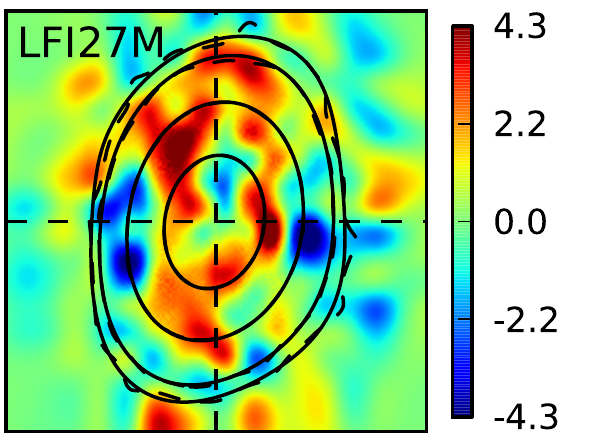} &
\includegraphics[width=4cm]{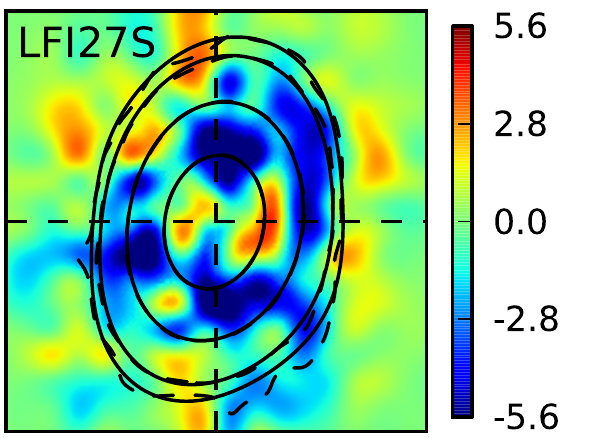} \\
\includegraphics[width=4cm]{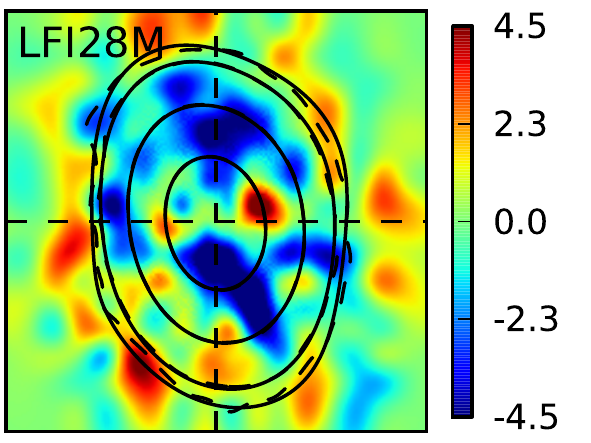} &
\includegraphics[width=4cm]{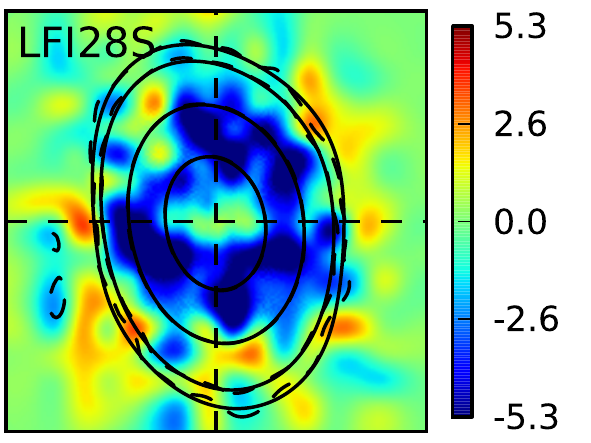} \\
\end{tabular}
\caption{Difference between measured (dashed line) scanning beams and simulated (solid line) beams (30\,GHz channel). 
The color scale is 2.25 times the rms of the beam difference and the units of the color bar are in thousandths of the peak height, i.e., 0.1\% of the beam maximum.
The contours correspond to --3, --10, --20, and --25 dB from the peak.
The size of each patch is 120'$\times$120', centered along the beam line of sight.}
\label{fig:comp30}
\end{figure}

In Fig.~\ref{fig:comp70} the maps obtained from the difference between measurements and simulations for the 70\,GHz beams are shown; the same comparison {is} plotted for the 44 GHz radiometers in Fig.~\ref{fig:comp44}, and for the 30\,GHz radiometers in Fig.~\ref{fig:comp30}.
The color scale spans 2.25 times the rms of the beam difference and the units of the color bar are in thousandths of the peak height, i.e., 0.1\% of the beam maximum. 
The color scale {is} symmetrized between the minimum and maximum values so that the zero level is shown as green in all the plots. 
The size of each patch is fixed: 120'$\times$120' for 30 and 44\,GHz and 50'$\times$50' for 70\,GHz.

Table \ref{mbe} reports the main beam efficiency of each LFI optical beam, together with the solid angles. 
The main beam efficiency is defined as:
\begin{equation}
\eta = \frac{\Omega_{{\rm MB}}}{{\bf\Omega_{{\rm A}}}}
\end{equation}
\noindent
{where $\Omega_{{\rm MB}}$ is the main beam solid angle and $\Omega_{{\rm A}}$ is the total antenna solid angle\footnote{In our simulations, since the pattern is normalized to the isotropic level, $\Omega_{{\rm A}} = 4\pi$ and the efficiency can be calculated integrating the pattern in the main beam region.}.}
In the same Table are also reported the main beam solid angles of the simulated and scanning beams; the values agree to better than 1\%.

\begin{table}[htpb]
\centering
\caption{Beam efficiency and solid angles computed {from} the optical beams and simulated beams. In the first column the main beam efficiency, $\eta$, derived from optical beams, is reported. {The second column reports} the percentage of the power entering the sidelobes ($f_{{\rm sl}}$): these values are directly computed as $1 - \eta$. In the last column the solid angles computed {from} the scanning beams are reported. {The solid angles (in arcmin$^2$) have been computed from the beams normalized to their maximum.} The 1\,$\sigma$ statistical error on the estimated solid angle from the scanning beams is about 0.2\%. The comparison between the simulated beams and the scanning beams shows that most of the solid angles agree to better than 1\%. Note, however, that only the simulated beams account for the cross-polarization response and low-level regions of the beams.  The averaged values of the simulated beams are 194, 850, and 1183 arcmin$^2$ at 70, 44, and 30\,GHz, respectively. The averaged values of the measured scanning beams are 193, 849, and 1182 arcmin$^2$ at 70, 44, and 30\,GHz, respectively.}
\begin{tabular}{c c c r@{.}l r@{.}l r@{.}l} 
\hline
\hline
\noalign{\vskip 2pt}
Beam &  $\eta$ & $f_{{\rm sl}}$ & \multicolumn{2}{c}{$\Omega_{{\rm opt}}$} & \multicolumn{2}{c}{$\Omega_{\rm sim}$} & \multicolumn{2}{c}{$\Omega_{\rm scn}$} \\
     &  (\%)   & (\%)    & \multicolumn{2}{c}{(arcmin$^2$)} & \multicolumn{2}{c}{(arcmin$^2$)} & \multicolumn{2}{c}{(arcmin$^2$)} \\
\hline
\noalign{\vskip 2pt}
\multicolumn{8}{l}{70 GHz} \\
\noalign{\vskip 4pt}
\texttt{18S} 	 & 99.34	& 0.66 & 198&10 &  203&28  &   205&81 \\    
\texttt{18M} 	 & 99.42	& 0.58 & 196&89 &  201&84  &   203&98 \\    
\texttt{19S} 	 & 99.29  & 0.71 & 188&65 &  193&34  &   193&51 \\  
\texttt{19M} 	 & 99.35	& 0.65 & 186&61 &  191&60  &   195&04 \\  
\texttt{20S} 	 & 99.18	& 0.82 & 181&21 &  185&63  &   185&51 \\  
\texttt{20M} 	 & 99.21	& 0.79 & 180&43 &  185&20  &   185&45 \\  
\texttt{21S} 	 & 99.20	& 0.80 & 182&50 &  186&94  &   186&63 \\  
\texttt{21M} 	 & 99.21	& 0.79 & 181&26 &  185&71  &   183&87 \\  
\texttt{22S} 	 & 99.27	& 0.73 & 188&18 &  193&07  &   190&22 \\  
\texttt{22M} 	 & 99.34	& 0.66 & 187&45 &  192&07  &   188&24 \\  
\texttt{23S} 	 & 99.35	& 0.65 & 199&95 &  204&84  &   200&91 \\    
\texttt{23M} 	 & 99.43	& 0.57 & 198&74 &  203&72  &   200&99 \\    
\hline                                                                                                           
\noalign{\vskip 2pt}                                                                                             
\multicolumn{8}{l}{44 GHz} \\                                                                                
\noalign{\vskip 4pt}                                                                                             
\texttt{24S}   & 99.84	& 0.16 & 576&85  & 590&99  &   591&86 \\
\texttt{24M} 	 & 99.79	& 0.21 & 589&99  & 602&42  &   594&76 \\
\texttt{25S} 	 & 99.80	& 0.20 & 1020&68 & 1041&63 &  1040&47 \\
\texttt{25M} 	 & 99.79	& 0.21 & 967&93  & 990&28  &   996&72 \\
\texttt{26S} 	 & 99.80	& 0.20 & 1006&67 & 1027&13 &  1019&03 \\
\texttt{26M} 	 & 99.79	& 0.21 & 967&93  & 989&89  &   993&56 \\
\hline\noalign{\vskip 2pt}                                                                                       
\multicolumn{8}{l}{30 GHz} \\                                                                                
\noalign{\vskip 4pt}                                                                                             
\texttt{27S} 	 & 99.33	& 0.67 & 1153&02 & 1181&94  &  1184&64 \\ 
\texttt{27M} 	 & 99.30	& 0.70 & 1158&00 & 1186&14  &  1174&48 \\ 
\texttt{28S} 	 & 99.34	& 0.66 & 1153&14 & 1180&99  &  1188&41 \\ 
\texttt{28M} 	 & 99.29	& 0.71 & 1152&56 & 1181&98  &  1179&34 \\ 
\hline
\end{tabular}
\label{mbe}
\end{table}

\subsubsection{Beam validation through deconvolution}

To test the goodness of the beam representation, the maps for each individual horn at 30\,GHz and 44\,GHz have been deconvolved using the {\tt ArtDeco} beam deconvolution algorithm described in \citet{keihanen2012}.  
The code takes as input the time-ordered data stream, along with pointing information and the harmonic representation of the simulated beam, to construct the harmonic $a_{slm}$ coefficients that represent the sky signal. 
From the harmonic coefficients we further construct a sky map, which is now free from the effects of beam asymmetry, assuming that our beam representation is correct.

Before deconvolution we ran the time-ordered data through the {\tt Madam} map-making code \citep{keihanen2010}, to remove low-frequency noise. We  saved the baselines that represent the correlated noise component, and subtracted  them from the original data stream. 
The cleaned data thus consist of signal with a residual noise component that is dominated by white noise. 
This {is what we } used as input to the deconvolution code.

\begin{figure}[hptb]
\centering
\begin{tabular}{cc}
\includegraphics[width=4cm]{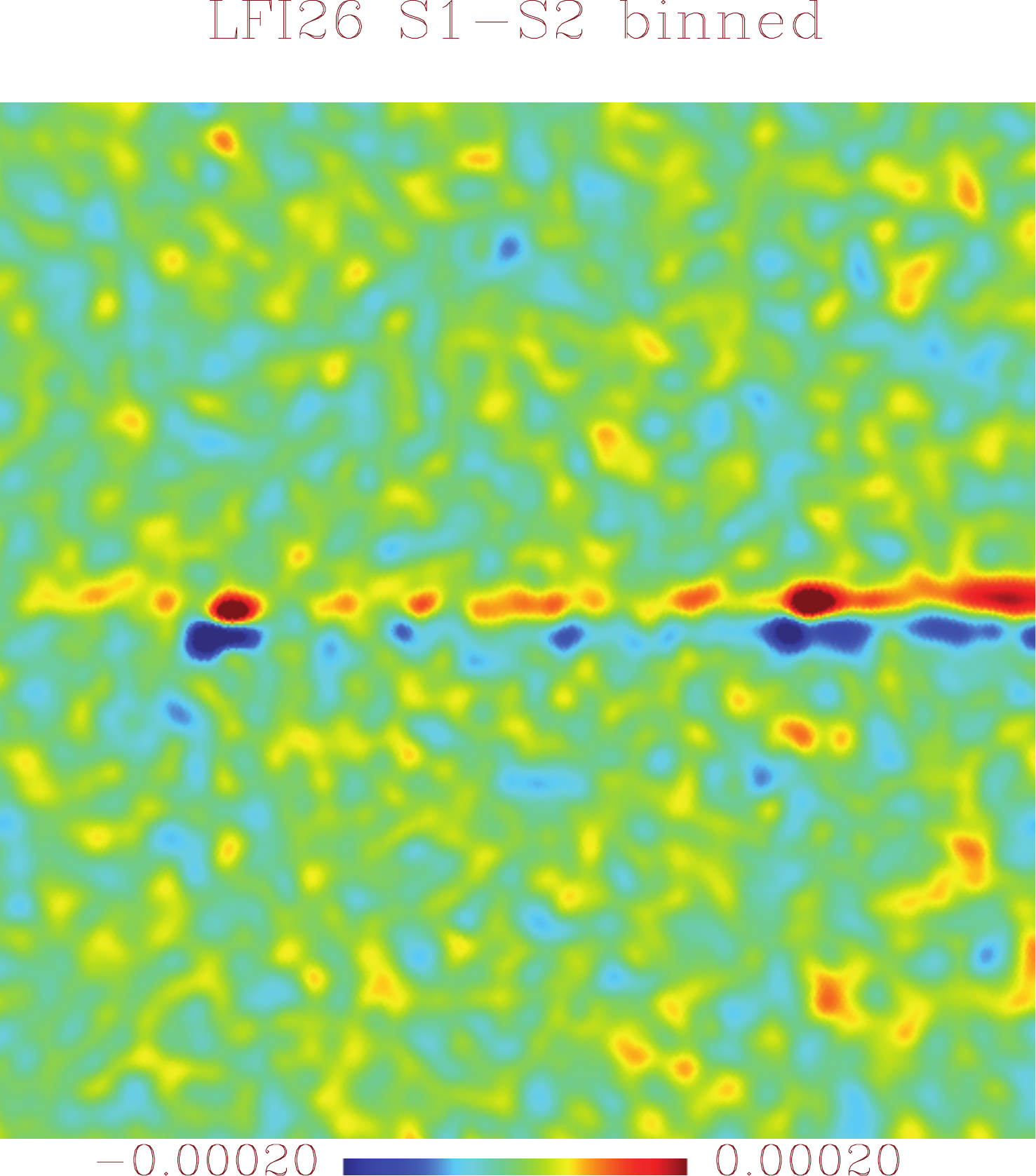} &
\includegraphics[width=4cm]{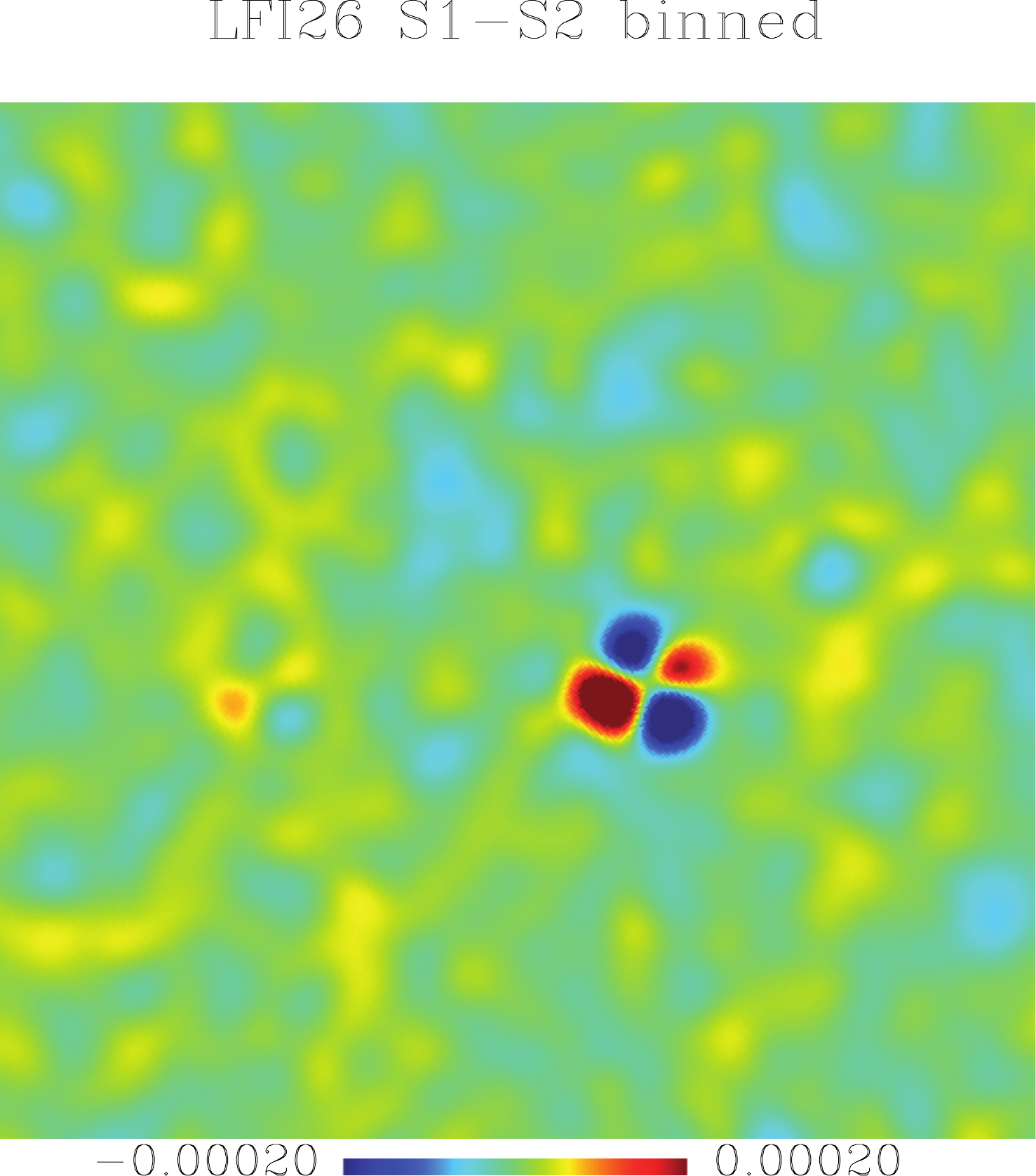} \\
\noalign{\vskip 2pt}
\includegraphics[width=4cm]{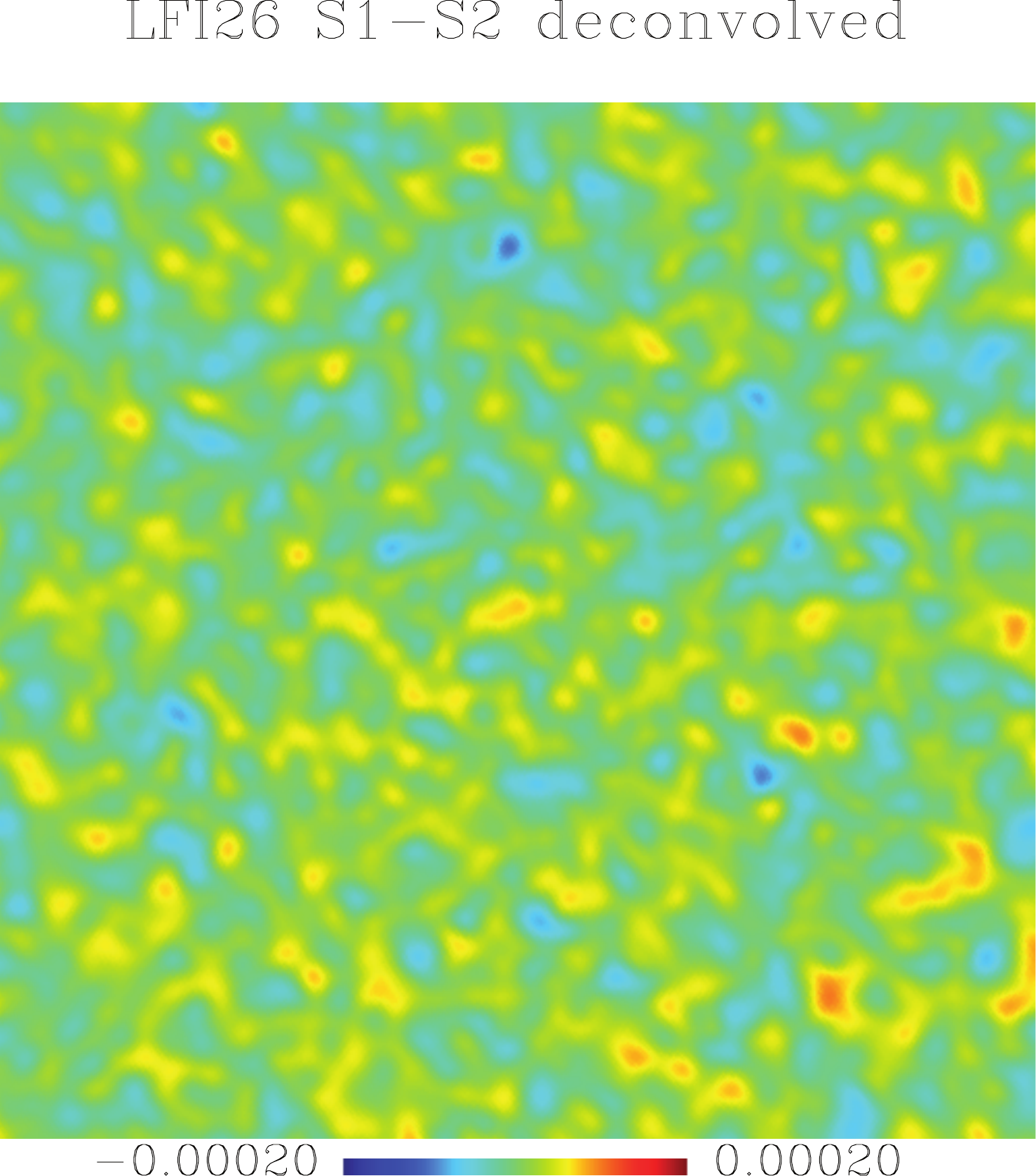} &
\includegraphics[width=4cm]{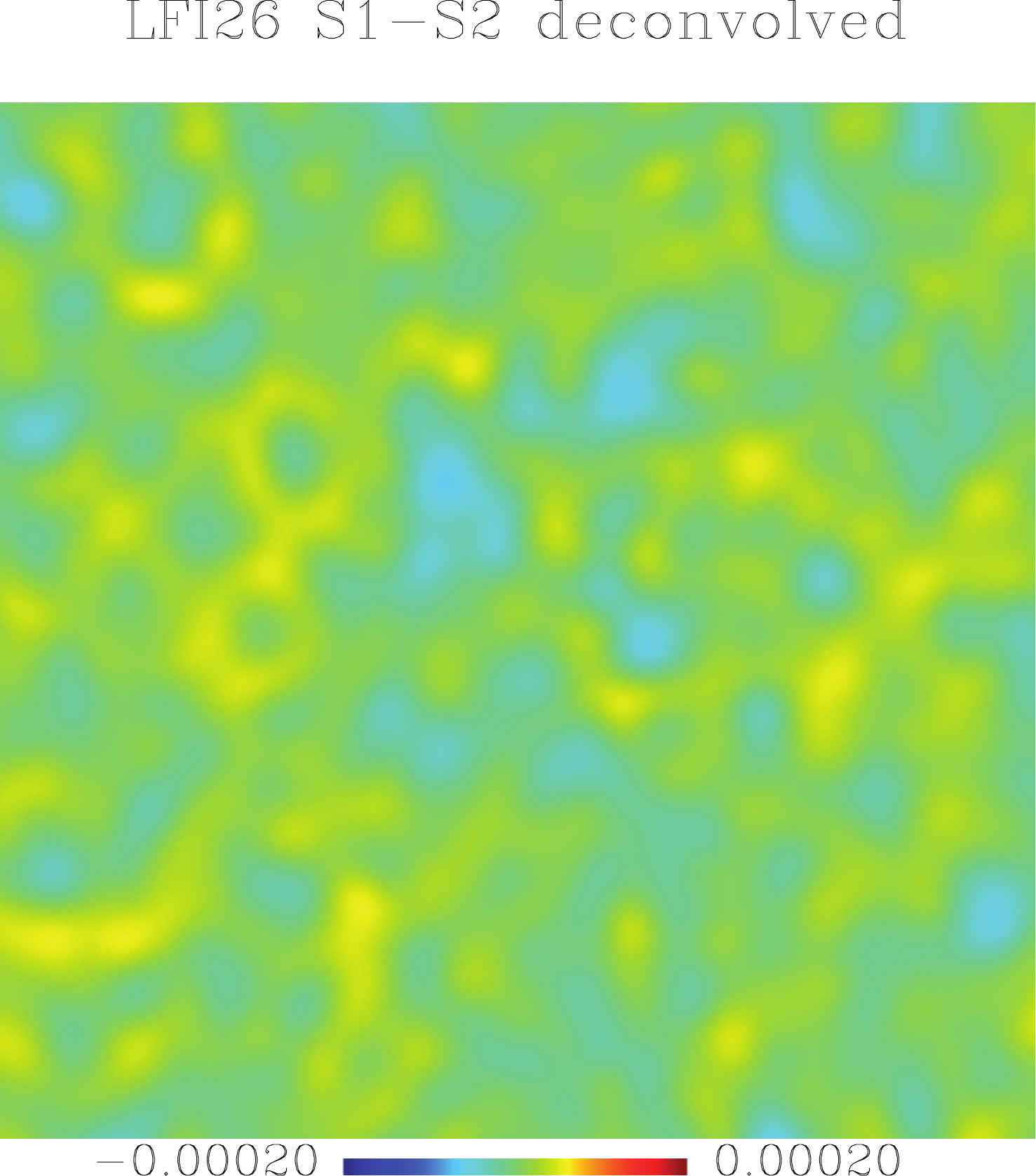} \\
\end{tabular}
\caption{Comparison between survey differences {(Survey 1 -- Survey 2)} of binned and deconvolved maps obtained using the simulated beam, \texttt{LFI26}.
The maps {are} smoothed to 1$^\circ$ resolution in order to suppress noise, and the units are Kelvin. 
The first row of each image corresponds to the binned map, showing a zoom into the Galactic region slightly left from the center (left), and into an unidentified point source {at a location near} (--90$^\circ$,0$^\circ$) (right).  The improvement in the deconvolved images is clear.}
\label{fig:deconvolution}
\end{figure} 

We ran the deconvoultion of data from each single survey {(where a survey is defined as a scan of the full sky)}, and looked for residual differences between single-survey maps.
Results for horn \texttt{LFI26} are shown in Fig.~\ref{fig:deconvolution}, where the difference between first ``S1'' and second ``S2'' survey maps is reported.
The left-hand column shows a zoom into the Galactic region at location {b $ = 40^\circ$, l $ = 0^\circ$}.
One image covers a square of width 13.3$^\circ$.
The right-hand column shows a zoom into a point source {at a location near} (--90$^\circ$,0$^\circ$). 
The width of this image is 16.7$^\circ$.
In the absence of beam asymmetry and other systematics the difference should be due to noise only.

{The top row of} Fig.~\ref{fig:deconvolution} shows, for comparison, the difference between binned maps. 
In this case, the maps were binned directly from the time-ordered data, without attempting to correct for beam effects. 
A given region on the sky is scanned with different beam orientations during the different surveys.
This gives rise to the residual signal that is evident in the top row images.

The maps were smoothed to a 1$^\circ$ (FWHM) resolution, in order to suppress noise.
In the case of binned maps this was achieved by smoothing with a symmetric Gaussian beam with FWHM of 50'. 
Combined with the width of the radiometer beam, this gives a total smoothing of approximately 1$^\circ$.

The bottom row shows the corresponding difference of the deconvolved maps.
We show the same regions as in the top row and with the same scaling. 
We smoothed the deconvolved harmonic coefficients with a 1$^\circ$ (FWHM) Gaussian beam, and constructed a sky map through harmonic expansion.
Deconvolution almost completely removes the Galactic residual, as well as the ``butterfly'' residual pattern of the point source.
This indicates that the simulated beams, based on the tuned optical model, are a good representation of the true beams.

The deconvolution is not part of the nominal pipeline but this test provides an important cross-check on the beam representation since it tests the beam model against the data in a way that is independent from the construction of the model. 

\subsubsection{Spectral dependence of beam geometry}

Throughout this work, we have assumed a monochromatic response at each LFI frequency.
In fact, the bandpasses are wide, and vary in detail from one radiometer to another, even within the same band.
The effective center frequency for each band used in this paper was calculated assuming a thermal (CMB) spectrum.
For different source spectra, the central frequency shifts. We must also take into account the fact that the beam pattern has some frequency dependence.
The geometry of the beams is characterized by three parameters described in the previous section: the FWHM; the ellipticity; and the orientation of the beam $\psi_{{\rm ell}}$.
We have investigated the effect on the LFI beams of assuming a power-law spectrum {$S_\nu =\nu^\alpha$ } with power index $\alpha$ ranging from $-6$  to $+6$, where $\alpha=2$ is representative of the CMB spectrum and $\alpha=0$ of a flat spectrum.
We start{ed} by generating {\tt GRASP} models of the main beam $\BeamNu(\Pointing)$ at a set of frequencies defined by splitting the bandpass into 15 equally spaced steps centered on the nominal central band frequency.
The 15 beam maps were then averaged by weighting each {\tt GRASP} map pixel by the bandpass $\bandpass(\nu)$ and the source spectrum $\nu^\alpha$, giving the effective beam pattern

\begin{equation}
  \BeamA(\Pointing) = N^{-1} \int d\nu \; \bandpass(\nu) \nu^\alpha \BeamNu(\Pointing),
\end{equation}

\noindent
with $N=\int d\nu \; \bandpass(\nu) \nu^\alpha$.
{Then we derived} the geometric beam parameters as a function of $\alpha$.
Since the telescope is achromatic, only slight variations of the geometric beam parameters are observed. 
In addition, the bandpass averaging process further reduces the variability with respect to the monochromatic case.
The most interesting result is that the three geometrical parameters vary nearly linearly with $\alpha$, with different slopes for each radiometer.
The most sensitive radiometer in FWHM is \texttt{LFI28-S}, {for which $d$FWHM/$d\alpha$ is about} $+3\times10^{-4}$ degrees.
Changing $\alpha$ from $+2$ to $-2$ {causes} a relative change of at most $0.2\%$, $0.3\%$ and $0.4\%$ respectively in the FWHM at 30, 44, and 70\,GHz, well below the 1\% level.
A similar range of relative variations occurs for the beam ellipticity. 
For the orientation parameter, $\psiell$, the amplitude of $d\psiell$/$d\alpha$ varies from a minimum value of $-2\times10^{-4}$ degrees (for the detector \texttt{LFI24-S}) to a maximum of 0.36 degrees (for \texttt{LFI26-M}), so that a change in the spectral index $\alpha$ from $-2$ to $+2$  {produces} a rotation {$\psiell$ } of the beam of 1.4 degrees at most, in one direction or the other.
Those values (assuming {a thermal CMB} spectrum) contribute to the overall calibration uncertainty \citep{planck2013-p02b}, and we emphasize that these uncertainties in the beam properties are completely independent of the color corrections needed to adjust intensity scales for sources with non-thermal spectra. 

{Whereas the impact of the main beam variation across the band is small, this is not true for the near and far sidelobes. For this reason, the variation of the sidelobes across the band has been taken into account in the error budget evaluation, as reported in Sect.~\ref{sec:errors}.}

\subsection{Sidelobes}

The response of the beam pattern outside the main beam needs to be carefully understood, as it may have significant impact on the \Planck\ data analysis.
Although a full physical optics computation could be developed to predict accurately the antenna pattern of the telescope, this is not {feasible} for the whole-spacecraft simulations since the physical optics approach {is very complicated} when multiple diffractions and reflections between scattering surfaces are involved. 
For this reason,  we have calculated the sidelobe patterns through the {\tt GRASP} multi-reflector geometrical theory of diffraction (MrGTD), which computes the scattered field from the reflectors by performing backward ray tracing. This represents a suitable method for predicting the full-sky radiation pattern of complex mm-wavelength optical systems in a reasonable time.
The MrGTD sequentially computes the diffraction fields from any reflector surfaces that are illuminated, starting from the feed horn.
The sequence of scatterers and the type of interaction (reflection or diffraction, occurring on each scatter) must be defined in the input to the simulation.
The simplest (first order) optical contributions producing significant power levels are reflections onto the sub-reflector, onto the main reflector, and onto the baffle, as well as diffractions by the sub-reflector, by the main reflector, and by the baffle. 
Other non-negligible contributions derive from two interactions with the reflectors (second order -- for example, rays reflected on the sub-reflector and then diffracted by the main reflector), three interactions (third order -- for example, rays reflected on the sub-reflector, diffracted by the main reflector, and then diffracted by the baffle) and so on.  
Although MrGTD is, in general, less time consuming than a full physical optics calculation, it should be applied in a rigorous way in order to obtain reliable results, especially at low power levels (down to --50 dBi). In addition, when many scattering surfaces are involved, the number of ray {tracings} needed may lead to unacceptable computational time, even with MrGTD.
Since our analysis requires the production of band-integrated patterns to account for the frequency-dependent beam responses and the radiometer bandpasses, for now the sidelobes simulations have been carried out only up to the first order plus two contributions at the second order (reflections and diffractions on the sub-reflector, and then diffracted by the main reflector){\bf: the final error budget will then also take into account this approximation.} 

{The contributions to beam solid angle found in this simulation of  the sidelobe region using the MrGTD up to the first order are about a factor two lower than the expected value derived from physical optics calculations (see $f_{{\rm sl}}$ reported in Table \ref{mbe}).
That suggests that} the first order approximation adopted in this subsection underestimates the actual integrated power in the sidelobes.
In the future, it will be necessary to take into account the impact of higher order contributions in combination with physical optics analyses.

\begin{figure}[!hp]
\centering
\begin{tabular}{c}
\includegraphics[width=8.5cm]{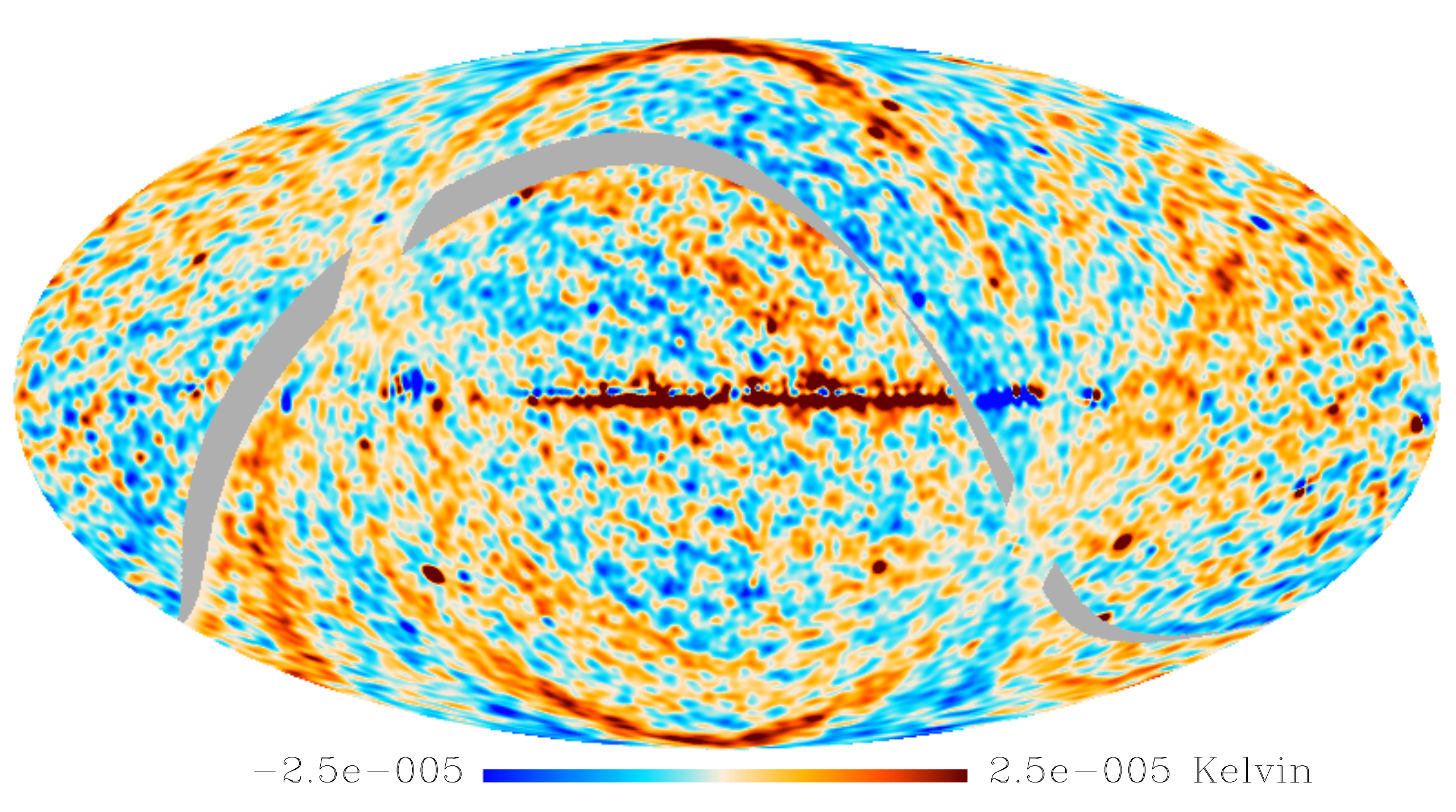} \\
\includegraphics[width=8.5cm]{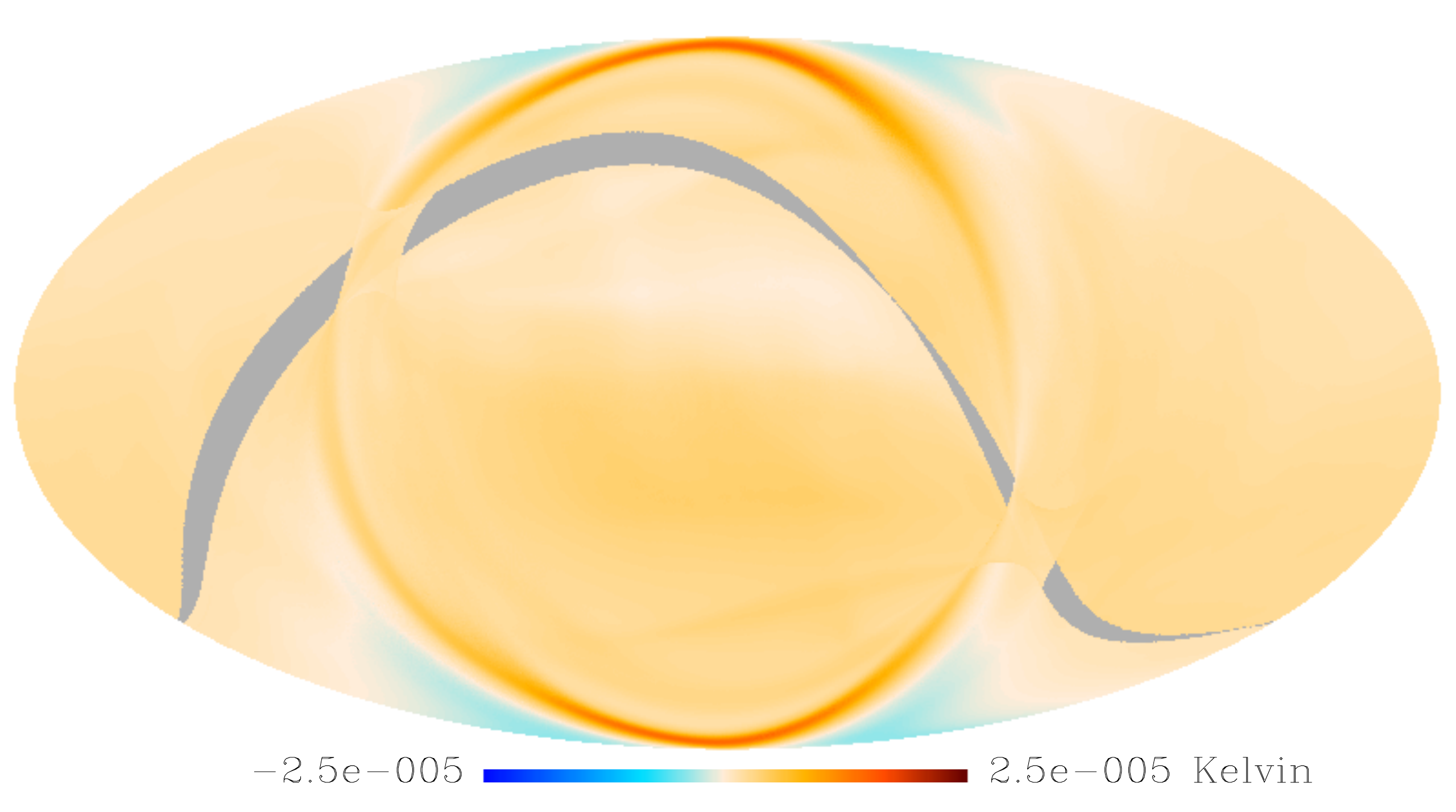} \\
\includegraphics[width=8.5cm]{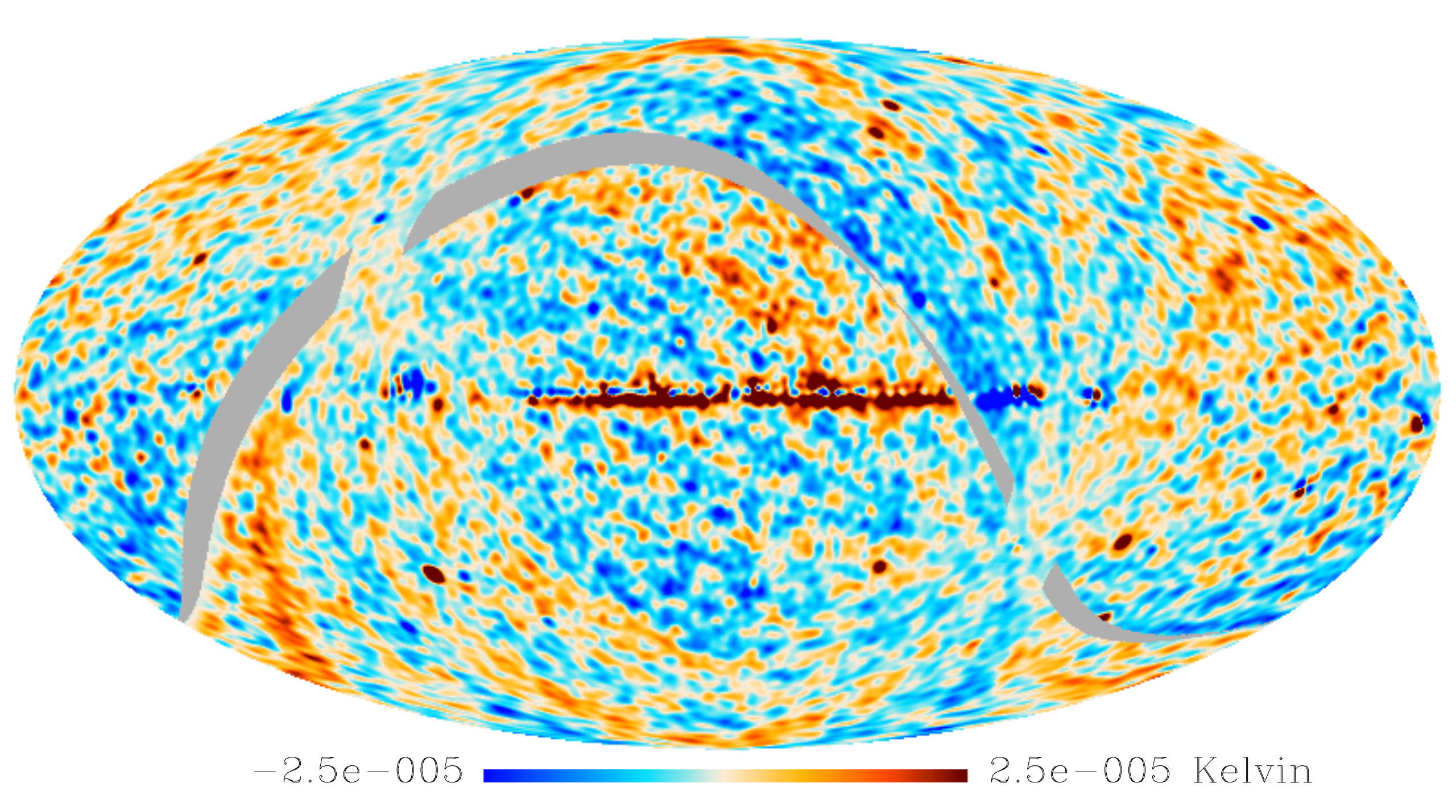} \\
\includegraphics[width=8.5cm]{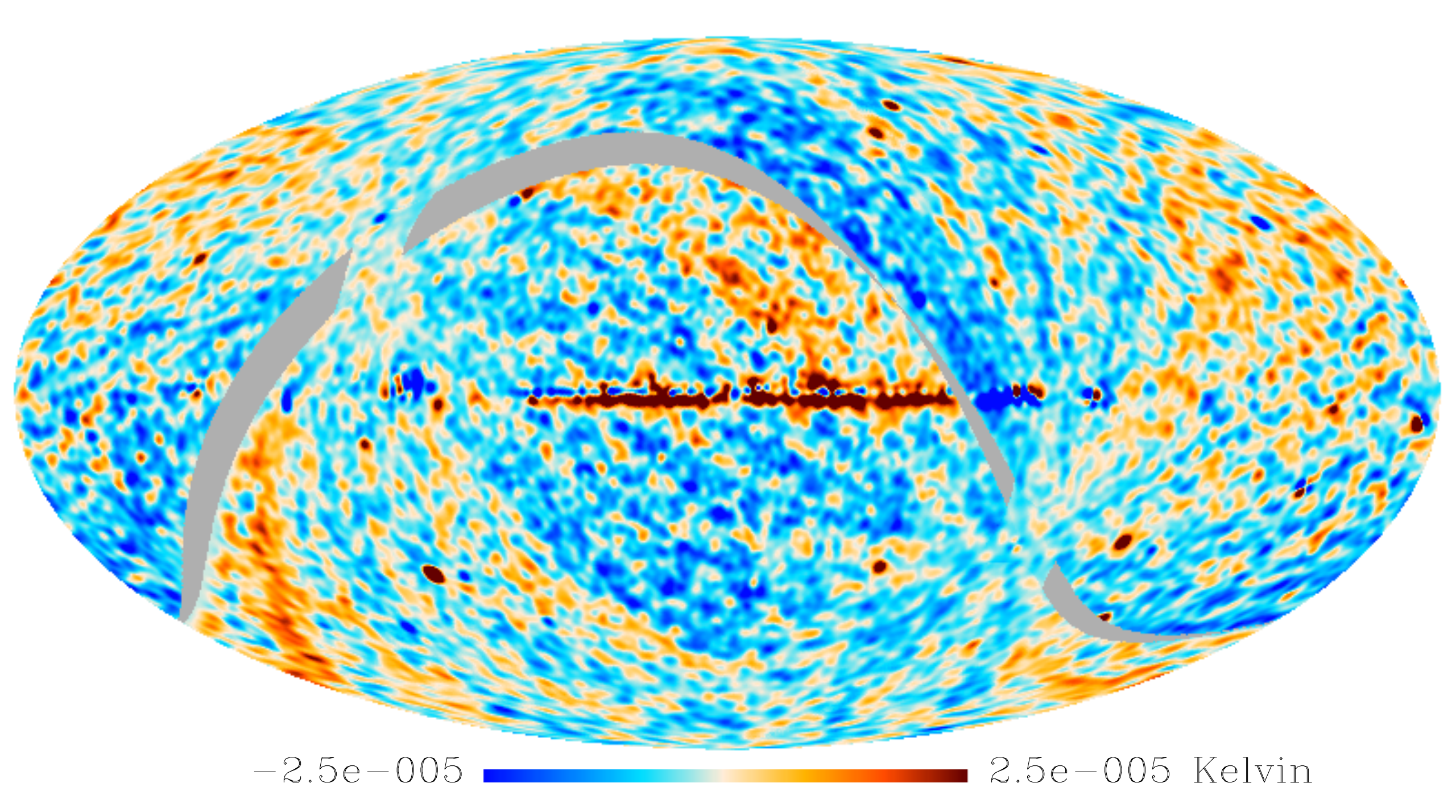} \\
\end{tabular}
\caption{
{In all the panels, Survey 2 -- Survey 1 difference maps are shown for the 30\,GHz channel.} 
{\it Top}: difference map S2--S1 for the real data, in $\mu$K. 
{\it Second from top}: difference map S2--S1 for simulated Galactic straylight. 
{\it Third from top}: simulated difference map subtracted from data difference map (difference between the first two panels). 
The ring still visible in the third panel suggests that the model should be improved by adding higher order contributions, possibly combined with physical optics analysis.
{\it Bottom}: difference S2--S1 between data and simulations, amplified by a factor equal to the ratio of the power missing the main beam ($f_{{\rm sl}}$) and the power entering the simulated sidelobes using the first order approximation (in this case, this ratio is about 1.93). The grey band shows areas not scanned.
}
\label{fig:sss}
\end{figure}

Careful analysis of the LFI 30\,GHz data {reveals} the imprint of Galactic radiation received through the far sidelobes. 
Such a detection is amplified when taking the difference between maps of even and odd surveys: {the different satellite orientation during odd and even surveys reverses the sidelobe pattern with respect to the Galactic radiation}.
A detailed discussion of the systematic effects introduced by sidelobe pickup at 30\,GHz is given in \citet{planck2013-p02a}. 
The expectations of the sidelobe pick-up based on the known level of Galactic emission (as measured by \Planck\ itself) and our sidelobe model, are in good qualitative agreement with the observed effect, as shown in Fig.~\ref{fig:sss}.
The residual ring clearly visible in the third panel of Fig.~\ref{fig:sss} demonstrates the need to improve  the sidelobe model with higher order contributions, possibly combined with full physical optics analyses. 
In the bottom panel of the Fig.~\ref{fig:sss} we show the difference between data and the simulations amplified by a factor equal to the ratio of the power entering the sidelobes (computed from the main beam efficiency) and the integral of the simulated sidelobes.
It is evident that, once the sidelobe amplitude {is re-normalized}, the ring artifact almost completely disappears.

It should be noted that, while the sidelobe effect introduces additional complication in the analysis, its detection at 30\,GHz provides an important validation of the simulated beams, which can be trusted even to very low power levels in the higher-frequency \Planck\ cosmological channels as well, for which the sidelobes signatures are not measurable. 
{Future analysis aimed at CMB polarization will make use of refined in-band integrated beams for each radiometer}.
{Other accurate studies of the beam systematics in different CMB experiments can be found in \cite{QUIET1}, \cite{QUIET2}, \cite{QUIETinstrument}, \cite{BICEP}, \cite{WMAPbeam}, and \cite{jarosik2010}.}

\section{Effective Beams} 
\label{effective_beams}
The effective beam is the average of all scanning beams that cross a given pixel of the sky map, given $Planck's$ scan strategy. 
The effective beams capture the pointing information about the difference between the true and observed images of the sky. 
They are, by definition, the objects whose convolution with the true CMB sky produce the observed sky map {in the absence of sidelobes}.
Similarly, the effective beam window functions capture the difference between the true and observed angular power spectra of the sky.
We compute the effective beam at each sky pixel for each LFI frequency scanning beam and scan history  using the {\tt FEBeCoP} \citep{mitra2010} method, {as in} \Planck's early release \citep{planck2011-1.7}. 

The pre-computation of the effective beams was executed at the National Energy Research Scientific Computing Center (NERSC) in Berkeley (California). 
The beam data were delivered to the \Planck\ data processing centers \citep{planck2013-p02,planck2013-p03} over the network, on tape and disk, and ingested into the Data Management Component (DMC). 
{\tt FEBeCoP} associated application software was developed and installed to use the effective beams, e.g., fast Monte Carlo full sky convolution codes.

In estimating the effective beams, a cut-off is applied to the input simulated beams. 
{The several tests performed converged} to a cut-off radius of 2.5 $\times$ FWHM.  
{The beam within this cut-off radius is named nominal beam and it is the portion of the beam used to create the beam window function.}
The beam efficiency of the simulated beams within this cut-off radius is reported in Table \ref{mbeco}.

\begin{table}[hptb]
\centering
\caption{Main beam efficiencies computed {from} the simulated beams using a cut-off radius of 2.5 $\times$ FWHM: {efficiencies from the OMT's} main and side arm are reported in the first two columns, the average value of the two arms is reported in the third column, and the difference between the two arms is reported in the last column.}
\begin{tabular}{c c c c r@{.}l} 
\hline
\hline
\noalign{\vskip 2pt}
 Horn  &   Main OMT   &  Side OMT    &  Mean   &   \multicolumn{2}{c}{(Diff) OMT}   \\
\hline
\noalign{\vskip 2pt}
\multicolumn{5}{l}{70\,GHz} \\
\noalign{\vskip 4pt}
\texttt{LFI-18} &  0.99345 & 0.99262 & 0.99304 & --0&00082  \\
\texttt{LFI-19} &  0.99270 & 0.99206 & 0.99238 & --0&00065  \\
\texttt{LFI-20} &  0.99111 & 0.99084 & 0.99098 & --0&00027  \\
\texttt{LFI-21} &  0.99115 & 0.99105 & 0.99110 & --0&00010  \\
\texttt{LFI-22} &  0.99259 & 0.99184 & 0.99222 & --0&00075  \\
\texttt{LFI-23} &  0.99360 & 0.99274 & 0.99317 & --0&00086  \\
\hline
\noalign{\vskip 2pt}
\multicolumn{5}{l}{44\,GHz} \\
\noalign{\vskip 4pt}
\texttt{LFI-24} &  0.99762 & 0.99826 & 0.99794 &   0&00064  \\
\texttt{LFI-25} &  0.99788 & 0.99792 & 0.99790 &   0&00005  \\
\texttt{LFI-26} &  0.99787 & 0.99793 & 0.99790 &   0&00006  \\
\hline
\noalign{\vskip 2pt}
\multicolumn{5}{l}{30\,GHz} \\
\noalign{\vskip 4pt}
\texttt{LFI-27} &  0.99247 & 0.99282 & 0.99264 &   0&00036  \\
\texttt{LFI-28} &  0.99230 & 0.99284 & 0.99257 &   0&00054  \\
\hline
\end{tabular} 
\label{mbeco}       
\end{table}     
 
For a detailed account of the algebra involving the effective beams for temperature and polarization see \citet{mitra2010}.
Here {the main results are summarized}. 

The observed temperature {map} $\widetilde{\vec{T}}$ is a convolution of the true {map} $\vec{T}$ and the effective beam $\vec{B}$,
%
%
\begin{equation}
\widetilde{\vec{T}} \ = \ \Delta\Omega \, \vec{B} \cdot \vec{T},
\end{equation}

\noindent
where the {elements of the} effective beam {matrix} ${B} $ can be written for the temperature in terms of the pointing matrix $\tens{A}_{ti}$ and the scanning beam $b(\hat{{\vec{r}}}_j,\hat{{\vec{p}}}_t)$ as
\begin{equation}
B_{ij} \ = \ \frac{\sum_t \tens{A}_{ti} \, b(\hat{\vec{r}}_j, \hat{\vec{p}}_t)}{\sum_t \tens{A}_{ti}} \, .
\label{eq:EBT}
\end{equation}
Here $t$ represents time samples, $\tens{A}_{ti}$ is 1 if the pointing direction falls in pixel number $i$, else it is 0; $\vec{p}_t$ represents the exact pointing direction (not approximated by the pixel center location), and $\hat{\vec{r}}_j$ is the center of pixel number $j$, where the scanning beam $b(\hat{\vec{r}}_j, \hat{\vec{p}}_t)$ is being evaluated (if the pointing direction falls within the cut-off radius of 2.5 $\times$ FWHM, for LFI channels).
An analogous formula can be written for the temperature + polarization effective beam, including the weight vector $\vec{w}_t \equiv [1,\gamma\cos(2\psi),\gamma\sin(2\psi)]$, as:
 
\begin{equation}
B_{ij} =\left[ \sum_{t} \tens{A}_{ti}\vec{w}_t \vec{w}_t^{\rm T} \right]^{-1}\sum_{t} \tens{A}_{ti} b(\hat{r}_j,\hat{p}_t)\vec{w}_t\vec{w}_t^{\rm T}.
\end{equation}

\begin{figure*}[htpb]
\centering
\includegraphics[width=17cm]{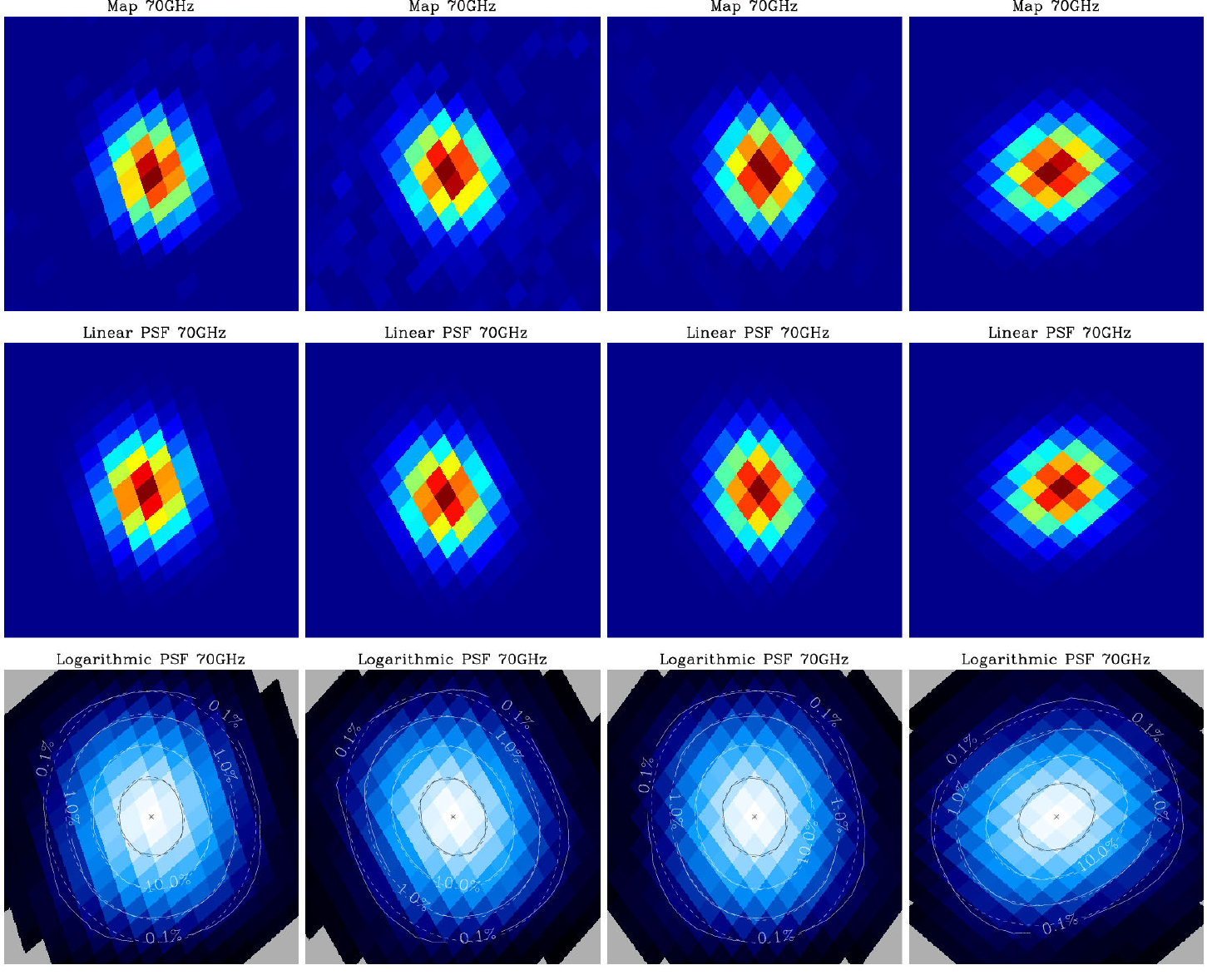}
\caption{Four ERCSC sources as seen by LFI 70\,GHz channel (upper panel); linear scale FEBeCoP {Point Spread functions (PSFs)} computed using input simulated beams (central panel); both in arbitrary units. Bottom panel: PSF iso-contours shown in solid line, elliptical Gaussian fit iso-contours shown in broken {lines}. PSFs are shown in log scale. {The galactic coordinates in degrees $ \ell, b $ of the four sources, from left to right, are, respectively, (306.1, 57.1), (86.1, -38.2), (290.0, 64.4), (184.5,-5.8)}.}
\label{fig:ploteffectivebeam70}
\end{figure*}

\begin{figure*}[htpb]
\centering
\includegraphics[width=17cm]{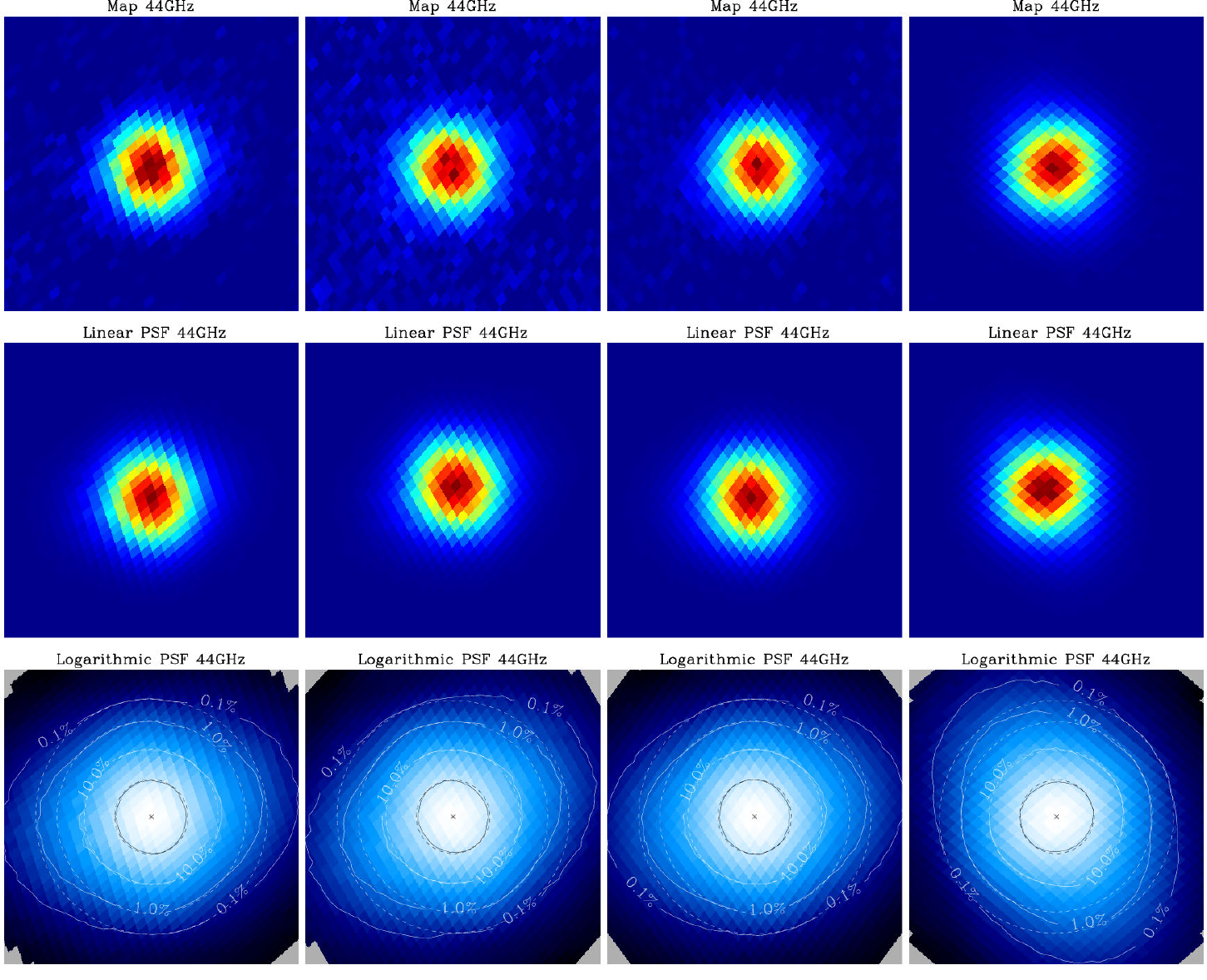}
\caption{The same four ERCSC sources as seen by LFI 44\,GHz channel (upper panel); linear scale FEBeCoP {Point Spread functions (PSFs)} computed using input simulated beams (central panel); both in arbitrary units. Bottom panel: PSF iso-contours shown in solid line, elliptical Gaussian fit iso-contours shown in broken {lines}. PSFs are shown in log scale.{The galactic coordinates are as in the previous figure.}}
\label{fig:ploteffectivebeam44}
\end{figure*}

\begin{figure*}[htpb]
\centering
\includegraphics[width=17cm]{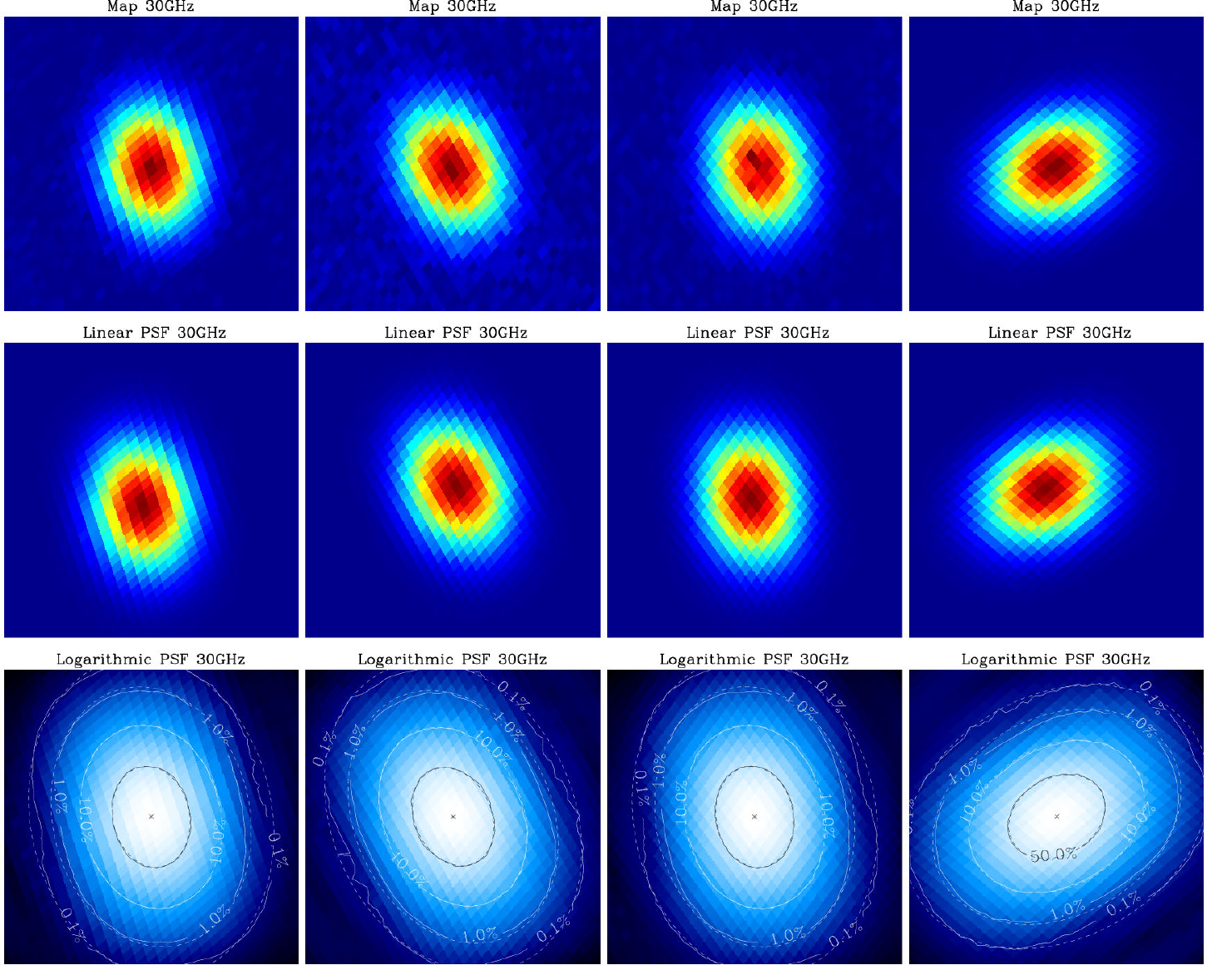}
\caption{As for previous two figures, but at 30\,GHz.}
\label{fig:ploteffectivebeam30}
\end{figure*}

As an example, { Figs.~\ref{fig:ploteffectivebeam70}, \ref{fig:ploteffectivebeam44}, and \ref{fig:ploteffectivebeam30}
 compare} images of four sources (assumed to be unresolved) from the \Planck\ Early Release Compact Source Catalogue (ERCSC) \citep{planck2011-1.10} and {\tt FEBeCoP} point spread functions (i.e., the transpose of the effective beam matrix) on the same patch of the sky for the LFI channel maps.
The galactic coordinates ($l,b$) of the four sources are shown under the color bar: in our sample, these are, respectively, (305.1$^\circ$, 57.1$^\circ$), (86.1$^\circ$, --38.2$^\circ$), (290.0$^\circ$, 64.4$^\circ$) and (184.5$^\circ$, --5.8$^\circ$), from left to right in the three figures.  

\begin{figure*}[htpb]
\centering
\includegraphics[width=17cm]{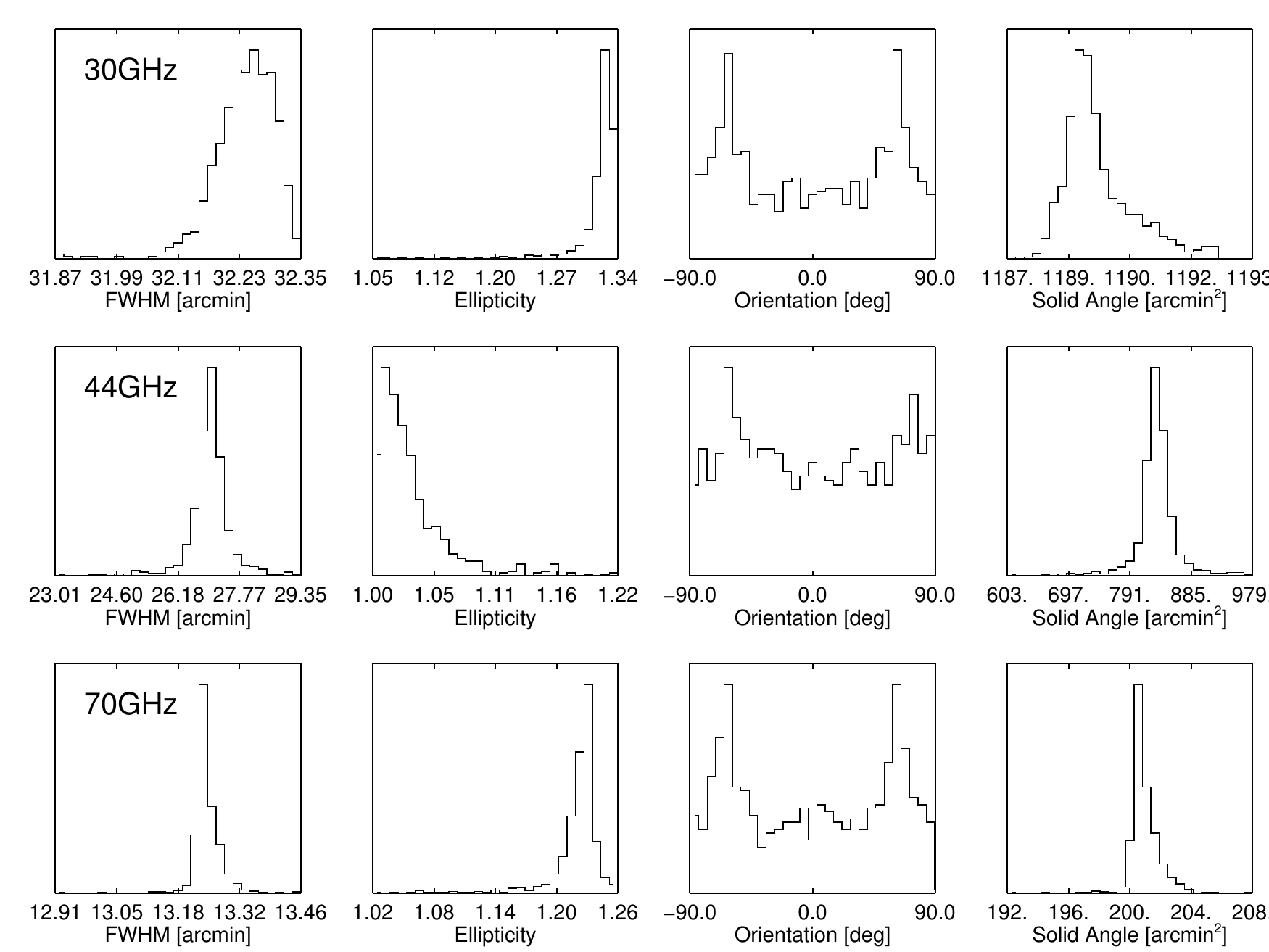} 
\caption{Histograms of the three fit parameters (beam FWHM, ellipticity, and orientation with respect to the local meridian) for the effective beams computed using {\tt FEBeCoP} with the simulated beams. {The sky is sampled } (fairly sparsely) {in }768 directions, chosen as {\tt HEALpix} N$_{\rm side}$=8 pixel centers to uniformly sample the sky.}
\label{fig:histoEB}
\end{figure*}

We then performed a 2D Gaussian fit of the effective beam at several positions of the sky and studied the distribution of the fitted parameters: beam FWHM; ellipticity; solid angle; and orientation with respect to the local meridian.
In order to perform such statistics, {the sky is sampled } (fairly sparsely) at 768 directions {chosen} as {\tt HEALpix} \citep{gorski2005} $N_{\rm side}$=8 pixel centers to uniformly sample the sky. 
The histograms of these quantities are shown in Fig.~\ref{fig:histoEB}.
From the histograms, we derive the statistical properties of these quantities (mean values and standard deviations), which are provided in Table \ref{tab:statistics}). 

\begin{figure}[htpb]
\centering
\vbox{
\includegraphics[width=8.5cm]{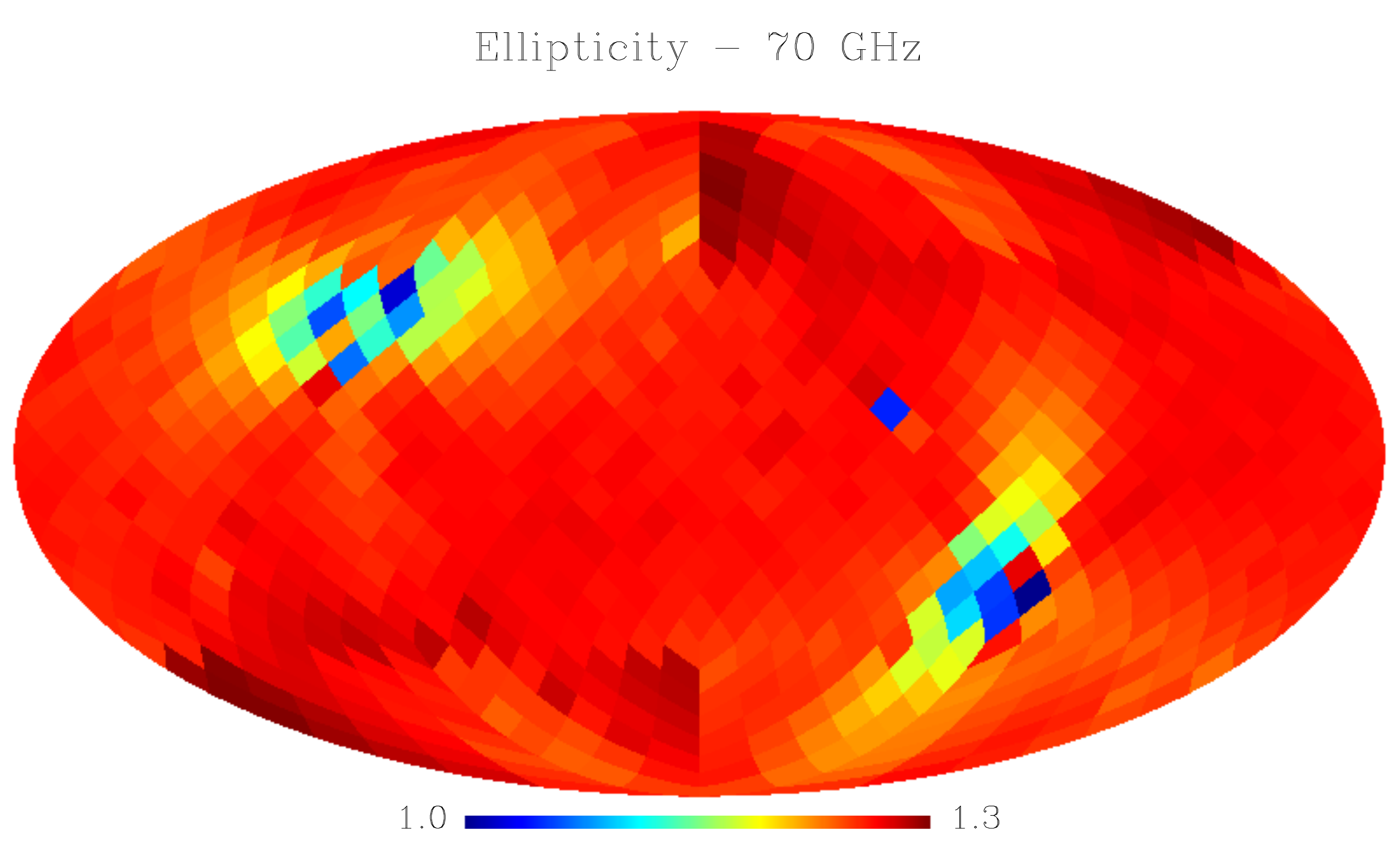}
\includegraphics[width=8.5cm]{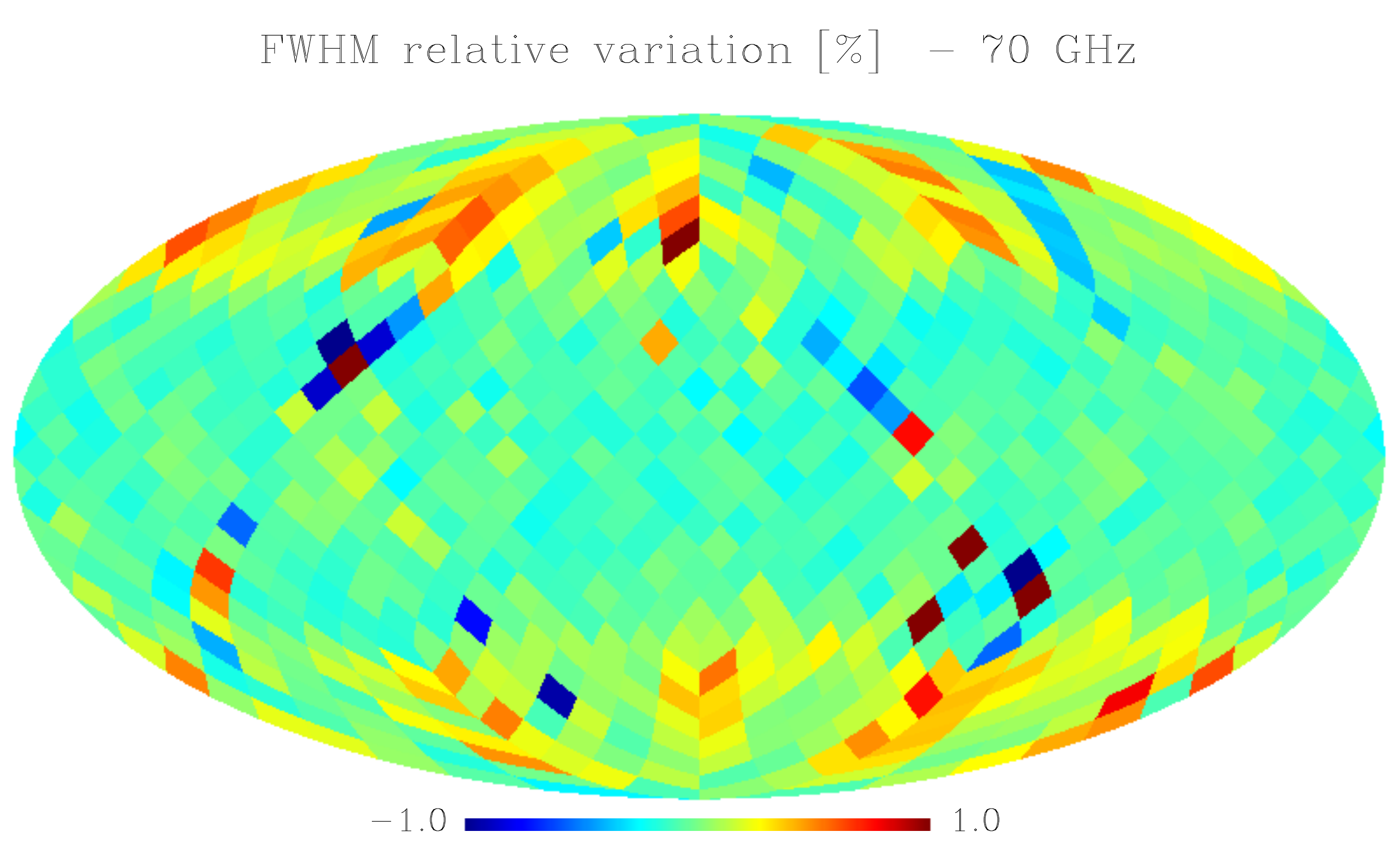}
\includegraphics[width=8.5cm]{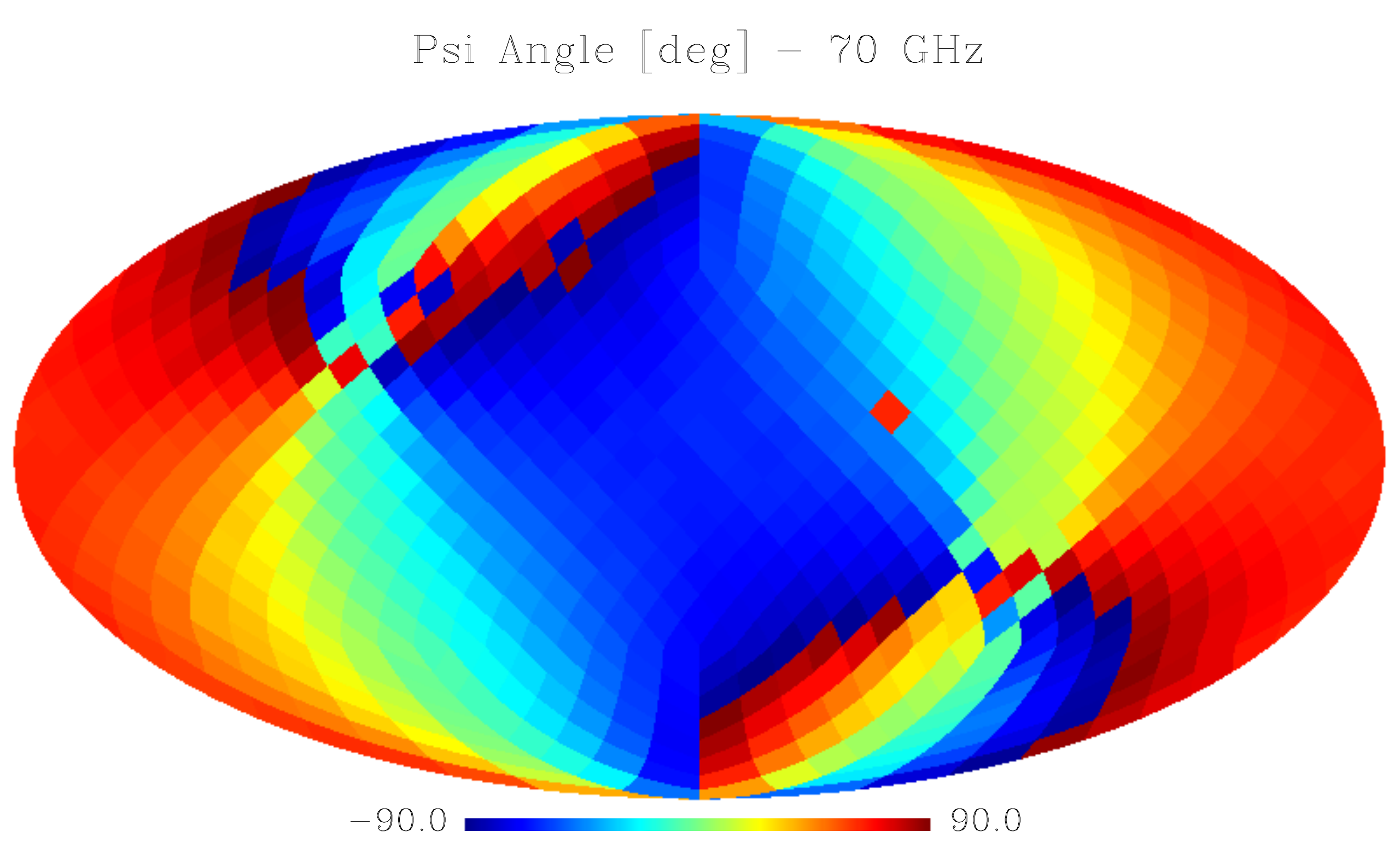}
\includegraphics[width=8.5cm]{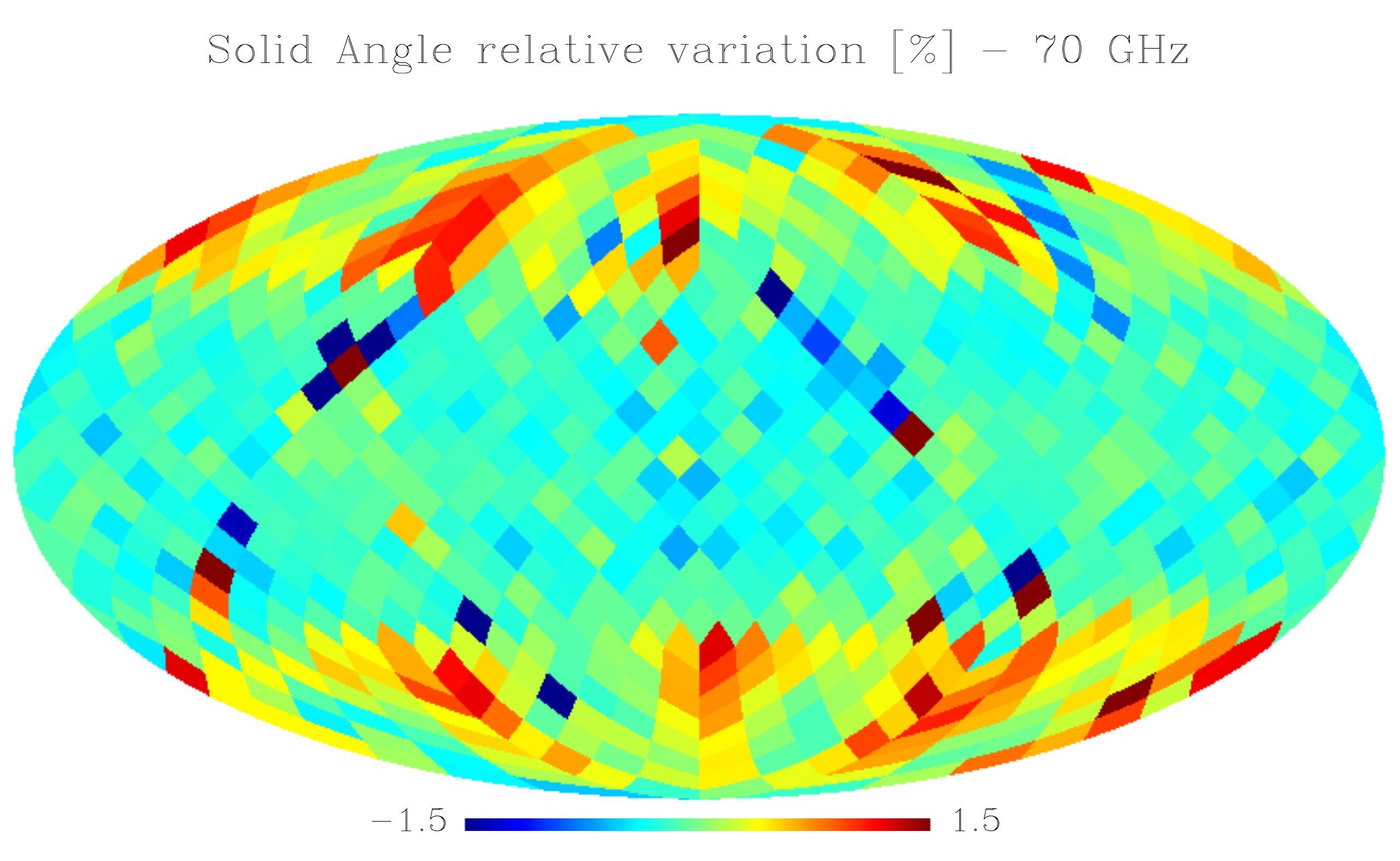}
}
\caption{Main parameters of the LFI effective beams: ellipticity (first row); FWHM (relative variation with respect to the FWHM of the scanning main beam, second row); {$\psi_{\ell}$} (third row); and beam solid angle (relative variation with respect to the scanning main beam solid angle reported in Table \ref{tab:imo}, fourth row), for the 70\,GHz channel. }
\label{fig:parameff}
\end{figure}

In Fig.~\ref{fig:parameff} we show the sky variation of ellipticity, FWHM (relative variation with respect to the FWHM of the scanning main beam), {$\psi_{ell}$} (orientation of the effective beam) and beam solid angle (relative variation with respect to the scanning main beam solid angle reported in Table \ref{tab:imo}) of the best-fit Gaussian to the effective beam at {\tt HEALpix} $N_{\rm side}$=16 pixel centers for 70\,GHz.
The effective beam is less elliptical near the ecliptic poles, where {the larger number of }scanning angles symmetrize the beam.  

\begin{table*}[ht]
\centering
\caption{Mean and standard deviation of FWHM, ellipticity, orientation, and solid angle of the {\tt FEBeCoP} effective beams computed with the simulated beams.
FWHM$_{{\rm eff}}$ is the effective FWHM estimated from the main beam solid angle of the effective beam, $\Omega_{{\rm eff}} =$ mean($\Omega$), under a Gaussian approximation.}
\begin{tabular}{ c c c r@{.}l  r@{.}l c}
\hline 
\hline
\noalign{\vskip 2pt}
Channel &  mean(FWHM) & mean($e$) & \multicolumn{2}{c}{mean($\psi$)} & \multicolumn{2}{c}{mean($\Omega$)}  & FWHM$_{{\rm eff}}$ \\ 
                 &        (arcmin)     &                      & \multicolumn{2}{c}{(deg)}                & \multicolumn{2}{c}{(arcmin$^{2}$)}       & (arcmin)  \\
\hline 
\noalign{\vskip 2pt}
70 &   13.252$\pm$0.033    &     1.223$\pm$0.026     &      0&587$\pm$55.066     &     200&742$\pm$1.027      &     13.31 \\ 
44 &    27.005$\pm$0.552   &     1.034$\pm$0.033     &      0&059$\pm$53.767   &      832&946$\pm$31.774   &      27.12  \\ 
30 &    32.239$\pm$0.013   &     1.320$\pm$0.031     &    --0&304$\pm$55.349   &     1189&513$\pm$0.842   &       32.34 \\ 
\hline 
\label{tab:statistics}
\end{tabular} 
\end{table*}
 
The main beam solid angle of the effective beam, $\Omega_{{\rm eff}}$, is estimated according to the definition: $4 \pi \sum(B_{ij}) / {\rm max}(B_{ij})$, i.e., as an integral over the full extent of the effective beam {(that is, within 2.5 FWHM)}. 
From the effective beam solid angle, we can estimate the effective FWHMs, {assuming it is Gaussian}: these  are tabulated in Table \ref{tab:statistics}. 
The reported FWHM$_{{\rm eff}}$ are derived from the solid angles, under a Gaussian approximation. 
The mean(FWHM) are the averages of the Gaussian fits to the effective beam maps. 
The former is best used for flux determination, the latter for source identification.

Note that the FWHM and ellipticity in Table \ref{tab:statistics} differ slightly from the values reported in Table~\ref{tab:imo}; this results from the different way
in which the Gaussian fit was applied. The scanning beam fit was determined by fitting the profile of Jupiter on timelines and limiting the fit to the data with a signal above the $3\sigma$ level from the noise, while the fit of the effective beam was computed on the maps of the simulated beams projected in several position of the sky; the latter are less affected by the noise.

In Table \ref{tab:deli2}, {$\Omega_{{\rm eff}}^{(1)}$ indicates } the beam solid angle estimated up to a radius equal to the FWHM$_{{\rm eff}}$ (as defined above), while $\Omega_{{\rm eff}}^{(2)}$ is the beam solid angle estimated up to a radius equal to twice the effective FWHM (FWHM$_{{\rm eff}}$). {The table also reports the standard deviation of the beam solid angle distribution in the sky}.
These were estimated according to the procedure followed in the aperture photometry code for the \Planck\ {Catalogue} of Compact Sources (PCCS) (i.e., if the pixel center does not lie within the given radius it is not included).  
These additional quantities were evaluated for the production of the PCCS \citep{planck2013-p05}.

\begin{table*}[!h]
\centering
\caption{Band averaged effective beam solid angles under a gaussian approximation.  {$\Omega_{{\rm eff}}$} is the beam solid angle estimated up to a radius equal to the $2.5\times$FWHM$_{{\rm eff}}$.  {$\Omega_{{\rm eff}}^{(1)}$ } is the beam solid angle estimated up to a radius equal to the FWHM$_{{\rm eff}}$ (see Table \ref{tab:statistics}), while $\Omega_{{\rm eff}}^{(2)}$ indicates the beam solid angle estimated up to a radius$=2\times$FWHM$_{{\rm eff}}$.}
\begin{tabular}{ c r@{.}l r@{.}l r@{.}l }
\hline 
\hline
\noalign{\vskip 2pt}
Channel & \multicolumn{2}{c}{$\Omega_{{{\rm eff}}}$} & \multicolumn{2}{c}{$\Omega_{{{\rm eff}}}^{(1)}$}  & \multicolumn{2}{c}{$\Omega_{{{\rm eff}}}^{(2)}$}  \\ 
                 & \multicolumn{2}{c}{(arcmin$^{2}$)}              & \multicolumn{2}{c}{(arcmin$^{2}$)}                          & \multicolumn{2}{c}{(arcmin$^{2}$)} \\
\hline 
\noalign{\vskip 2pt}
70 &    200&74$\pm$1.03      &    186&26$\pm$2.30     &     200&59$\pm$1.03 \\
44 &    832&95$\pm$31.77    &    758&68$\pm$29.70   &     832&17$\pm$31.81 \\
30 &    1189&51$\pm$0.84    &    1116&49$\pm$2.27   &     1188&95$\pm$0.85 \\
\hline 
\label{tab:deli2}
\end{tabular} 
\end{table*}

\section{Beam Window Function}
\label{window_function}
CMB temperature anisotropies are a scalar random field on a sphere, and can be decomposed in spherical harmonic coefficients:
\begin{equation}
a_{\ell m} = \int d\Omega T(\hat{n}) Y^{*}_{\ell m}(\hat{n}),  \quad T(\hat n)=\sum_{\ell m} a_{\ell m} Y_{\ell m} ,
\end{equation}
where
\begin{equation}
\langle a_{\ell m}\rangle = 0, \quad \langle a_{\ell m} a^{*}_{\ell ^{\prime} m^{\prime } } \rangle
= \delta _{\ell \ell ^{\prime }}\delta _{mm^{\prime }}C_{\ell } .
\end{equation}%

The finite angular resolution of an instrument $b(\hat{n}, \hat{n^{\prime}})$ can be described {by} a convolution in real space:
\begin{equation}
T_{{\rm obs}}(\hat{n})=\int d\Omega_{\hat{n^{\prime}}} b(\hat{n}, \hat{n^{\prime}}) T(\hat{n^{\prime}}) ,
\end{equation}
which is equivalent to a low-pass filter in harmonic space, and whose effective action on the power spectrum can be written as:
\begin{equation}
\label{eq:wl}
C_{\ell }^{{\rm obs}} = W_{\ell} C_{\ell} ,
\end{equation}
where $W_{\ell}$ is the beam window function.

As discussed in the previous section, a basic symmetric Gaussian approximation is not a good description of \Planck\ effective beams. 
Moreover, the combination of intrinsic beam asymmetry and scanning strategy produces effective beams that vary significantly over the sky. 
Therefore, in order to produce accurate estimates of the beam window functions, {we have to} to use detailed Monte Carlo simulations. 
This has been implemented using two appreoaches: first, full timeline-to-map simulations, where the CMB signal is convolved with realistic scanning beams in harmonic space, and then projected into a {time ordered data (TOD)} through the \Planck\ scanning strategy and processed in the same way as real data; and second, pixel space convolution of CMB signal-only maps using the effective beams derived with {\tt FEBeCoP}. 

In principle, for full-sky maps the effective azimuthally averaged beam window function can be estimated directly from Eq.~\ref{eq:wl}:
\begin{equation}
\label{eq:wlfullsky}
W_{\ell} = \langle C_{\ell }^{{\rm obs}} \rangle / C_{\ell } ,
\end{equation}
where $C_{\ell }^{{\rm obs}}$ is the power spectrum of simulated CMB-only maps, $C_{\ell }$ is the fiducial model used as input, and the ensemble average is taken over the Monte Carlo simulations.
However, in a realistic case {those regions of the sky that are contaminated by foreground are masked out}, and the above equation no longer applies. 
Instead, using the same {notation as in} \citet{master}:
\begin{equation}
\label{eq:wlmask}
\langle C_{\ell }^{{\rm obs}} \rangle= \sum_{\ell^{\prime}}M_{\ell \ell^{\prime}}W_{\ell^{\prime}} \langle C_{\ell^{\prime}} \rangle ,
\end{equation}
where the coupling kernel $ M_{\ell \ell^{\prime}}$ encodes the geometric mode-mode coupling effect introduced by masking the sky.
However, we have verified that for the Galactic mask used for power spectrum estimation \citep{planck2013-p02,planck2013-p08} 
the differences between full-sky and cut-sky window functions are marginal with respect to the error envelopes discussed in Sect.~\ref{error_propagation}.
Therefore, the full-sky approximation is used hereafter.

\subsection{Timeline-to-map Monte Carlo window functions}

Signal-only timeline-to-map Monte Carlo (MC) simulations are produced using Level-S \citep{reinecke2006} and {\tt HEALpix} subroutines and the {\tt Madam} map-maker \citep{kurki-suonio2009, keihanen2010} on the \emph{Louhi} supercomputer at \emph{CSC-IT Center for Science} in Finland; see Appendix B for details. 

Starting from a fiducial CMB power spectrum, we have generated a set of sky $a_{\ell m}$ realizations of this $C_\ell$  that are convolved with the beam $b_{\ell m}$ obtained from the simulated scanning beams. Note that the main beams do not collect the full power of the signal, since a small part of the signal spills outside the main beam to form sidelobes. In this MC just the main beam up to 4 FWHM was simulated, not the sidelobes, so the calculated signal values were missing that part of the power that goes to the sidelobes. This was taken into account at the map-making stage. Note that the main beam definition used here (4 FWHM) differs from that adopted in the effective beams computation ({2.5 $\times$ FWHM}). The consequences are discussed in Sect.~\ref{error_propagation}.

The CMB {TODs} for each realization were produced according to the detector pointing for each radiometer, and maps were made from these CMB timelines with {\tt Madam}.
The same {\tt Madam} parameter settings were used as for the flight maps \citep{planck2011-1.6,planck2013-p02}. 
The calibration step was not simulated, as the simulated signal was constructed as already calibrated, except for the effect of power lost to the sidelobes.  
The impact of sidelobes on the calibration of flight data is discussed in \citet{planck2013-p02b}, while its effect on the beam window function will be discussed in Sect.~\ref{error_propagation} of this paper.
For the MC, we assumed that the calibration compensates for the missing power in the main beams, according to the discussion in Sect.~2.2 of \citet{planck2013-p02b}. 

We produced in this way 30\,GHz, 44\,GHz, and 70\,GHz frequency maps, and the ``horn-pair'' maps for 70\,GHz 18/23, 19/22, and 20/21 from the 15.5 month nominal survey. The computational cost of producing one realization of this set was about 2000 CPUh. Given this relatively large computational cost, we have generated only 102 CMB realizations. Although this leaves some residual scatter in the estimated beam window functions especially at low multipoles, these maps have been generated mostly as a consistency check with respect to the {\tt FEBeCoP} approach as described below, and therefore the number of simulations is adequate for this purpose.

\begin{figure}[!h]
\centering
\includegraphics[width=8.5cm]{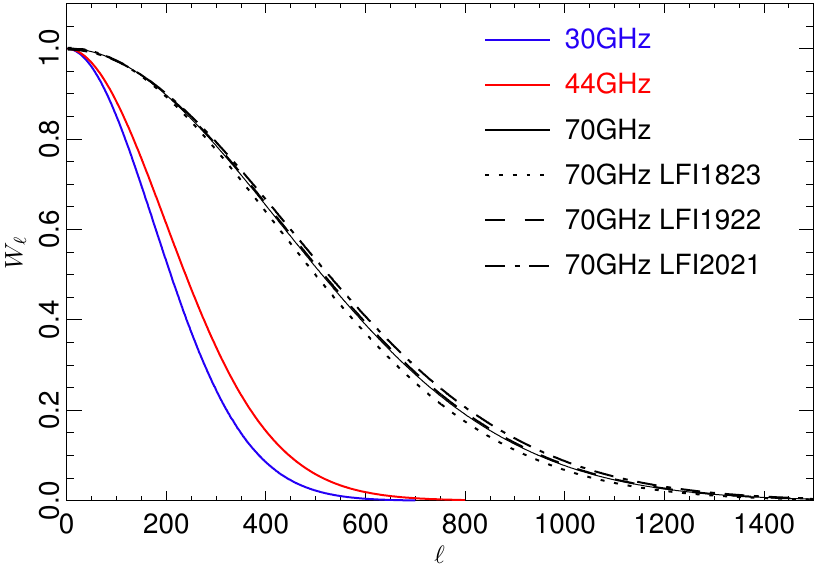}
\caption{Timeline-to-map, Monte Carlo-based beam window functions for $Planck$ 30, 44, and 70\,GHz frequency maps. For 70\,GHz, {also shown are} the beam window functions for a subset of paired horns, namely \texttt{LFI18-23}, \texttt{LFI19-22}, and \texttt{LFI20-21}.}
\label{fig:wfmc}
\end{figure}

Full-sky, timeline-to-map Monte Carlo based beam window functions are shown in Fig.~\ref{fig:wfmc} for 30, 44, and 70\,GHz frequency maps. For 70\,GHz we also show the beam window functions obtained considering only subsets of detectors , namely \texttt{LFI18-23}, \texttt{LFI19-22}, and \texttt{LFI20-21}. 

\subsection{FEBeCoP window functions}
\begin{figure}[!h]
\centering
\includegraphics[width=8.5cm]{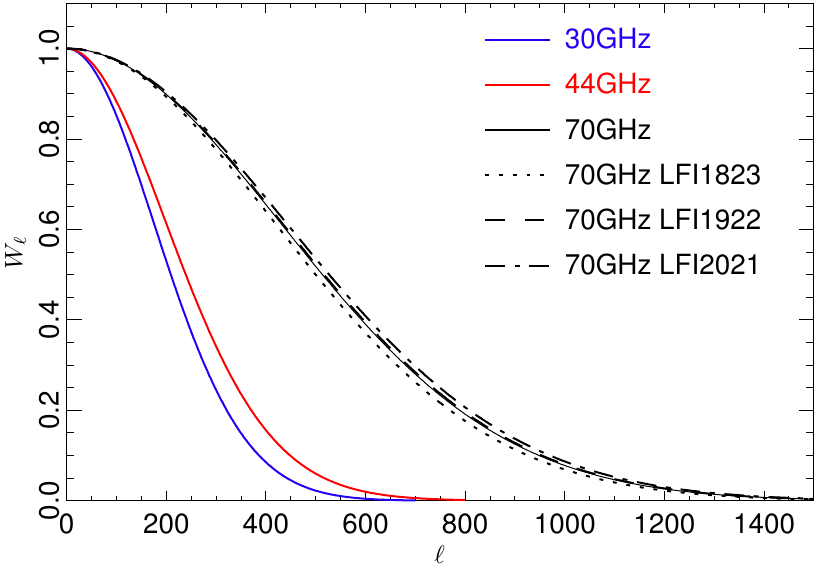}
\caption{{\tt FEBeCoP} beam window functions for $Planck$ 30, 44, and 70\,GHz frequency maps. For 70\,GHz, we also show  the beam window functions for a subset of detectors, namely \texttt{LFI18-23}, \texttt{LFI19-22}, and \texttt{LFI20-21}.}
\label{fig:wf}
\end{figure}

{\tt FEBeCoP} beam window functions are shown in Fig.~\ref{fig:wf} for 30, 44, and 70\,GHz frequency maps. For 70\,GHz we also show the beam window functions obtained considering only subsets of detectors, namely \texttt{LFI18-23}, \texttt{LFI19-22}, and \texttt{LFI20-21}. These are computed using the effective beams obtained from the simulated scanning beams with a cutoff radius of {2.5 $\times$ FWHM}. The resulting window functions {using} full sky approximation are obtained by averaging Eq.~\ref{eq:wlfullsky} over 1000 signal only simulations, where every simulated CMB maps is convolved with the effective beams described in Sect.~\ref{effective_beams}.

\begin{figure*}[!p]
\centering
\includegraphics[width=17cm]{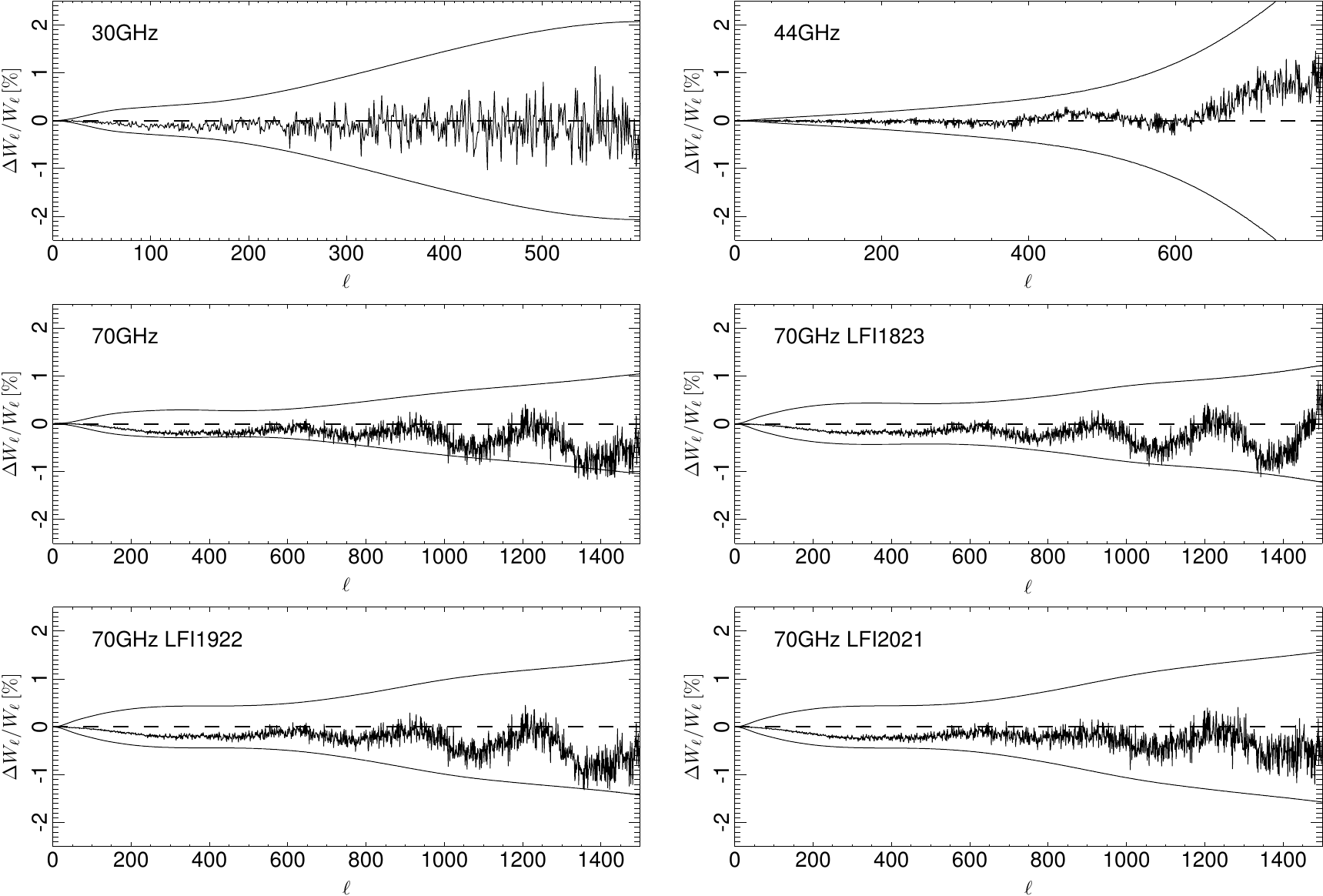}
\caption{Comparison between timeline-to-map {Monte Carlo} -based and {\tt FEBeCoP} beam window functions. {Also shown,}  the $\pm 1\sigma$ error envelopes obtained in Sec.~\ref{error_propagation}.}
\label{fig:wfcomp}
\end{figure*}

{Fig.~\ref{fig:wfcomp} shows} a comparison between MC-based and {\tt FEBeCoP} beam window functions. 
Although there are some high-$\ell$ discrepancies at 70\,GHz, these are located at $\ell \gtrsim$ 1300 where the {the amplitude of the }beam $W_\ell$s drop below 0.01. 

In addition, however, as { explained} in the next section, {it is necessary to} account for the effect of the different choices for the cutoff radius between the two methods. The {\tt FEBeCoP} calculations used a {2.5 $\times$ FWHM} cutoff radius for the main beam, while the timeline-to-map Monte Carlo window functions are derived using a 4 FWHM cutoff. In order to quantify the agreement, {we show in Fig.~\ref{fig:wfcomp}} the $\pm 1 \sigma$ error envelopes that will be discussed in the next section.
{In the} "region of interest" the two methods agree to {within} 1\% level. 
Since the {\tt FEBeCoP} algorithm is faster than the timeline-to-map Monte Carlo for convolutions that include only the main beam, it allows {a} significantly larger number of simulations (1000 vs 102), resulting in a more accurate estimation of the window functions.
For the same reason, {\tt FEBeCoP} also allows {us} to perform a robust error assessment,  {presented} in the next section. Hence, the {\tt FEBeCoP} window functions will be used for the power spectrum analysis \citep{planck2013-p02,planck2013-p08}, and will be distributed within the data release.

With {\tt FEBeCoP}  we also estimate the level of contamination of the transfer functions due to a non-uniform sky sampling within the pixels, comparing the ideal {\tt HEALpix} pixel window function, which is derived under the assumption of uniform coverage, with the true one computed with {\tt FEBeCoP}. 
This effect acts as a noise term in the maps and it becomes important only at very high $\ell$. 
An analytic treatment of the contamination on the maps is described in Appendix F of the HFI Beam paper \citep{planck2013-p03c}. 
{The level of distortion of the window function is}: $0.1\%$ at $\ell=600$ for 30\,GHz, $0.4\%$ at $\ell=800$ for 30\,GHz, and $0.5\%$ at $\ell=1400$ for 70 GHz, in all cases within the error bars.

\section{Error Budget}
\label{error_propagation}
We discuss here the main sources of uncertainties in the window functions that have proven to be relevant for the LFI.

\subsection{Main beam knowledge}

The propagation of the uncertainties in the beam knowledge to the window function has been carried out with a dedicated MC pipeline on the $Planck$ optics. 
The tuned optical model \citep{planck2013-p28} was used as the basis to run MC simulations with about 500 realizations of the $Planck$ optics. 
More specifically, the wavefront at the aperture of the telescope has been artificially distorted by adding to the primary reflector randomly-varying amplitude distortions described as modes of Zernike polynomials, up to the fifth order. 

The idea behind this assumption is that the true flight beam comes from a true flight field distribution at the telescope aperture that gives the true wavefront. 
Any small difference between our telescope model and the real one can be mapped by aberrations {of} the aperture wavefront. 
For each wavefront, we {used GRASP to simulate the corresponding beam, and only the beams with parameters (angular resolution, ellipticity, and beam orientation)
in agreement with those measured in flight within 3$\sigma$ have been selected. The errors associated to each parameter are those reported} in Table \ref{tab:imo}. 
We repeated {this procedure} for all the twenty-two LFI beams producing a set of 3036 beams (corresponding to 138 slightly different optical models). 
{The product is} a set of beams calculated from plausible optical models of the telescope {whose} parameters are in agreement with the parameters {measured in flight}. 
Then this set of beams was used as input to {\tt FEBeCoP} to compute the corresponding window functions. 

\begin{figure}[htpb]
\centering
\includegraphics[width=8.5cm]{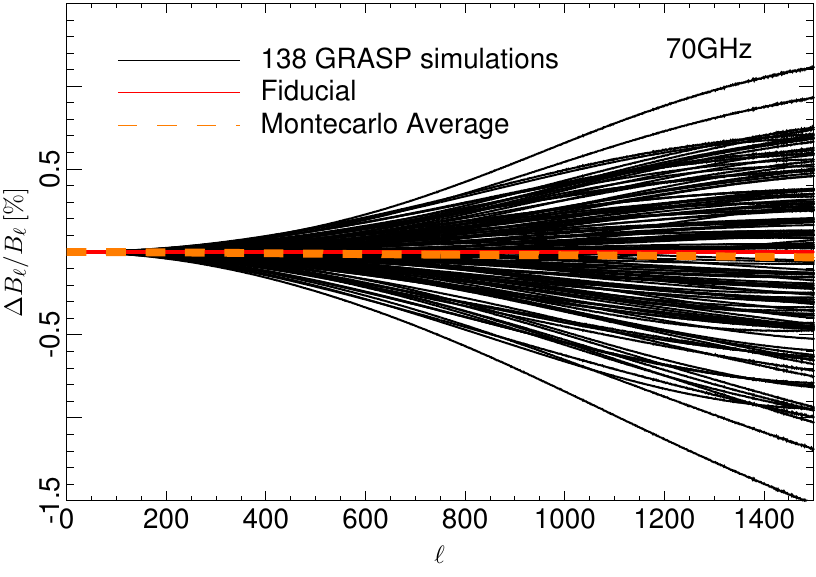}
\caption{{Relative differences of the GRASP beams with respect to the fiducial. Their ensemble average, represented by the dashed orange line, is very close to the $B_{\ell}$ of the fiducial (red line).}}
\label{fig:ratios}
\end{figure}

The three parameters used in the comparison between simulations and measurements (angular resolution, ellipticity, and beam orientation) are strongly correlated and this original method to obtain the errors on the window function using a MC pipeline on the optics takes this correlation properly into account, avoiding unphysical solutions in which no correlation is assumed. In Fig.~\ref{fig:ratios} we show the beam window functions at 70\,GHz for all the 138 simulated optical models. 
Window functions are normalized to the fiducial for the 70\,GHz channel.

\subsection{Cutoff radius in the main beam computation and impact of sidelobes}\label{sec:errors}

The impact of sidelobes on the calibration has been discussed in \citet{planck2013-p02b}. 
The main result for the discussion presented here is that the gain values are unbiased, and this imposes a constraint on the dipole term of the window function, i.e., $W_1 = 1$, which fixes the normalization. 
In principle, in order to fully account for beam effects in the window function, one should perform a computation of the window function including the full beam pattern, {generated from} either {\tt FEBeCoP} or the timeline-to-map Monte Carlo. 
However, this would have a huge computational cost making it unfeasible in a Monte Carlo approach. 
As a result, LFI beam window functions are derived from Monte Carlo simulations including the main beam only, and therefore it is important to assess the effect of neglecting sidelobes.

\begin{figure}[htpb]
\centering
\includegraphics[width=8.5cm]{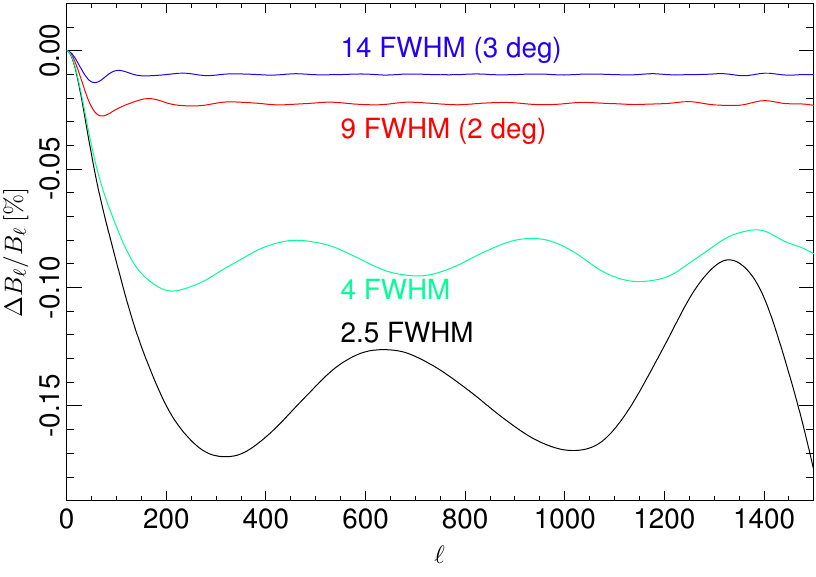}
\caption{Relative difference between $B_{\ell}$s computed for various cutoff values and the one with full integration. All the functions have been computed using eq.~\ref{eq:wfanalytic} for a symmetrized version of the \texttt{LFI18M} beam profile.}
\label{fig:impactofsidelobe}
\end{figure}

We have done a preliminary evaluation {using an  analytical approach to the window function calculation}. 
In fact, for a given azimuthally symmetric beam profile $b_{\rm s}(\theta)$, the corresponding $\ell$-space function $B_{\ell}$ can be computed using the Legendre transform:
\begin{equation}\label{eq:wfanalytic}
B_{\ell} = \Omega_{\rm B} b_{\ell} = 2\pi \int d \cos(\theta) b_{\rm s}(\theta)P_{\ell}(\theta) ,
\end{equation}
and $W_{\ell}=B^2_{\ell}$.
{For one case, the symmetrized beam profile for the \texttt{LFI18M} detector, we have computed the corresponding $B_{\ell}$. 
We terminated the integration at different angles, namely 2.5 $\times$ FWHM, 4 $\times$ FWHM, 2$^\circ$, and 3$^\circ$ from the center, and imposed the normalization constraint at $\ell=1$}.
{Fig.~\ref{fig:impactofsidelobe} shows} the relative difference between $B_{\ell}$s obtained for the four cutoff values and the one resulting from full integration. 
As expected, {a small $\ell$-dependent correction is affecting} mostly the large angular scales. 
In particular, extending the calculation up to the near sidelobes makes this effect negligible.

\begin{figure}[htpb]
\centering
\includegraphics[width=8.5cm]{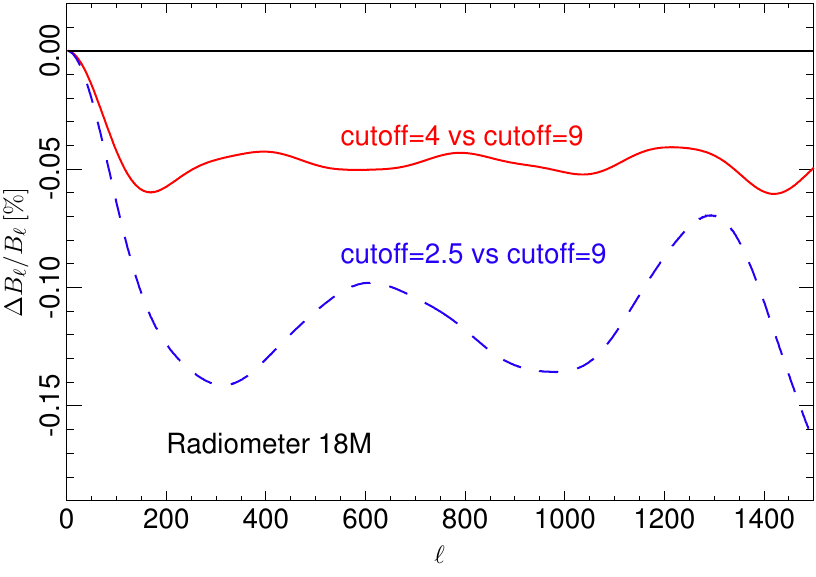}
\caption{Same as Fig.~\ref{fig:impactofsidelobe}, but the window functions have been computed with {\tt FEBeCoP}.}
\label{fig:febecopcutoff}
\end{figure}

As already stated in Sect.~\ref{window_function}, the combination of intrinsic asymmetries in the $Planck$ beam and scanning strategy forced us to discard a simple, {symmetric beam approximation}. The same argument applies here, especially considering that near and far sidelobes are even more asymmetric than the main beam. Therefore, we have extended {\tt FEBeCoP} calculation for the \texttt{LFI18M} detector to cutoff radii of 4 FWHM and 3$^\circ$ (corresponding to $\sim$ 9 FWHM). In this case,  we have used the latter window function as a reference to compare with when computing the relative difference.
Results are reported in Fig.~\ref{fig:febecopcutoff}, confirming that similar conclusions can be drawn for a realistic case as well.

A further improvement in the assessment of the sidelobes effect on the window function has been carried out considering the variation across the band of the sidelobes themselves.
Whereas the impact of the main beam variation across the band is small, this is not true for the near and far sidelobes.
The $4\pi$ beams of the radiometer \texttt{LFI18M} (main beam, near and far sidelobes) were computed at about twenty frequencies across the radiometer bandpass and they were averaged taking into account the radiometer bandshape. The resulting averaged beam has been used to evaluate the impact on the beam window function using the analytical approach described above.
The shape of the bias is very close to that reported in Fig.~\ref{fig:impactofsidelobe} but the amplitude is slightly different with respect to the monochromatic beam.
In the error budget {the worst case is considered}, in order to be {as conservative as possible}.  


\subsection{Total error budget on window functions}\label{sec:toterrb}

\begin{figure}[hptb]
\includegraphics[width=8.5cm]{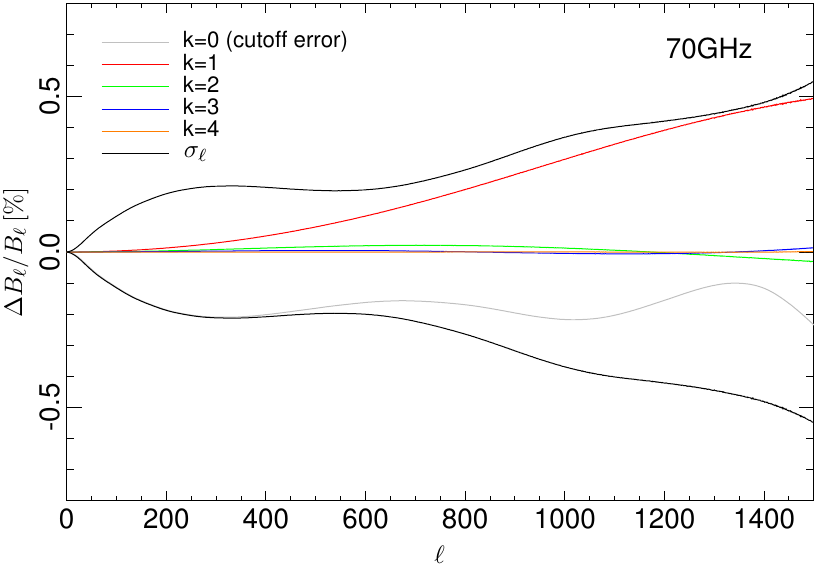}
\caption{{Error budget for the 70\,GHz channel relative to the fiducial $B_{\ell}$. The grey curve represents the error between the full beam and the cutoff approximation used in the window function computations. The colored lines represent the first four modes used in the current beam error model, as described in section  \ref{sec:toterrb}. The black line is the $one-\sigma$ error obtained by adding the cutoff error and the squared first four eigenmodes.}   }
\label{fig:decomposition70}
\end{figure}

\begin{figure}[hptb]
\includegraphics[width=8.5cm]{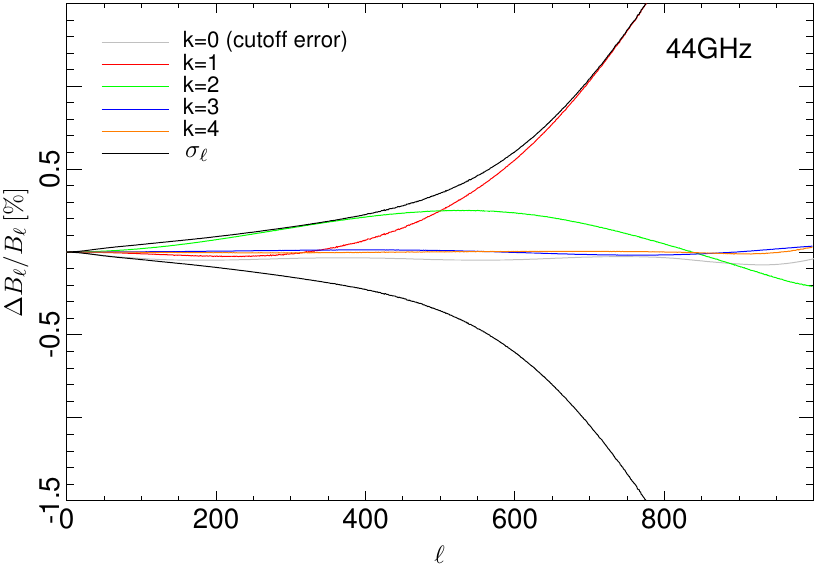}
\caption{First four eigenmodes of the covariance matrix of the 44\,GHz channel.}
\label{fig:decomposition44}
\end{figure}

\begin{figure}[hptb]
\includegraphics[width=8.5cm]{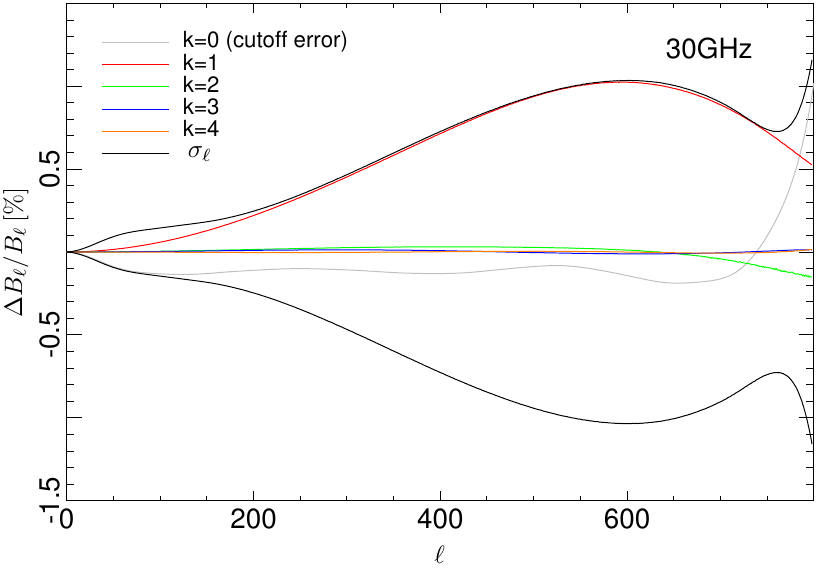}
\caption{First four eigenmodes of the covariance matrix of the 30\,GHz channel.}
\label{fig:decomposition30}
\end{figure}

Using the set of simulated beam window functions, {we have built} the covariance matrix $\bf C$ in $\ell$-space computing:
\begin{equation}
{C}_{\ell\ell'} = \left\langle (W_{\ell}- \left\langle W_{\ell} \right\rangle)(W_{\ell'}-\left\langle W_{\ell'} \right\rangle ) \right\rangle ,
\end{equation}
where the averaging is performed on the 138 simulations.
Then we have decomposed the covariance matrix in eigenvalues ($\Lambda_k$) and eigenvectors (${V}_k$)  (see, e.g., \cite{Bond1995} for detailed discussion). 
{Fig.~\ref{fig:decomposition70} shows} the first eigenmodes for the 70\,GHz channel. All the error content is substantially encompassed by the first two eigenvalues, that account for cutoff radius and main beam uncertainties respectively.
Figures \ref{fig:decomposition44} and \ref{fig:decomposition30} show the eigenmodes for the 44 and 30\,GHz, respectively.
The eigenmodes can be used as input of the Markov Chain Beam Randomization (MCBR) marginalization code to account for beam errors in cosmological parameter estimation \citep{rocha2010a}. We apply the MCBR procedure is applied to a simulated 70\,GHz dataset, {finding} that the parameters mostly affected are $n_{\rm s}$, $\Omega_{\rm b} h^2$, $\Omega_{\rm c} h^2$, and $A_{\rm s}$; the increase of the errors can be quantified respectively as $12\%$ of $\sigma$ for the first and less then $8\%$ of $\sigma$ for the others. 

\section{Conclusions}
\label{conclusions}
The optics and electronics of the \Planck\ detectors, combined with the satellite motion, determine the instrumental angular response to a sky signal.
An accurate characterization and a thorough understanding of the LFI beam patterns is the key to determining their imprint on the transfer function from the observed to the true sky anisotropy spectrum.  
In this paper we discussed the algorithms used to reveal the most significant LFI beam features that impact the exploration of the underlying cosmology. 
The in-flight assessment of the LFI main beams relied mainly on the measurements performed during four Jupiter crossings. 
The calibrated data from the Jupiter scans were used to determine the \textit{scanning beams}: the signal-to-noise ratio for this data is such as to make it possible to follow the LFI beams profile down to --20 dB from the peak, corresponding to distances from the beam line of sight of about 1.25 FWHM, i.e., the inner parts of the main beams. 
Fitting the main beam shapes with an elliptical Gaussian, we could express the uncertainties of the measured scanning beam in terms of statistical errors for the Gaussian parameters: ellipticity; orientation; and FWHM. 
While this method allows the accurate in-flight measurement of the LFI main beams, the (lower) angular response at larger distances from the beam centroid (near and far sidelobes) cannot be directly measured from a single point source signal, mostly because of the noise and background dominance, so it must be modelled differently. 
Therefore, a further step was taken to build an optimal model for the full LFI beams profile.
We developed a tuned optical model such that the simulated beams would provide the best fit to the available measurements of the LFI main beams from Jupiter: we found that this model represents all the LFI beams with an accuracy of 1\%, which has been considered in the propagation of the uncertainties at the window function level.
The corresponding simulated sidelobes are, in turn, consistent with the effect induced by the Galactic spillover as observed in survey maps differences. 
This model, together with the pointing information derived from the focal plane geometry reconstruction \citep{planck2013-p02}, gives the most advanced and precise noise-free representation of the LFI beams.  These were also independently cross-checked through a beam deconvolution test. 
The simulated beams were the input to calculate the effective beams, which take into account the specific scanning strategy to include any smearing and orientation effects on the beams themselves. The approach was validated by comparing the effective beam Point Spread Functions with images from the Planck Catalogue of Compact Sources. 

To evaluate the beam window function, we adopted two independent approaches, both based on Monte Carlo simulations. In one case, we convolved a fiducial CMB signal with realistic scanning beams in the harmonic space to generate the corresponding timelines and maps; in the other case, we convolved the fiducial CMB map with effective beams in the pixel space. The two methods agree to 1\% level. 

To evaluate the error on the resulting window functions, we took into account the fact that they were calculated assuming full-power main beams. 
Thus, part of the error budget comes from the propagation of the main beam uncertainties throughout the analysis, while another contribution comes from neglecting near and far sidelobes in the Monte Carlo simulation chain. 
We found that the two error sources have different relevance depending on the angular scale. Ignoring the near and far sidelobes is the dominant error at low multipoles, while the main beam uncertainties dominate the total error budget at $\ell\geq 600$. Representative values of the total error, for scales of cosmological interest, range from 0.3\% ($\ell \approx 200$) to about 0.8\% ($\ell \approx 1200$). 
The total uncertainties in the effective beam window functions are: 2\% and 1.2\% at 30 and 44\,GHz, respectively (at $\ell \approx 600$); and 0.7\% at 70\,GHz at $\ell \approx 1000$.

\begin{acknowledgements}

  \Planck\ is too large a project to allow full acknowledgement of all
  contributions by individuals, institutions, industries, and funding
  agencies. The main entities involved in the mission operations are
  as follows. The European Space Agency (ESA) operates the satellite via its
  Mission Operations Centre located at ESOC (Darmstadt, Germany) and
  coordinates scientific operations via the Planck Science Office
  located at ESAC (Madrid, Spain). Two Consortia, comprising around 50
  scientific institutes within Europe, the USA, and Canada, and funded
  by agencies from the participating countries, developed the
  scientific instruments LFI and HFI, and continue to operate them via
  Instrument Operations Teams located in Trieste (Italy) and Orsay
  (France). The Consortia are also responsible for scientific
  processing of the acquired data. The Consortia are led by the
  Principal Investigators: J.L. Puget in France for HFI (funded
  principally by CNES and CNRS/INSU-IN2P3-INP) and N. Mandolesi in Italy
  for LFI (funded principally via ASI). NASA US Planck Project,
  based at JPL and involving scientists at many US institutions,
  contributes significantly to the efforts of these two Consortia. The
  author list for this paper has been selected by the Planck Science
  Team, and is composed of individuals from all of the above entities
  who have made multi-year contributions to the development of the
  mission. It does not pretend to be inclusive of all contributions.
  The \Planck -LFI project is developed by an International Consortium
  lead by Italy and involving Canada, Finland, Germany, Norway, Spain,
  Switzerland, UK, USA. The Italian contribution to \Planck\ is
  supported by the Italian Space Agency (ASI) and INAF. 
  This work was supported by the Academy of Finland grants 253204, 256265, and 257989. 
  This work was granted access to the HPC resources of CSC made available within the Distributed European 
  Computing Initiative by the PRACE-2IP, receiving funding from the European Community's Seventh Framework Programme 
  (FP7/2007-2013) under grant agreement RI-283493.  
  We thank CSC -- IT Center for Science Ltd (Finland) for computational resources.  
  We acknowledge financial support provided by the Spanish Ministerio
  de Ciencia e Innovaci{\~o}n through the Plan Nacional del Espacio y
  Plan Nacional de Astronomia y Astrofisica.  
  We acknowledge the Max Planck Institute for Astrophysics Planck Analysis Centre (MPAC),
  funded by the Space Agency of the German Aerospace Center (DLR)
  under grant 50OP0901 with resources of the German Federal Ministry
  of Economics and Technology, and by the Max Planck Society. This
  work has made use of the Planck satellite simulation package
  (Level-S), which is assembled by the Max Planck Institute for
  Astrophysics Planck Analysis Centre (MPAC) \cite{reinecke2006}. We
  acknowledge financial support provided by the National Energy
  Research Scientific Computing Center, which is supported by the
  Office of Science of the U.S. Department of Energy under Contract
  No. DE-AC02-05CH11231. Some of the results in this paper have been
  derived using the HEALPix package \cite{gorski2005}.
The development of \Planck\ has been supported by: ESA; CNES and CNRS/INSU-IN2P3-INP (France); ASI, CNR, and INAF (Italy); 
NASA and DoE (USA); STFC and UKSA (UK); CSIC, MICINN, JA and RES (Spain); Tekes, AoF and CSC (Finland); DLR and MPG (Germany); 
CSA (Canada); DTU Space (Denmark); SER/SSO (Switzerland); RCN (Norway); SFI (Ireland); FCT/MCTES (Portugal); and PRACE (EU). 
A description of the Planck Collaboration and a list of its members, including the technical or scientific activities in which 
they have been involved, can be found at \url{http://www.sciops.esa.int/index.php?project=planck&page=Planck_Collaboration}.
 
\end{acknowledgements}

\appendix
\section{LFI beams notation}
In table \ref{tab:symbols} we report a selection of the most important symbols used in this paper. 
 
\begin{table*}[!h]
\centering
\caption{Selected LFI beams analysis notation.}
\begin{tabular}{ l l  }
\hline
\hline
\noalign{\vskip 2pt}
{Symbol} &  {Description} \\ 
\hline 
\noalign{\vskip 2pt}
M, S  & Main and Side radiometer arm \\
LOS frame & Telescope's Line of Sight reference frame \\
$\theta, \phi$ & Polar coordinates in the LOS frame  \\
$\theta_{{\rm MB}}, \phi_{{\rm MB}}$ & Polar coordinates in the main beam frame \\
u, v & Cartesian dimensionless coordinates in the LOS frame \\
\textit{e} & Beam ellipticity \\
FWHM & Full width half maximum \\ 
$\psi_{ell}$ & Beam orientation defined with respect to the x-axis of the LOS frame \\
$\psi$ & Polarization angle (angle between the detector's polarization axis and the local meridian) \\
$\sigma_{{\rm max}}^{{\rm b}}, \sigma_{{\rm max}}^{{\rm b}}$ & Standard deviation of the elliptical Gaussian \\
$\eta $ & Main beam efficiency \\
$\tau(\nu)$ & Bandpass \\
$T_{\rm{A}}^{{\rm M}}$    ($T_{{\rm A}}^{{\rm S}}$)   & Detector output in antenna temperature for the M (S) radiometer in the main beam frame  \\
$E_{{\rm cp}}^{{\rm M}}$   ($E_{{\rm cp}}^{{\rm S}}$)  & Co-polar electric field component of the beam in the M (S) radiometer  \\
$E_{{\rm xp}}^{{\rm M}}$   ($E_{{\rm xp}}^{{\rm S}}$)  & Cross-polar electric field component of the beam in the M (S) radiometer  \\
$\chi^{{\rm omt}}$ & Cross-polarization of the orthomode transducer \\
$\hat{\mathbf{ \vec p}}_{\rm t}$ & Pointing direction for time sample \textit{t} \\
$\hat{\mathbf{ \vec r}}_{\rm i}$& \textit{i}-pixel center direction \\
$A_{ \rm {ti}}$ & Pointing matrix for pixel \textit{i} and time sample \textit{t}  \\
$B_{ij} \equiv  \mathbf{ \vec B}$ & Effective beam for pointing pixel \textit{i} and beam pixel \textit{j} \\
$b(\hat{\mathbf{\vec r}},\hat{\mathbf{\vec p}})$ & Scanning beam at a direction $\hat{\mathbf{r}} \equiv \left[ \theta, \phi\right]$ with the pointing angles $\hat{\mathbf{p}}$ \\
$b_{ \rm opt}(\theta)$ & Optical beam profile \\
$\gamma$ & Polarization efficiency  $\gamma = (1-\epsilon)/(1+\epsilon)$, being $\epsilon$ the cross-polar leakage\\
$\mathbf{\vec w}$ &  Polarization weight factor \\
$\Omega_{{\rm scn}}$ & Solid angle of the scanning beam \\
$\Omega_{{\rm opt}}$ & Solid angle of the optical beam \\
$\Omega_{{\rm sim}}$ & Solid angle of the simulated beam \\
$\Omega_{{\rm eff}}$ & Solid angle of the effective beam \\
FWHM$_{{\rm eff}}$ & Effective beam full width half maximum  \\
$W_{\ell}$ & Beam window function \\
$\Lambda_k$ & Eigenvalues of the covariance matrix \\
\textbf{\vec V}$_k$ & Eigenvectors of the covariance matrix \\
\hline
\label{tab:symbols}
\end{tabular} 
\end{table*}

\section{Impact of polarization of Jupiter}

In this appendix we estimate the effects of a partially polarized point source in beam measurements.  We define the following normalized patterns for the M and S radiometers,

\begin{eqnarray}
P_{\rm n}(\theta,\phi)_{\rm cp}^{\rm M} &=& {| E(\theta,\phi)_{\rm cp}^{\rm M} |^2 \over | E(0,0)_{\rm cp}^{\rm M} |^2} \\
P_{\rm n}(\theta,\phi)_{\rm xp}^{\rm M} &=& {| E(\theta,\phi)_{\rm xp}^{\rm M} |^2 \over | E(0,0)_{\rm cp}^{\rm M} |^2}
\end{eqnarray}

\begin{eqnarray}
P_{\rm n}(\theta,\phi)_{\rm cp}^{\rm S} &=& {| E(\theta,\phi)_{\rm cp}^{\rm S}|^2 \over | E(0,0)_{\rm cp}^{\rm S} |^2} \\
P_{\rm n}(\theta,\phi)_{\rm xp}^{\rm S} &=& {| E(\theta,\phi)_{\rm xp}^{\rm S} |^2 \over | E(0,0)_{\rm cp}^{\rm S} |^2}
\end{eqnarray}
where $P_n(\theta,\phi)_{\rm cp}$ and $P_n(\theta,\phi)_{\rm xp}$ are the 3rd Ludwig co-polar and the cross-polar components both normalized at the co-polar peak coincident with the direction $(\theta,\phi) = (0,0)$. When scanning a point-like source, the pattern shape is determined by the normalization of the signal as a function of boresight angles $(\theta, \phi)$. 
The measured pattern is derived from equations \ref{eq:TM} and \ref{eq:TS} by the following equations where the apex $\star$ denotes the pattern derived from measurements:

\begin{eqnarray}
\label{eq:pnmeasM}
P_{\rm n}^{\star\rm M} &=& {T_{\rm A}(\theta,\phi)^{{\rm M}} \over \max\left[T_{\rm A}(\theta,\phi)^{\rm M}\right]} \nonumber \\
&\propto&  P_{\rm n}(\theta,\phi)^{\rm M}_{\rm cp} + P_{\rm n}(\theta,\phi)^{\rm M}_{\rm xp} + \\ \nonumber
&+& \chi^{{\rm OMT}} \cdot \left[P_{\rm n}(\theta,\phi)^{\rm S}_{\rm cp}  + P_{\rm n}(\theta,\phi)^{\rm S}_{\rm xp} \right]   
\end{eqnarray}
\begin{eqnarray}
\label{eq:pnmeasS}
P_{\rm n}^{\star\rm S} &=& {T_{\rm A}(\theta,\phi)^{{\rm S}} \over \max\left[T_{\rm A}(\theta,\phi)^{\rm S}\right]} \nonumber \\
&\propto& P_{\rm n}(\theta,\phi)^{\rm S}_{\rm cp} + P_{\rm n}(\theta,\phi)^{\rm S}_{\rm xp} +  \\ \nonumber
&+&\chi^{{\rm OMT}} \cdot \left[   P_{\rm n}(\theta,\phi)^{\rm M}_{\rm cp} + P_{\rm n}(\theta,\phi)^{\rm M}_{\rm xp}  \right]  
\end{eqnarray}
The equations \ref{eq:pnmeasM} and \ref{eq:pnmeasS} are valid under the hypothesis that, for each Radiometer Chain Assembly (RCA), the maximum of the co-polar level of the M-beam equals the maximum of the co-polar level of the correspondant S-beam. This hypothesis is true because any difference is absorbed by the calibration.  
According to \cite{kraus1986radio} 
the observed temperature along the line of sight $(\theta_0,\phi_0)$ is 

\begin{equation}
\label{eq:kraus}
T(\theta_0,\phi_0) = {1\over\Omega_A} \int{T_{\rm s}(\theta-\theta_0,\phi-\phi_0)\cdot P_{\rm n}(\theta,\phi)~d\Omega}
\end{equation}
If the source is point-like, $T_s = T_s\cdot\delta(\theta-\theta_0,\phi-\phi_0)$ so that the 
recorded temperature as a function of $(\theta_0,\phi_0)$ is proportional to the the antenna radiation pattern: 
\begin{equation}
T(\theta_0,\phi_0)  \propto  T_{\rm s}\cdot P_{\rm n}(\theta_0,\phi_0)
\end{equation}
Depending on the polarization of the source, the telescope response may be different. For unpolarized source the response is derived directly form equation \ref{eq:pnmeasM} and \ref{eq:pnmeasS} and can be written as: 
\begin{equation}
\label{eq:pnMunpol}
P_{\rm n}^{\rm M}(\theta,\phi)  = P_{\rm n}(\theta,\phi)_{\rm cp}^{\rm M}  + P_{\rm n}(\theta,\phi)_{\rm xp}^{\rm M}  + \chi^{{\rm OMT}} \cdot \left [ P_{\rm n}(\theta,\phi)_{\rm cp}^{\rm S} + P_{\rm n}(\theta,\phi)_{\rm xp}^S\right ] 
\end{equation}

\begin{equation}
P_{\rm n}^{\rm S}(\theta,\phi)  = P_{\rm n}(\theta,\phi)_{\rm cp}^{\rm S}  + P_{\rm n}(\theta,\phi)_{\rm xp}^{\rm S}  + \chi^{{\rm OMT}} \cdot \left [ P_{\rm n}(\theta,\phi)_{\rm cp}^{\rm M} + P_{\rm n}(\theta,\phi)_{\rm xp}^{\rm M}\right ]
\end{equation}
If the source is assumed to be completely linearly polarized along the polarization of the M-radiometer, the response of the telescope is then as follows: 
\begin{equation}
\label{eq:pnMpol}
P_{\rm n}^{\rm M}(\theta,\phi) = P_{\rm n}(\theta,\phi)_{\rm cp}^{\rm M}  + \chi^{{\rm OMT}} \cdot \left [ P_{\rm n}(\theta,\phi)_{\rm xp}^S\right ]
\end{equation}

\begin{equation}
P_{\rm n}^{\rm S}(\theta,\phi)  =  P_{\rm n}(\theta,\phi)_{\rm xp}^{\rm S}  + \chi^{{\rm OMT}} \cdot \left [ P_{\rm n}(\theta,\phi)_{\rm cp}^{\rm M} \right ]
\end{equation}
The response of the telescope to a partially polarized source is a combination of the two responses.  We define the source temperature as the sum of unpolarized ($T_{\rm s}^{\rm U}$) and M-aligned polarized ($T_{\rm s}^{\rm M}$) contributions, so that:  
\begin{equation}
\label{eq:tsum}
T_{\rm s}(\theta,\phi) = T_{\rm s}^{\rm U}(\theta,\phi) + T_{\rm s}^{\rm M}(\theta,\phi) 
\end{equation}
The observed temperature is obtained inserting equation \ref{eq:pnMunpol}, \ref{eq:pnMpol}, and \ref{eq:tsum} in equation \ref{eq:kraus}.  Taking into account the polarization coupling factor, that is $1/2$ for coupling with unpolarized signal and $1$ for coupling with polarized signal, we have: 
\begin{eqnarray}
T(\theta_0,\phi_0) &\propto& {1\over 2} \cdot T_{\rm s}^U \cdot  \nonumber\\
&\cdot& \left\{ P_n(\theta_0,\phi_0)_{\rm cp}^{\rm M}  + P_n(\theta_0,\phi_0)_{\rm xp}^{\rm M} +  \chi^{{\rm OMT}} \cdot  P_n(\theta_0,\phi_0)_{\rm cp}^S \right\} +\nonumber\\
&+& T_{\rm s}^{\rm M} \cdot P_n(\theta_0,\phi_0)_{\rm cp}^{\rm M}
\label{eq:measurement}
\end{eqnarray}
The term $\chi^{{\rm OMT}} \cdot \left [ P_n(\theta,\phi)_{xp}^S\right]$ is negligible, being a systematic 2nd order effect. Equation \ref{eq:measurement} can be rewritten as follows: 
\begin{eqnarray}
T(\theta_0,\phi_0) &\propto& 
\left (T_{\rm s}^{\rm M} + {1\over 2} T_{\rm s}^U\right ) \cdot \nonumber \\
&\cdot& \left\{ P_n(\theta_0,\phi_0)_{\rm cp}^{\rm M}  + P_n(\theta_0,\phi_0)_{\rm xp}^{\rm M} + \chi^{{\rm OMT}} \cdot  P_n(\theta_0,\phi_0)_{\rm cp}^S \right\} +  \nonumber \\
&-& T_{\rm s}^{\rm M} \cdot \left\{ P_n(\theta_0,\phi_0)_{\rm xp}^{\rm M}  +\chi^{{\rm OMT}} \cdot  P_n(\theta_0,\phi_0)_{\rm cp}^S \right\}
\end{eqnarray}
The measured normalized pattern for the observation of a partially polarized point-like source is then: 
\begin{eqnarray}
\label{eq:pnMgeneral}
P^\star(\theta_0,\phi_0&)& = \\  
&=& \left\{ P_n(\theta_0,\phi_0)_{\rm cp}^{\rm M}  + P_n(\theta_0,\phi_0)_{\rm xp}^{\rm M} + \chi^{{\rm OMT}} \cdot  P_n(\theta_0,\phi_0)_{\rm cp}^S \right\} +  \nonumber \\
&-& {T_{\rm s}^{\rm M} \over \left (T_{\rm s}^{\rm M} + {1\over 2} T_{\rm s}^U \right ) } \cdot \left\{ P_n(\theta_0,\phi_0)_{\rm xp}^{\rm M}  +\chi^{{\rm OMT}} \cdot  P_n(\theta_0,\phi_0)_{\rm cp}^S \right\}\nonumber
\end{eqnarray}
The comparison between eq. \ref{eq:pnMgeneral} and eq. \ref{eq:pnMunpol} gives 
the error in the pattern determination which is caused when treating a partially-polarized source as fully unpolarized; the error, $\epsilon(\theta_0,\phi_0)$ is dependent on the beam region and polarization factor $\wp = T_{\rm s}^{\rm M}/T_{\rm s}^{\rm U}$
\begin{equation}
\epsilon({\theta_0,\phi_0}) ={\wp\over\left (\wp + {1\over 2} \right)} \cdot \left\{ P_n(\theta_0,\phi_0)_{\rm xp}^{\rm M}  +\chi^{{\rm OMT}} \cdot  P_n(\theta_0,\phi_0)_{\rm cp}^S \right\}
\end{equation}
As expected the error is maximum when the source is totally polarized, so that ${\wp \over \wp + {1\over 2}} = 1$ for which the error is just the sum of the M-side cosspolar response and the S-side co-polar response multiplied by the OMT isolation. For Planck LFI these are both below the noise level. For a source with 1\% of polarization, a $\chi^{{\rm OMT}} = -25~$dB, a cross-polar peak $-25~$dB with respect to the co-polar peak, the effect is at level of about $-40$ dB, well below the level of beam measurements. 

\section{Timeline-to-map Monte Carlo simulations}

Signal-only timeline-to-map Monte Carlo (MC) simulations are produced using Level-S \citep{reinecke2006} and Healpix \citep{gorski2005} subroutines, and the {\tt Madam} map-maker \citep{kurki-suonio2009, keihanen2010} on the \emph{Louhi} supercomputer at \emph{CSC-IT Center for Science} in Finland. 

Starting from a fiducial CMB power spectrum with $\ell_\mathrm{max}=3000$, we used {\tt syn\_alm\_cxx} to generate a set of sky $a_{\ell m}$ realizations of this $C_\ell$.  
Starting from the simulated scanning beams we calculated their beam $a_{\ell m}$ using {\tt beam2alm} with {\tt beam\_lmax} = 5400 and {\tt beam\_mmax} = 14.  
The sky $a_{\ell m}$ were convolved with the beam $a_{\ell m}$ using {\tt conviqt\_v3} (\cite{prezeau2010}) with {\tt conv\_lmax} = 3000, {\tt lmax\_out} = 3000, {\tt beammmax} = 14. The output is a ``ringset'' table for each realization, i.e., a grid of observed sky signal values for 6001 values of $\phi$, 3002 values of $\theta$ and for 29 beam orientations ($\psi$).  Note that the main beams do not collect the full power of the signal, since a small part of the signal spills outside the main beam to form sidelobes. In this MC just the main beam was simulated, not the sidelobes, so the calculated signal values were missing that part of the power that goes to the sidelobes. This was taken into account at the map-making stage.

The CMB timelines for each realization were produced with {\tt multimod}.  The detector pointing ($\phi$, $\theta$, $\psi$) for each radiometer was reconstructed internally using satellite pointing information and the focal plane geometry. The observed CMB signal for each sample was interpolated in $\theta$ and $\phi$ from the ``ringset'' table for its pointing using {\tt interpol\_order} = 9 (the effect of beam orientation $\psi$ is solved exactly by {\tt multimod} for the {\tt beammmax} representation of the beam) and output to CMB timeline files.  The detector pointing reconstructed by {\tt multimod} was also output to disk for the use of the {\tt Madam} map-maker.  The reconstructed pointing was compared to 
the pointing used at the LFI DPC for the flight data maps to check that their agreement was satisfactory.  There is an option in {\tt multimod}, ``sampler'', to simulate the scanning motion of the radiometer during measurement of one observation sample.  This, however, increases the computational cost so much that we turned this sampler off, and simulated this scanning motion by using the scanning beams (see above) instead of the optical beams.

Maps were made from these CMB timelines with {\tt Madam} using the reconstructed pointing.  The same {\tt Madam} parameter settings were used as for the flight maps, see \cite{planck2011-1.6} and \citet{planck2013-p02}. As already discussed in Sect.~\ref{window_function}, the calibration step was not simulated.

\label{appendix}

\allearlypapers

\bibliographystyle{aa}
\bibliography{Planck_bib,LFI_beams_bib,conviqt_bib}
\raggedright 
\end{document}